\documentclass[sigconf, 10pt]{acmart}
\renewcommand\footnotetextcopyrightpermission[1]{}
\renewcommand\footnotetextcopyrightpermission[1]{}

\setcopyright{none}

\acmConference[Conference'17]{}{xx}{xx}

\usepackage{algorithm}
\usepackage[noend]{algpseudocode}

\usepackage{graphicx}
\usepackage{subfigure}
\usepackage{subcaption}
\usepackage{textcomp}
\usepackage{tabularx}
\usepackage{tabu}
\usepackage{longtable}
\usepackage{enumitem}
\usepackage{makecell}
\usepackage{multirow}
\usepackage{xspace}
\usepackage{url}

\usepackage{booktabs}
\usepackage{graphicx}

\usepackage{balance}

\makeatletter
\def\do@url@hyp{\do\-\do\_}
\makeatother

\newcommand{\PHM}[1]{\noindent\textbf{#1}} 

\newcommand{\SysName}{\texttt{$\lambda$Scale}\xspace}
\newcommand{\AlgoName}{\texttt{$\lambda$Pipe}\xspace}

\newcommand{\revise}[1]{{#1}}

\newif{\ifSubmit}
\ifSubmit
\newcommand{\ruicomment}[1]{}
\else
\newcommand{\ruicomment}[1]{\noindent\textcolor{teal}{Rui: #1}}
\fi

\usepackage[font=bf]{caption}

\graphicspath{{figures/}}

\begin{document}

\title{\SysName: Enabling Fast Scaling for Serverless Large Language Model Inference}

\author{
    \normalsize
    \textrm{Minchen Yu$^{\dag}$\textsuperscript{*}, Rui Yang$^{\ddag}$\textsuperscript{*}, Chaobo Jia$^{\dag}$, Zhaoyuan Su$^{\ddag}$, Sheng Yao$^{\S}$, Tingfeng Lan$^{\ddag}$, Yuchen Yang$^{\S}$, \\ Zirui Wang$^{\ddag}$, Yue Cheng$^{\ddag}$,  Wei Wang$^{\S}$, Ao Wang$^{\diamond}$, and Ruichuan Chen$^{\triangle}$}\\
	$^{\dag}$The Chinese University of Hong Kong, Shenzhen \quad $^{\ddag}$University of Virginia  \\\quad $^{\S}$Hong Kong University of Science and Technology  \quad $^{\diamond}$Alibaba Group   \quad $^{\triangle}$Nokia Bell Labs 
}

\begin{abstract}
Serverless computing has emerged as a compelling solution for cloud-based model inference. 
However, as modern large language models (LLMs) continue to grow in size, existing serverless platforms often face substantial model startup overhead.
This poses a significant challenge in efficiently scaling model instances to accommodate dynamic, bursty workloads commonly observed in real-world inference services.
In this paper, we introduce \SysName, an efficient serverless inference system to achieve fast model scaling.
The key idea behind \SysName is to leverage high-speed RDMA networks between GPU nodes for fast model multicast, while enabling distributed inference execution during model transmission---referred to as ``execute-while-load''.
\SysName proposes an efficient model scaling scheme, \AlgoName, which supports adaptive model multicast and dynamically constructs execution pipelines across receiving nodes for collaborative, distributed inference.
Additionally, \SysName supports efficient model management across GPU and host memory, allowing fast scaling for models across different storage tiers. 
Evaluation results show that \SysName enables fast model scaling and effectively handles load spikes, achieving up to 5$\times$ tail-latency improvement and 31.3\% cost reduction compared to state-of-the-art solutions on real-world LLM inference traces. 

\end{abstract}
\maketitle
\renewcommand{\thefootnote}{*}
\footnotetext{Co-first authors.}
\renewcommand{\thefootnote}{\arabic{footnote}} 

\section{Introduction}
\label{sec:intro}



Recent advancements in machine learning (ML) have fueled a surging demand for cloud-based ML inference
services~\cite{zhang_shepherd_nodate,zhang_mark:_2019,shen_nexus_2019,choi_serving_2022,gujarati_serving_2020}. Serverless computing offers a compelling cloud model for inference
serving that can effectively handle dynamic request patterns~\cite{ali_fc,yang_infless_2022, yu_gillis_icdcs, ali_batch_nodate}. 
In this approach, users simply publish ML models, expose service endpoints, and delegate resource provisioning and scaling responsibilities to the cloud platform. 
Serverless computing is also economically appealing as users pay only for actual resource usage, eliminating the resource idling cost.

However, current serverless inference platforms suffer from significant cold-start problems, especially as modern models grow increasingly large and resource-intensive (e.g., large language models, or LLMs).
The cold-start process, involving remote model loading and initialization, can result in long startup delays up to several minutes~\cite{fu_serverlessllm_2024}. 
Such high latencies prevent platforms from rapidly scaling out 
to handle highly dynamic workloads, making it impractical to support online inference services with stringent latency requirements on the order of milliseconds or seconds~\cite{fu_serverlessllm_2024,zhang_mark:_2019,zhong_distserve_2024,agrawal_taming_2024}.


To mitigate cold starts, existing platforms typically cache
inference models in local GPUs or host memory during idle periods and resume execution upon request arrivals~\cite{yang_infless_2022,ali_fc, bai_pipeswitch_nodate,faaswap,fu_serverlessllm_2024}. 
However, these approaches have limited scalability and incur high resource costs. In multi-tenant serverless environments, workers typically need to handle multiple large models simultaneously, but storing all these models locally is infeasible and/or cost prohibitive due to limited GPU and host memory capacity. 
While recent works explore the use of local SSDs to expand storage capacity~\cite{fu_serverlessllm_2024}, the data transfer between SSDs and GPUs introduces significant delays under dynamic workloads. As a result, platforms either suffer from unacceptable startup costs or resort to overprovisioning GPU workers to maintain active model replicas at all times, leading to high resource costs.


An efficient serverless inference platform should not settle for the fundamental tradeoff between startup latency and resource overprovisioning. 
Ideally, it should scale out rapidly in response to load spikes without incurring additional resource costs. 
\emph{Fast serverless model scaling is achievable through two key insights.}
First, modern GPU clusters employ high-speed network interconnects (e.g., 400Gbps with RDMA capability)~\cite{acme_nsdi24, kundu_llm_analysis_arxiv24, azure_ai_cluster_url}, providing opportunities for efficient model multicast and fast scaling. 
Second, model inference can begin before a node receives all model parameters, enabling collaborative, distributed inference execution across multiple nodes during model loading. 
This ``execute-while-load'' approach, combined with low-latency model loading, substantially improves system scalability under spikes. 


Following these insights, we propose \SysName, a scalable serverless inference platform that delivers fast and resource-efficient model scaling. 
\SysName enables distributed inference execution during cross-node model loading (i.e., ``execute-while-load''), leveraging high-speed RDMA networks and the GPUDirect RDMA (GDR)~\cite{gdr}.
However, realizing this approach poses both algorithmic and system-level challenges.

A key challenge for \SysName is to develop efficient solutions for cross-node model loading and distributed execution.
\SysName requires to accommodate varying model scaling demands and enable collaborative, distributed model execution in a dynamic environment where the model is being loaded.
Existing distributed inference solutions do not meet such requirements, which typically lack support for dynamic, on-demand model scaling and rely on static resource allocations for distributed model execution~\cite{li_alpaserve_2023,zhang_shepherd_nodate,yang_infless_2022}.
To address this challenge, \SysName introduces an abstraction of an \emph{execution pipeline} and a novel scheme, \emph{Dynamic Pipelined Execution with Multicast} (\AlgoName).
This scheme partitions the model into fine-grained blocks for efficient multicasting while dynamically constructing execution pipelines --- complete model instances spanning multiple receiving nodes for collaborative, distributed inference execution.


\AlgoName comprises three key designs. 
\revise{First, \AlgoName designs an adaptive model-multicast mechanism based on 
the binomial pipeline multicast algorithm originally described in~\cite{rdmc, binomial-pipe}.}
This mechanism optimizes the granularity and transfer order of the model blocks to enable the rapid construction of execution pipelines, thus improving overall inference performance (referred to as \textbf{\emph{Adaptive Model Multicast}}).
Second, \AlgoName judiciously groups nodes into execution pipelines at runtime to improve overall resource efficiency while minimizing data transfers during distributed inference.
This strategy effectively improves system throughput and reduces request queueing under load spikes (referred to as \textbf{\emph{Pipelined Inference Execution}}). 
Furthermore, once model loading is complete, \AlgoName allows workers to seamlessly switch to local execution mode without incurring cross-node overheads (referred to as \textbf{\emph{Mode Switching}}).

In addition to \AlgoName, \SysName must efficiently manage models cached across various media (e.g., GPU memory and host memory) to facilitate rapid model scaling from different storage tiers.
To address this challenge, \SysName proposes two key system designs.
First, \SysName employs a holistic model startup mechanism that adapts to varying model locality. It also enables model instances stored in both GPU and memory to collectively improve scaling performance (referred to as \textbf{\emph{Locality-driven Model Startup}})
Second, \SysName implements a memory management system that seamlessly consolidates data within model blocks for efficient model transmission and supports GPU memory pre-allocation for improved efficiency (referred to as \textbf{\emph{Efficient Memory Management}}). 
\revise{We have implemented \SysName by partially leveraging and extending some open-source projects such as Derecho~\cite{dere-github} and Meta's Llama inference framework~\cite{metainf}.} 
We evaluated \SysName against state-of-the-art solutions, including \textit{ServerlessLLM}~\cite{fu_serverlessllm_2024}, \textit{FaaSNet}~\cite{wang_faasnet_nodate}, and \textit{NCCL}~\cite{nccl}.
\SysName efficiently handles load spikes, leading to a 2.4$\times$ to 5$\times$ improvement in 90$^{th}$ tail latency while reducing resource costs by 17.8\% to 31.3\% compared to the baselines on real-world LLM inference traces~\cite{burstGPT_arxiv24}.  
Microbenchmark results further show that \SysName delivers fast model transmission and completes the scaling of Llama-13B across 8 nodes in less than 1 second, outperforming \textit{NCCL} by up to 1.5$\times$. 

\section{Background and Motivation}
\label{sec:background}

\subsection{LLM Inference}
\label{sec:background-llm-inference}
LLM inference autoregressively generates text output token by token from a user input (prompt) until it reaches the end-of-sentence (EoS) token or the maximum token limit. Each token depends on prior context, causing response times to vary. To speed up token generation, the model typically uses a KV cache~\cite{vllm} to cache the context from previous computations, reducing redundant work. Tokens are streamed to the user as they are produced, enabling real-time interaction. Performance is typically measured by \emph{time-to-first-token latency} or TTFT (time to generate the first token) and \emph{tokens-per-second} or TPS (token generation throughput).  


\subsection{Serverless Inference and Requirements}
\label{sec:background-serverless}

In recent years, serverless computing has emerged as a compelling option for hosting ML inference services---often termed ``serverless inference''~\cite
{yang_infless_2022,yu_gillis_icdcs,zhang_mark:_2019, ali2022optimizing_serverless_inference, ali_batch_nodate, hong2024optimus, fu_serverlessllm_2024, romero_infaas_nodate, wang_faasnet_nodate}.
This approach allows users to simply publish models with inference code as functions, while cloud providers automatically manage resource provisioning, auto-scaling, scheduling, and fault tolerance. 
Serverless inference platforms offer fine-grained, pay-per-use billing~\cite
{aws_lambda,azurefunc,alibaba_serverless_gpu}, enabling substantial cost savings given the dynamic nature of inference workloads~\cite
{shen_nexus_2019,zhang_shepherd_nodate,gujarati_serving_2020,han_microsecond-scale_2022,lee_pretzel,kosaian_parity_2019,romero_infaas_nodate,
crankshaw_clipper,choi_serving_2022}.  
These advantages have led to a growing trend of serverless-based LLM inference services~\cite{fu_serverlessllm_2024,hf_serverless,together_ai}, where end-users simply query a backend LLM and receive its output tokens as responses. 

\begin{figure}
\vspace{-8pt} 
    \centering
    \includegraphics[scale=0.125]{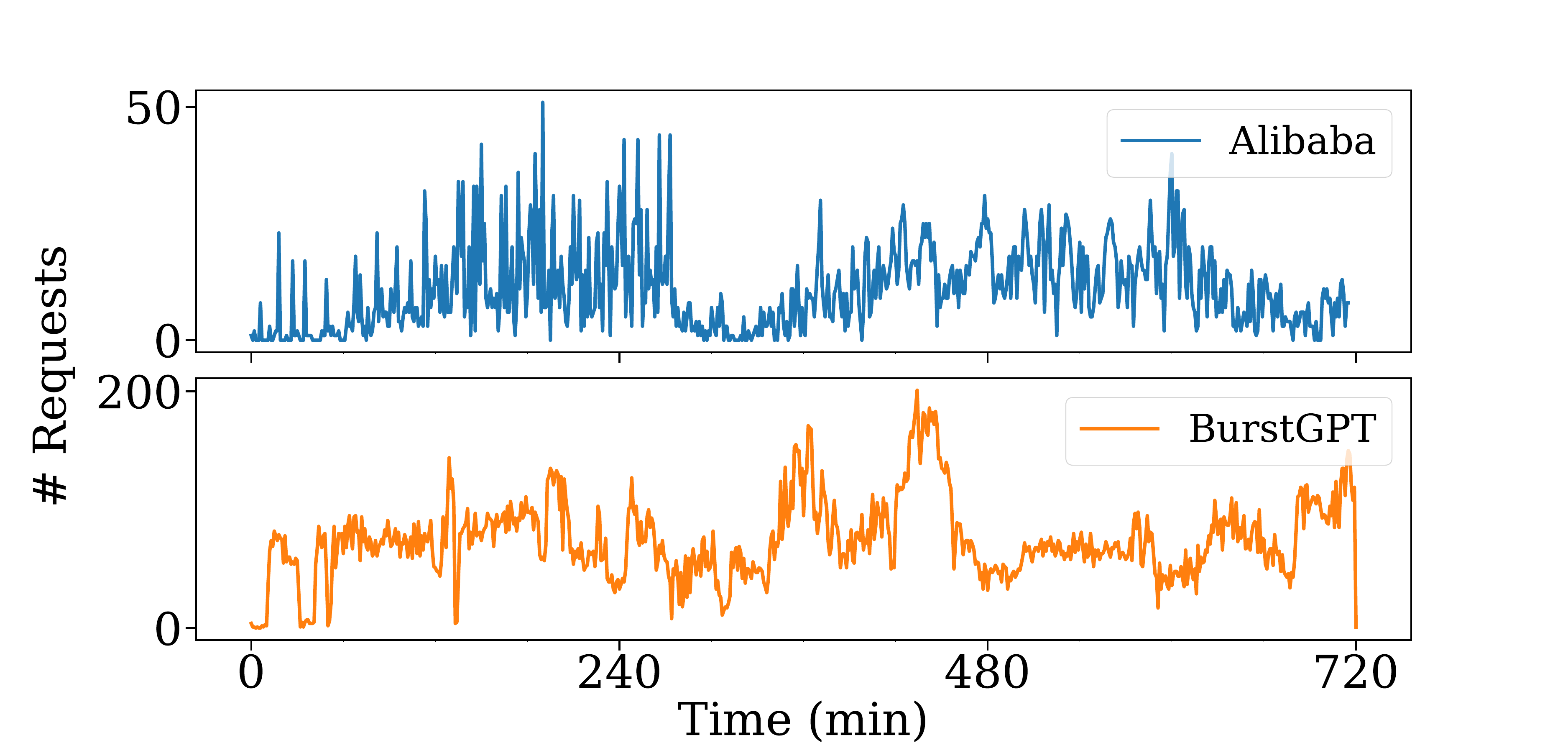}
    \vspace{-10pt}
    \caption{Normalized request rates of two representative serverless inference services.
    \textit{\textmd{\texttt{Trace 1} (top): a 12-hour serverless inference trace collected from Alibaba Cloud. \texttt{Trace 2} (bottom): a 12-hour trace from a real-world LLM workload~\cite{burstGPT_arxiv24}.}} 
    }
    \label{fig:serverless_trace}
    \vspace{-5pt}
\end{figure}

Key to serverless inference is to enable rapid scaling of model serving instances to effectively handle real-world dynamic workloads.
Fig.~\ref{fig:serverless_trace} shows 12-hour normalized request rates of two representative LLM inference workloads from two production serverless platforms: a serverless inference service at Alibaba Cloud~\cite{alibaba_serverless_gpu} and a regional Azure OpenAI GPT service~\cite{burstGPT_arxiv24}.
These traces reveal highly bursty request arrival patterns, with request rates surging by more than one order of magnitude within just a few minutes. 
Such bursty requests can overwhelm existing serving instances, leading to violations of the latency requirements defined by the Service-Level Objectives (SLOs) of inference services. 
Therefore, serverless inference platforms must rapidly scale out model serving instances to accommodate load spikes, ensuring low inference latency and SLO attainment. 
We detail the request serving results under these traces in \S\ref{sec:background-limitations}.

\subsection{Inefficiency of Existing Solutions}
\label{sec:background-limitations}

We next discuss existing serverless inference platforms and their issues in achieving fast model scaling.

\PHM{Solution\#1: Loading remote models.}
In serverless inference platforms, a typical scaling process involves retrieving models from remote model registries (e.g., Hugging Face) or storage services (e.g., S3) to GPU nodes, which then launch the model serving instances. 
As models continue to grow in size---particularly LLMs with tens or hundreds of billions of parameters---fetching them across the network becomes increasingly time-consuming. 
For instance, it takes over 18 minutes to load a Llama-70B model (140~GB) over a 1~Gbps network; and the overhead is further exacerbated by parallel model loading due to network bandwidth contention and registry throttling~\cite{wang_faasnet_nodate}.
Worse, initializing and loading large models within GPU nodes can take anywhere from tens of seconds to minutes to complete~\cite{faaswap,fu_serverlessllm_2024}. 
These significant startup overheads are incompatible with the requirements of real-time inference services, making the remote loading approach inefficient. 

\PHM{Solution\#2: Overprovisioning GPUs.}
To mitigate the cold-start problem, existing serverless inference solutions opt to maintain a sufficient number of active function instances with reserved GPUs, even when these instances are idling~\cite{aws_provisioned,yang_infless_2022,fc_billing}. 
This approach, however, leads to significant resource waste due to the dynamic nature of inference workloads. 
In particular, eliminating cold starts requires the platform to provision excessive GPU resources to accommodate peak loads (e.g., spikes in Fig.~\ref{fig:serverless_trace}), which in turn results in substantial GPU idling during periods of low demand. Consequently, the extremely low GPU utilization directly contradicts the pay-per-use model fundamental to serverless computing\cite{faaswap}.  

\PHM{Solution\#3: Caching models in host memory and SSDs.}
Recent studies propose to cache models in host memory and then load them into GPU upon request arrivals~\cite{bai_pipeswitch_nodate,deepplan_2023,fu_serverlessllm_2024,faaswap}. 
While this approach reduces the reservation cost compared to GPU overprovisioning, it still fails to achieve fast scaling in large-scale, multi-tenant serverless GPU clusters due to three key factors.
1) The dynamic and bursty nature of inference workloads often requires concurrent executions of a model on many nodes, sometimes involving the entire cluster. 
2) Large-scale platforms typically host thousands of large models, resulting in massive host memory consumption. 
3) Each node has limited host memory (e.g., up to 100s of GBs), which is often insufficient to accommodate even a few large models. 
Consequently, excessive host memory model caching becomes impractical, hindering the ability to achieve efficient model scaling. 

Compared to host memory, SSDs offer larger capacities and can store more models~\cite{fu_serverlessllm_2024}.
However, loading models from SSDs to GPUs remains time-consuming due to the limited SSD bandwidth. 
Our measurements (see testbed in \S\ref{sec:evaluation:setup}) show that loading a Llama-70B model from an SSD to a GPU takes over 30 seconds even with optimized implementations --- an order of magnitude slower than loading from host memory. 
This delayed loading hampers the ability of serverless inference platforms to scale models quickly on demand. 

\begin{figure}[t]
    \centering
    \begin{minipage}{0.225\textwidth}
        \centering
        \includegraphics[scale=0.125]{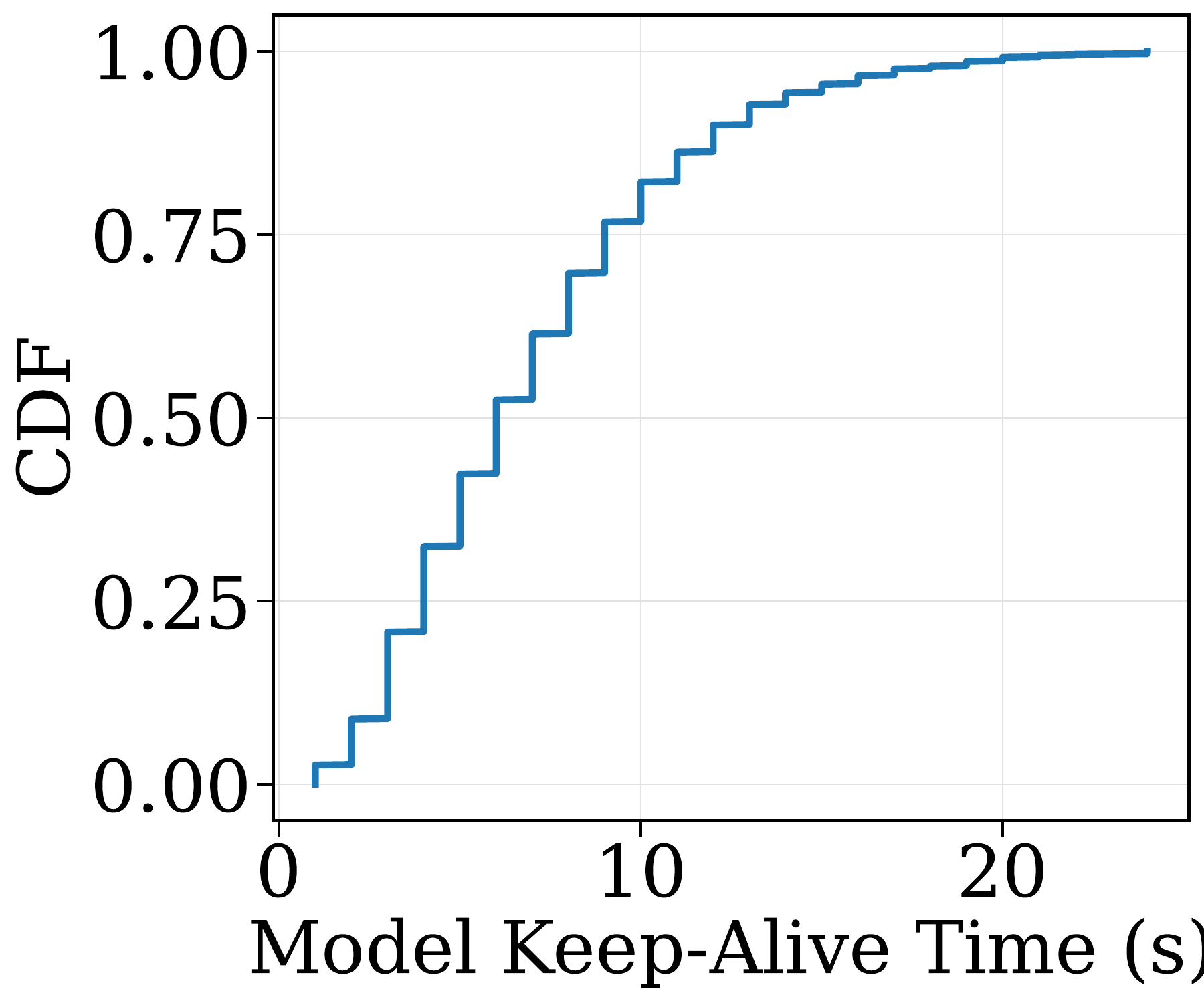} 
        \vspace{-1em}
        \caption{Distribution of models' keep-alive time in memory.}  
        \label{fig:memory_alive_cdf}
    \end{minipage}\hfill
    \begin{minipage}{0.225\textwidth}
        \centering
        \includegraphics[scale=0.125]{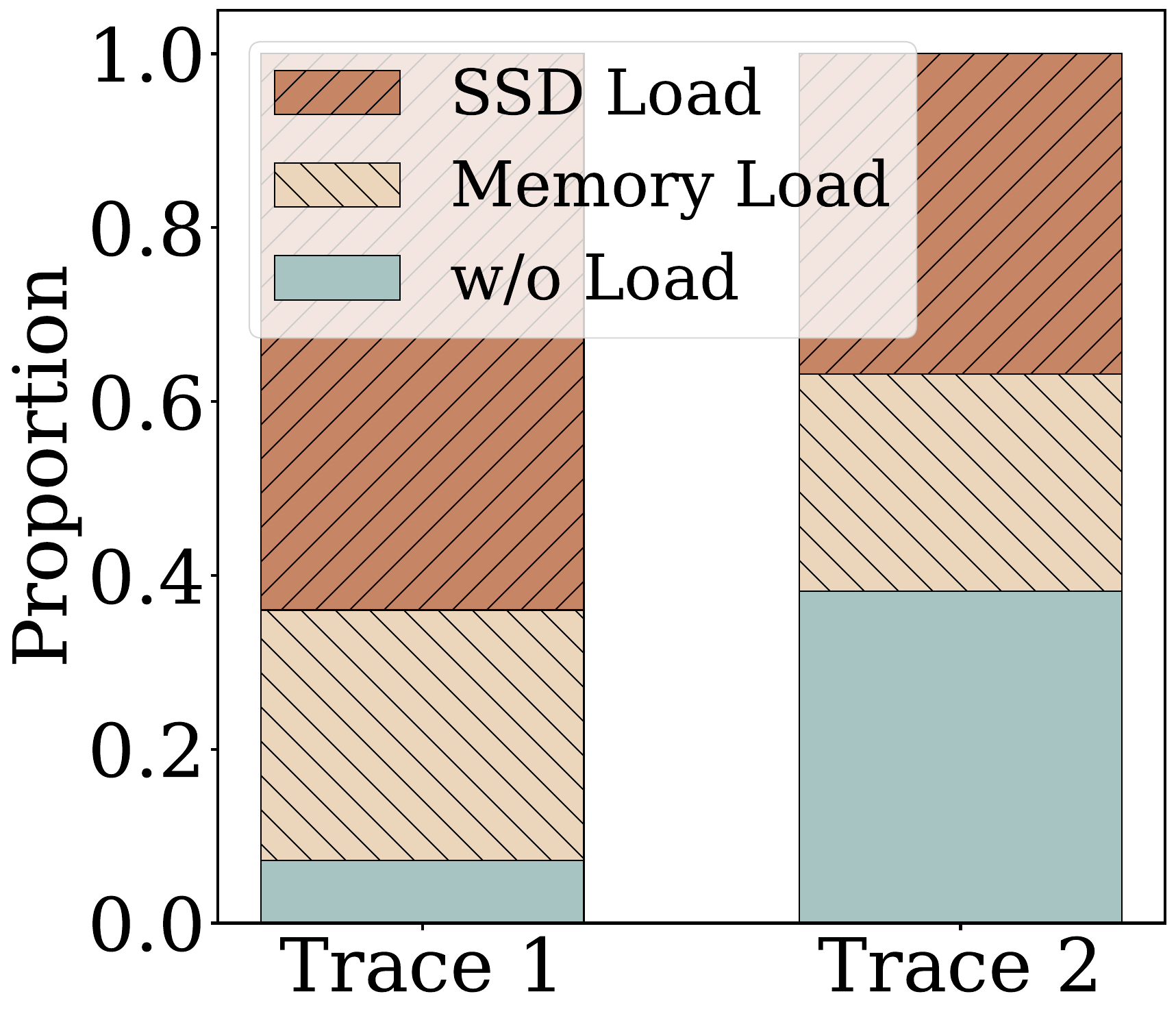}
        \vspace{-1em}
        \caption{Proportion of the 3 types of model loading.}
        \label{fig:load_prop}
    \end{minipage}
    \vspace{-2em}
\end{figure}

To illustrate the inefficiency of this approach, we conduct two simulations based on real-world workload traces.

\PHM{Short model keep-alive time in memory.} 
We first examine how long a model instance remains in memory before being evicted in a multi-tenant inference platform. 
Each node's memory is configured to hold up to 3 models, while 12 models are stored in SSDs. This configuration is based on the size of Llama-70B and the hardware specifications of our testbed (see \S\ref{sec:evaluation:setup}). 
We set each model's per-node request rate to 1 per minute, which reflects a typical request pattern in production serverless inference platforms~\cite{faaswap}. 
We use the LRU eviction policy and depict the distribution of model's keep-alive times in Fig.~\ref{fig:memory_alive_cdf}.
As we can see, models are frequently reloaded and evicted from memory, with over 95\% of them staying in memory for fewer than 15 seconds before being evicted. 

\PHM{High cache miss ratio.}
Next, we measure the cache miss ratio for models cached in memory 
by replaying the two traces shown in Fig.~\ref{fig:serverless_trace}.
Based on the findings of Fig.~\ref{fig:memory_alive_cdf}, we set models' keep-alive time to 15 seconds---the tail of the distribution.
Fig.~\ref{fig:load_prop} shows the proportion of three loading cases across the two traces: model load from memory, model load from SSD, and a hot start (w/o load). 
We observe that SSD loads (i.e., cache misses) account for 64\% and 36\% across the two traces, respectively.
This indicates that relying on memory caching alone is inadequate, resulting in a significant portion of slow model loads from SSDs or even remote storage, which severely impacts scalability and user experience. 


In summary, existing solutions either suffer from long startup times or incur large GPU and memory costs in order to keep models active, forcing a rigid tradeoff between scaling efficiency and resource costs. 
This inherent limitation prevents the platform from achieving fast model scaling, a critical requirement for serverless inference.


\subsection{Key Insights and Challenges}
\label{sec:challenges}

\PHM{Key insights.}
To achieve fast model scaling, we present three key observations. 
\textbf{First}, modern GPU clusters increasingly adopt high-speed interconnects between GPU nodes (e.g., 100-400Gbps with RDMA capability)~\cite{acme_nsdi24, kundu_llm_analysis_arxiv24, azure_ai_cluster_url} and support advanced data transfer techniques such as GPUDirect RDMA or GDR~\cite{gdr}. 
\textbf{Second}, multicast-based collective communication techniques are inherently well-suited for cross-node model scaling, enabling rapid distribution of model instances across receiving nodes. 
\textbf{Third}, models can begin computation before being fully loaded, enabling an ``execute-while-load'' approach. This allows inference tasks to be distributed and executed across multiple nodes in parallel during model loading. 

Following these observations, we advocate that a scalable serverless inference platform should synergize the ``execute-while-load'' approach with efficient cross-node model loading. 
Specifically, models should be rapidly transferred and replicated across worker nodes, while new worker nodes perform collaborative, distributed inference execution. 
This approach allows the platform to handle load spikes on demand, maintaining low request queueing delays without the need to keep models long-lived in GPUs or host memory.  


\PHM{Challenges.}
Realizing the idea above presents three key challenges.
\textbf{\emph{C1: Low-Latency Model Loading.}}
The platform must handle varying scaling demands while consistently achieving high scaling performance.
Hence achieving efficient, low-latency model loading across nodes remains a challenge.
\textbf{\emph{C2: Dynamic Model Execution.}}
Unlike existing distributed inference solutions designed for static environments~\cite{zhang_shepherd_nodate,li_alpaserve_2023}, the ``execute-while-load'' approach requires collaborative, distributed inference among nodes when the model is being loaded.
This introduces a challenge for the platform to dynamically configure and optimize inference execution plans at runtime.
\textbf{\emph{C3: Efficient Model Management across Storage Tiers.}}
Maintaining models in host memory is a prevalent practice to accelerate model loading~\cite{faaswap,fu_serverlessllm_2024,bai_pipeswitch_nodate}.
Therefore, the platform should efficiently manage models across various storage tiers, including GPU memory and host memory, while delivering fast model scaling.

In the following sections, we present \SysName, a scalable serverless inference platform to address the aforementioned challenges.
We provide an overview of \SysName in \S\ref{sec:overview} and delve into its detailed designs in \S\ref{sec:algo} and \S\ref{sec:system}.

\section{\SysName Overview}
\label{sec:overview}

\PHM{Design choices.}
The key to achieving fast model scaling is efficient cross-node communication.
While collective communication libraries such as {\it NCCL}~\cite{nccl} are widely adopted in GPU clusters, they are typically optimized for long-running, static environments such as large-scale model training. 
For example, we find that {\it NCCL} lacks the flexibility to handle frequent reconfigurations of communication groups when scaling model-serving instances under dynamic workloads~\cite{nccl_cold_start_issue}, which introduces additional overhead that delays end-to-end model loading (see \S\ref{subsec:model_transfer_performance}).

Therefore, we opt to implement a lightweight, yet efficient, multicast framework in \SysName based on the binomial pipeline algorithm~\cite{binomial-pipe,rdmc} to achieve scalable model inference. 
This approach is well-suited to address the challenges outlined in \S\ref{sec:challenges}. 
Compared to other multicast solutions such as binary tree~\cite{wang_faasnet_nodate}, the binomial pipeline generally delivers superior multicast performance (see \S\ref{sec:evaluation}). 
By leveraging fast data transfer techniques like RDMA and GDR, \SysName can efficiently implement and optimize the binomial-pipeline-based multicast, achieving low-latency model loading (\textbf{\emph{C1}}). 
Additionally, this approach provides high flexibility to orchestrate fine-grained execution tasks at runtime and manage models across storage tiers, which facilitates dynamic model execution (\textbf{\emph{C2}}) and efficient model management (\textbf{\emph{C3}}).


\begin{figure}
    \centering
    \includegraphics[width=0.4\textwidth]{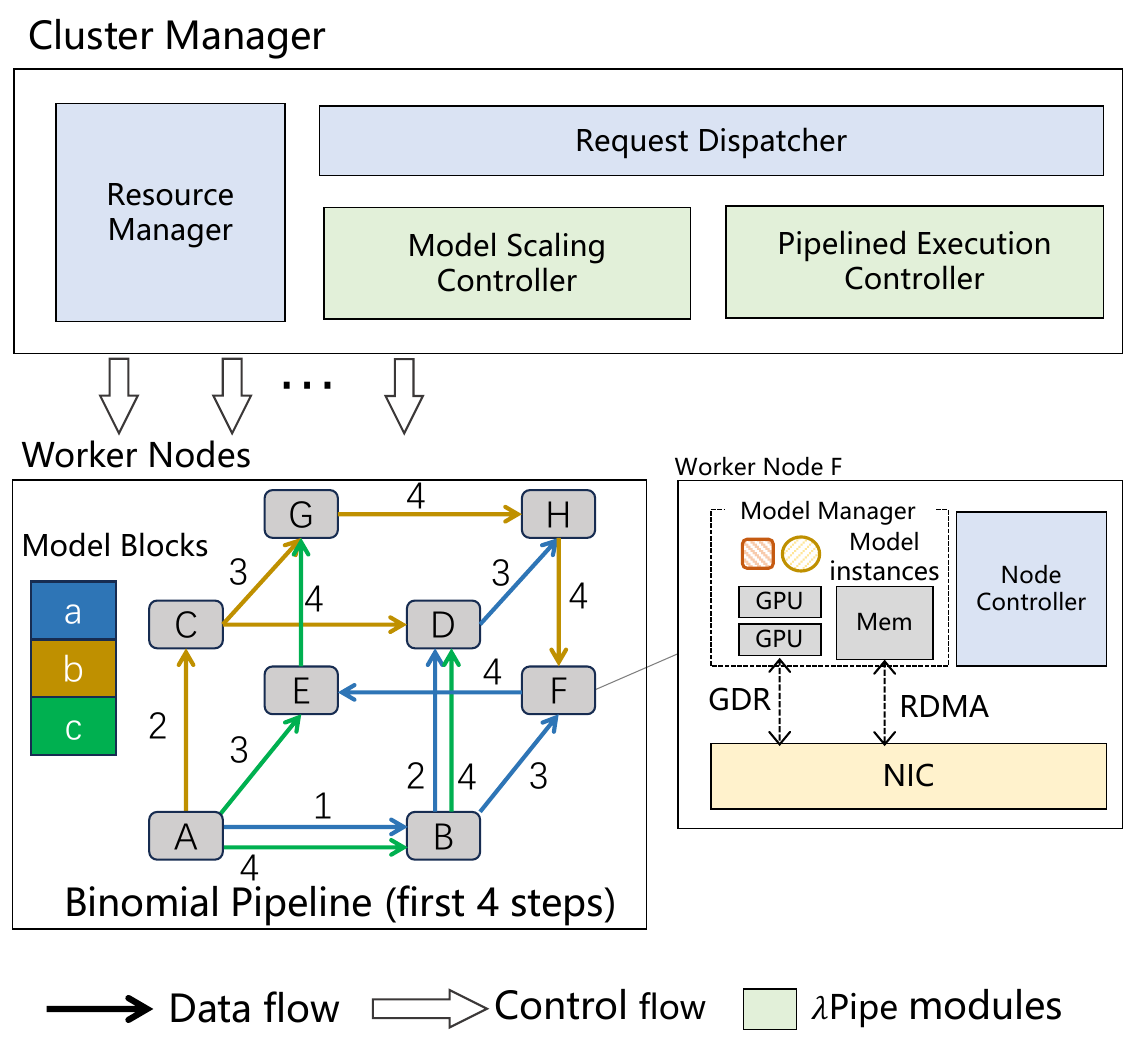} 
    \vspace{-10pt} 
    \caption{\SysName architecture overview. 
    \textit{\textmd{\revise{In this example, Node A initiates a binomial pipeline multicast per~\cite{rdmc, binomial-pipe} for a model partitioned into three model blocks. For a detailed hypercube binomial pipeline illustration, please refer to RDMC~\cite[Fig.~3]{rdmc}.}
    Each participating worker node transmits a model block in a sequential step (indicated by numbered labels along the data flow arrows). 
    A receiver node forwards the blocks it has received to its neighbors (e.g., Node B forwards block \textbf{a} at steps 2 and 3).
    The color-coded model blocks correspond to the data flow paths.}}
    }
    \label{fig:overview}
    \vspace{-10pt} 
\end{figure}

\PHM{Overview.}
Fig.~\ref{fig:overview} illustrates an architecture overview of \SysName. 
\SysName runs a cluster manager to dispatch end-user queries to worker nodes, manage global resources, and coordinate model scaling and pipeline execution.
\SysName implements an efficient model scaling scheme---\AlgoName---on the model scaling and pipeline execution controllers. 
Specifically, the model scaling controller coordinates fine-grained model distribution across participating nodes, in which a model is partitioned into blocks and transmitted using a binomial-pipeline-based approach.
In this approach, the nodes are organized into a hypercube communication topology~\footnote{This is achievable in real-world GPU clusters, which often use optimized network topologies such as fat trees~\cite{fattree} to ensure efficient communication.}, where each node transmits model blocks to its adjacent nodes.
\revise{The number of time steps required to complete block transmission is proven to be optimal (see both RDMC's system realization~\cite{rdmc} and Ganesan-Seshadri~\cite{binomial-pipe} for the optimality analysis).}
Additionally, the pipeline execution controller is responsible for distributing inference execution across nodes during model scaling.
It follows optimized block transfer orders to judiciously distribute block-level inference tasks among participating nodes to improve overall performance. (see details in \S\ref{sec:algo}).

Each worker node deploys user-provided models as model-serving instances and operates a node controller that synchronizes with the cluster manager, reports local status, and coordinates node-level tasks.
Additionally, \SysName runs a model manager at each node to track local resources such as GPUs and host memory and manage model instances.
The model manager is responsible for model execution and transmission tasks according to the instructions of the node controller.
It leverages GDR to efficiently exchange data across GPUs on different nodes, bypassing the data movement through the host to GPUs.
It also supports direct access to models stored in remote memory via RDMA.
These designs effectively improve network resource efficiency and distributed inference performance.

We next describe how \AlgoName achieves efficient model loading and dynamic model execution (\textbf{\emph{C1 and C2}}) in \S\ref{sec:algo}, and then discuss \SysName's model management (\textbf{\emph{C3}}) in \S\ref{sec:system}.


\section{\AlgoName Design}
\label{sec:algo}

\subsection{Design Rationale and Overview}

\AlgoName aims to efficiently support collaborative, distributed inference execution before nodes receive the entire model.
The key objective is to maximize the overall system throughput ---measured in tokens generated per second for LLM inference
---which in turn reduces request queueing under load spikes.
We therefore design \AlgoName based on \emph{two key principles}. 
1) Minimizing delays in assembling complete model instances across nodes enables distributed inference to begin as early as possible.
2) Reducing data movement (e.g., KV caches in LLMs) across nodes enhances overall efficiency.

\AlgoName introduces the abstraction of an \emph{execution pipeline} to facilitate distributed inference.
An execution pipeline serves as a model-serving instance spanning a group of nodes that collectively maintain a complete model and jointly perform pipeline parallelism (Fig.~\ref{fig:scale_example}).
Inference requests are assigned to specific execution pipelines, and the designated pipeline iteratively computes all output tokens (Fig.~\ref{fig:pipeline} (a)).
This minimizes the amount of intermediate results exchanged between nodes and eliminates the need to transfer the KV cache, ensuring efficient distributed execution.

\AlgoName provides efficient supports for execution pipelines through three key designs.
First, \AlgoName extends the binomial pipeline\revise{~\cite{rdmc, binomial-pipe}} to enable adaptive model multicast, supporting fast model distribution under various scaling scenarios and minimizing the time required to generate execution pipelines. 
Second, \AlgoName adopts an efficient strategy to generate execution pipelines and fully leverage the advantages of adaptive model multicast, which improves overall inference performance.
Finally, \AlgoName allows participating nodes to switch to local execution mode once they have received the full model replica.
In addition to cross-node scaling, execution pipelines can also be applied in memory-based model loading, which we describe in \S\ref{sec:system}.

\subsection{Adaptive Model Multicast}
\label{subsec:model_multicast}

\PHM{Multicast and sub-groups.}
We begin by modeling the multicast process in \AlgoName.
Consider a cluster of $N$ nodes and a scaling scenario where the source node holds a model instance and needs to distribute it to the remaining $N - 1$ destination nodes.
We denote this scaling process as $1 \rightarrow N$.
Assume the size of model is $M$ and it is partitioned into $b$ blocks for multicast. 
\revise{As shown in RDMC~\cite{rdmc} and Ganesan-Seshadri~\cite{binomial-pipe}, $1 \rightarrow N$ scaling completes in $b + \lceil \log N \rceil - 1$ steps when the model of size $M$ is partitioned into $b$ blocks. We refer readers to~\cite{rdmc, binomial-pipe} for full scheduling analysis.
Let $t$ be the time required for each step and $T$ be the total end-to-end multicast time.
We have $t \propto M/b$, i.e., in proportion to the amount of data exchanged between each pair of nodes in one step, and $T \propto M (1 + \log N / b) $.
}

More generally, we consider a scaling operation $k \rightarrow N$, i.e., $k$ nodes distributing the model to the remaining $N - k$ nodes (where $1 \le k < N$)~\footnote{$k \geq 1$ can be easily met in practice, such as by maintaining at least one model replica in host memory across the cluster.}.
\AlgoName evenly divides the $N$ nodes into $k$ sub-groups.
Each sub-group consists of $L$ nodes, where $L$ is either $\lfloor N/k \rfloor$ or $N\%k$, and each sub-group performs a $1 \rightarrow L$ scaling. 
This process supports $k \rightarrow N$ scaling for arbitrary values of $k$ and $N$.

\PHM{Selective block sizes.}
Determining the block size $b$ is critical to the performance of model multicast.
\AlgoName selectively configures this parameter to balance the transmission performance and execution efficiency.
We note that coarse-grained model partitioning (i.e., a small $b$) often results in longer end-to-end transmission times (i.e., a large $T$), while increasing $b$ generates more model blocks and leads to additional communication overhead for intermediate results during pipeline execution.
According to our modeling, $T$ exhibits an "elbow point" with respect to $b$: increasing the block size initially enhances transmission performance, but the benefits diminish beyond a certain threshold (detailed in Fig.~\ref{fig:ablation_num_blocks})
Therefore, \AlgoName configures this point as $b$, achieving good transmission performance while minimizing the additional overhead in pipeline execution.
Notably, configuring $b$ requires only offline profiling and has no impact on runtime performance.

\begin{algorithm}[tb]
    \caption{$k$-Way Transmission Strategy}
    \small
    \label{alg:k_way}
    \begin{algorithmic}[1]
    \Statex \textbf{Input:}
    \Statex \quad -- $k$ sub-groups $\{G_0, \dots, G_{k-1}\}$
    \Statex \quad -- $b$ ordered model blocks $\{M_0, \dots, M_{b-1}\}$
    \Statex \textbf{Output:}
    \Statex \quad -- Block transfer orders for $k$ sub-groups $\{O_0, \dots, O_{k-1}\}$
    
    \State $l \gets \lceil b / k \rceil$ \Comment{Size of block chunks}
    \State $\mathcal{S} \gets \{ \{M_j\ \mid j \in [l \cdot i , \min (l \cdot (i +1), b) - 1]\}  \mid i \in [0, k-1]\}$ \Comment{Partition blocks into $k$ chunks}

    \For{$i \in [0, k-1]$} \Comment{Generate $O_i$ via circular shift}
        \State $O_i \gets \biguplus_{j=0}^{k-1} \mathcal{S}_{(i+j) \bmod k}$
    \EndFor

    \end{algorithmic}
\end{algorithm}

\begin{figure}
    \centering
    \includegraphics[width=0.475\textwidth]{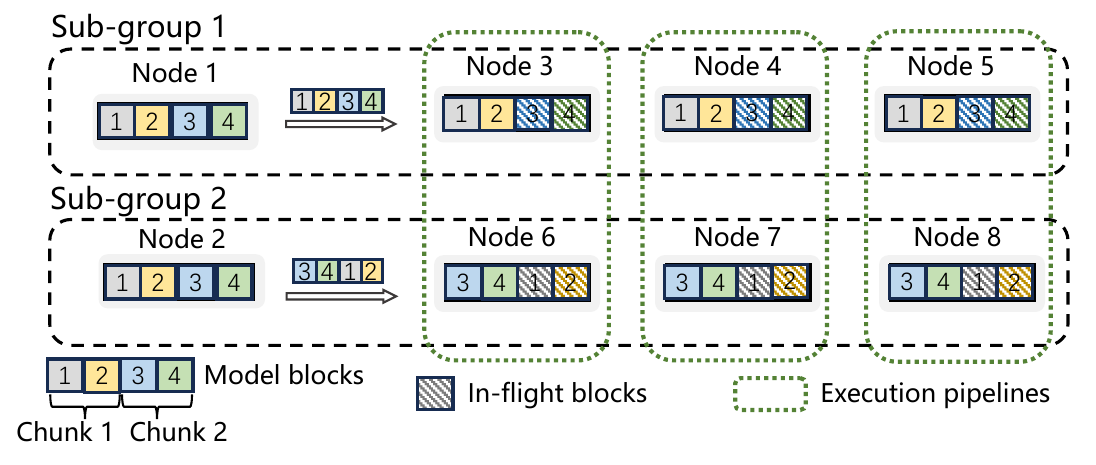}
    \vspace{-20pt}
    \caption{Example of $2 \rightarrow 8$ scaling.
    \textit{\textmd{Each sub-group transfers model chunks in circularly shifted order in parallel, to construct three pipeline parallel inference execution flows (Node 3 and 6, Node 4 and 7, and Node 5 and 8). This strategy allows multiple execution pipelines to be started as soon as enough model blocks are distributed while enabling non-blocking binomial pipeline multicast\revise{~\cite{rdmc, binomial-pipe}}. 
    }}
    }
    \vspace{-5pt}
    \label{fig:scale_example}
\end{figure}

\PHM{Optimized transfer order.}
We propose a $k$-way transmission strategy to optimize the order of model block transfers across $k$ sub-groups, which enables \AlgoName to minimize the time required for assembling the complete model.
Algorithm~\ref{alg:k_way} outlines the $k$-way transmission strategy. 
It first partitions the model blocks into $k$ equal-sized chunks (lines 1-2) and then generates the block transfer orders for each sub-group by circularly shifting these chunks (lines 3-4). 
This design ensures that the sub-groups work in tandem, with the first complete model instance becoming available after just $b/k$ time steps. 
Fig.~\ref{fig:scale_example} illustrates a $2 \rightarrow 8$, 2-way transmission scaling scenario, where two source nodes (Node 1 and Node 2) distributed model blocks to six destination nodes across two sub-groups. The model is partitioned into four blocks, which are grouped into two equal-sized chunks: Blocks 1-2 and Blocks 3-4. To improve the overall throughput, each sub-group follows a {\bf circular shifting strategy} when transferring model blocks. Specifically, Sub-group 1 (Node 1 as the source) starts by sending model blocks in the order of Blocks 1-2 and Blocks 3-4 to its destination nodes (Node 3-5). Meanwhile, Sub-group 2 (Node 2 as the source) follows a complementary pattern, transferring model blocks in the {\bf circular shifted order} (Blocks 3-4 followed by Blocks 1-2) to its destination nodes (Node 6-8). 
This circular shifting of block chunks ensures that different groups of destination nodes receive different parts of the model concurrently, allowing them to collaboratively construct three model inference execution pipelines using pipeline parallelism. As the multicast progresses, the destination nodes incrementally collect and assemble complete model instances, enabling efficient and scalable model execution. 

\begin{algorithm}[h]
    \caption{Execution Pipeline Generation Strategy}
    \small
    \label{alg:pipeline_gen}
    \begin{algorithmic}[1]
    \Statex \textbf{Input:}\quad -- $k$ sub-groups $\{G_0, \dots, G_{k-1}\}$
    \Statex \quad -- $L_i$: the number of nodes in $G_i$
    \Statex \quad -- $n_i^j$: the $j^{th}$ node in $G_i$
    \Statex \textbf{Output:}\quad -- All generated execution pipelines $\mathcal{P}$

    \State $\mathcal{G} \gets$ sub-groups with unassigned nodes
    \While{$|\mathcal{G}| > 0$}
        \If{$|\mathcal{G}| = 1$} \Comment{A pipeline within a single sub-group}
            \State $P$ $\gets$ ordered nodes in $\mathcal{G}_0$
            \State $\mathcal{P} \gets$ append($\mathcal{P}$, $P$)
        \Else \Comment{Generate pipelines across sub-groups}
            \State $a \gets \min_{G_i \in \mathcal{G}} L_i$     
            \For{$t \in [0, a-1]$} 
                \State $P$ $\gets$ []
                \For{$G_i \in \mathcal{G}$}
                    \State $P$ $\gets$ append($P$, $n_i^t$)
                \EndFor
               \State $\mathcal{P} \gets$ append($\mathcal{P}$, $P$)
            \EndFor
        \EndIf
        \State {Update $\mathcal{G}$}
    \EndWhile
\end{algorithmic}
\end{algorithm}


\subsection{Pipelined Inference Execution}

\begin{figure}
    \centering
    \includegraphics[width=0.45\textwidth]{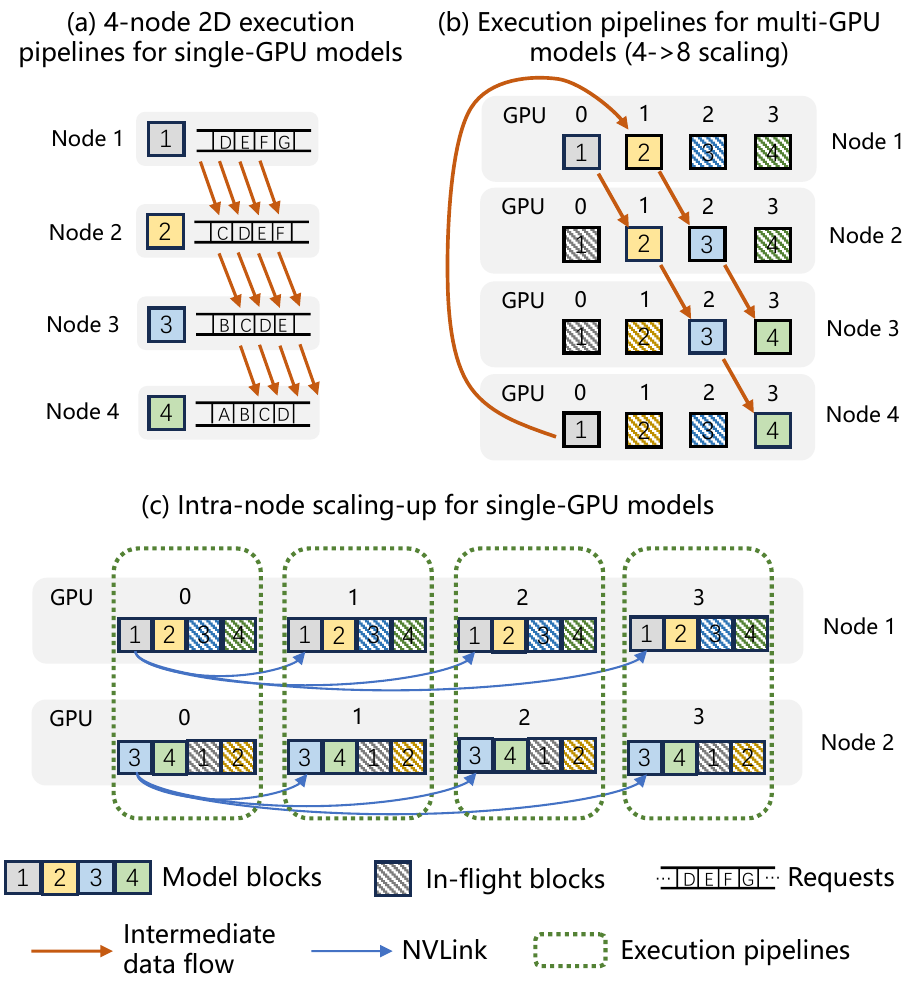}
    \vspace{-15pt} 
    \caption{Example of execution pipelines. 
    \textit{\textmd{(a)~A 4-node execution pipeline where each node executes its associated model block for multiple in-flight requests in parallel. 
    (b)~Execution pipelines across multi-GPU nodes where each pipeline contains GPUs from different nodes.
    (c)~Intra-node model replication to scale up the inference performance, where each local model block replica forms a separate cross-node execution pipeline.}}
    }
    \label{fig:pipeline}
    \vspace{-1em}
\end{figure}

\PHM{Generating execution pipelines.}
Building on the optimized block transfer order, we propose an efficient execution pipeline generation strategy outlined in Algorithm~\ref{alg:pipeline_gen}. 
The key idea is to construct execution pipelines from as many sub-groups as possible, maximizing the benefits of $k$-way transmission. 
Specifically, when the remaining unassigned nodes belong to only one sub-group, \AlgoName directly forms an execution pipeline using these nodes (lines 3-5). Otherwise, it selects one node from each available sub-group to construct the pipeline (lines 6-12). 
\SysName prioritizes pipeline configurations where each sub-group has an equal number of nodes (i.e., $N$ mod $k = 0$). 

\PHM{2D execution pipelines.}
Fig.~\ref{fig:pipeline} (a) illustrates how a 4-node execution pipeline processes multiple batches (one or multiple requests) of inference requests in parallel using a 2-dimensional pipelining strategy. Along the first dimension, each node is assigned a specific model block and computes a different batch of requests. Once a node completes processing a batch for its assigned block, it passes the intermediate result to the next node and starts computing the next batch along the second dimension. 
This 2D pipeline strategy efficiently utilizes resources distributed along the multicast route, enabling rapid scaling and processing of accumulated requests during spikes. 
\SysName schedules requests across multiple pipelines based on their available resources to improve overall resource efficiency.

\PHM{Models on multiple GPUs.}
When a model spans multiple GPUs, \AlgoName generates execution pipelines across GPUs within the same node and/or from nodes to minimize performance loss during model scaling. 
Fig.~\ref{fig:pipeline} illustrates the multi-GPU model scaling scenario, where each model instance is distributed across multiple GPUs. During scaling, as model blocks partially arrive, \AlgoName dynamically selects one of the three execution strategies based on model size and resource availability (i.e., whether a node has multiple GPUs). 
{\bf Case 1: Cross-node execution pipeline for single-GPU models} (models fitting in a single GPU): This is the default execution strategy already described in \S\ref{subsec:model_multicast}.
{\bf Case 2: Cross-node execution pipeline for multi-GPU models} (models that do not fit in a single GPU): GPUs that have received complete blocks can immediately begin forming execution pipelines across nodes to support distributed, pipelined inference for large models, without waiting for the full model to load (see Fig.~\ref{fig:pipeline} (b)). 
{\bf Case 3: Intra-node scaling-up for single-GPU models}: \AlgoName can opportunistically leverage multiple local GPUs on the same node to accelerate scaling if there are available local GPU resources on the same node. As soon as the first GPU receives model blocks, it can quickly replicate them to other local GPUs using high-speed node-local communication mediums like NVLink, which offers bandwidth up to an order of magnitude higher than RDMA networks (see Fig.~\ref{fig:pipeline} (c)). This fast intra-node replication enables rapid local model scaling-up. Replicated model blocks can then form cross-node execution pipelines (Case 1).   
This hybrid approach maximizes resource utilization and enhances inference performance by opportunistically reducing data movement overhead. 

\subsection{Mode Switching}

After model scaling is complete, each node maintains a full model instance and directly serves inference requests locally\footnote{Current \SysName supports models that fit within the GPU capacity of a single multi-GPU node (e.g., up to 640~GB of collective GPU memory for an 8-A100 node). We leave the support for larger models spanning multiple nodes for future work.}. 
A key challenge is to ensure that each node keeps corresponding runtime states (i.e., KV caches) for the active requests it serves.
\AlgoName addresses this challenge through \emph{KV cache recomputation}. 
It evenly distributes incomplete requests of an execution pipeline among all participating nodes, and then each node recomputes its assigned requests using available tokens generated by distributed inference during the multicast phase. 
This approach generally incurs lower overhead compared to transmitting existing KV caches, which requires costly all-to-all communication across participating nodes.
This mechanism ensures \AlgoName to seamlessly switch from pipelined execution mode to local execution mode while maintaining low inference latency.  


\section{Efficient Model Management}
\label{sec:system}


\SysName supports efficient model management in both host memory and GPUs through two key system designs.

\PHM{Locality-driven model startup.}
\SysName introduces a multi-level, locality-driven model startup scheme to efficiently support model instances stored in various storage tiers. \SysName optimizes locality using several startup strategies: 
\textbf{(1) \texttt{GPU}} (\emph{hot start}): The model is fully loaded into GPU memory, enabling fast, local execution.
\textbf{(2) \texttt{Memory}} (\emph{warm start}): The model is cached in host memory. \SysName directly loads host-memory-cached models into GPUs for inference execution, and before models are fully loaded, constructs an execution pipeline across multiple nodes of this kind for enhanced inference performance (\AlgoName in \S\ref{sec:algo}). 
\textbf{(3) \texttt{Null}} (\emph{cold start}): The model is neither cached in GPU or host memory, requiring \SysName to perform cold-start scaling by directly retrieving model blocks from remote \texttt{GPU} and/or \texttt{Memory} nodes. 

\PHM{Efficient memory management.}
\SysName employs two strategies to manage GPU and host memory for model blocks. 
\textbf{(1) Tensor packing:} \SysName maps each model block to a contiguous memory region for enhanced transmission efficiency. 
By consolidating all tensor data associated with a single model block into a contiguous memory chunk, \SysName enables bulk transfer of entire blocks, improving bandwidth efficiency.
Notably, the tensor memory layout optimization has no impact on inference execution. 
\textbf{(2) GPU memory pre-allocation:} \SysName pre-allocates memory chunks for model blocks and intermediate results, as their sizes remain consistent across requests during pipeline execution. 
Runtime states with dynamic memory requirements (e.g., KV caches) are internally managed by inference engines (e.g., vLLM~\cite{kwon2023efficient}). 
This design ensures memory efficiency while minimizing memory allocation overhead at runtime. 

\section{\SysName Implementation}
\label{sec:implementation}
We have implemented \SysName in 10K lines of Python and 4K lines of C++, structured into two core components: the cluster manager and worker nodes. \SysName's source code will be released upon the acceptance of the paper.  

The {\bf Cluster Manager} is implemented in Python (see Fig.~\ref{fig:overview}). 
Each worker node contains a {\bf Model Manager}, which consists of two key modules: (1) an \emph{inference module},  responsible for executing both local inference (within a GPU) and distributed inference (across nodes) and (2) a \emph{transfer module}, which implements GDR/RDMA-based model block transfer. For the inference module, we extend the codebase of Meta's Llama inference framework~\cite{metainf} to support both local and distributed inference using Python, which is open-source at~\cite{code_inf}. 
\revise{We build the transfer module on top of Derecho’s RDMC~\cite{rdmc, dere-github} (version 2.4.0), where we reuse its resource initialization and RDMA queue pair/connection management components, and extend support to enable one-sided RDMA and GPUDirect RDMA. 
Specificlly, we reuse \(\sim\)470 lines from RDMC and add \(\sim\)520 lines for one-sided RDMA and memory-region extenstions.\footnote{Primarily RDMC’s \texttt{rdmc.cpp}, \texttt{verbs\_helper.cpp}, and \texttt{group\_send.cpp} are reused for resource/connection setup and polling.}  
} 
The binomial pipeline algorithm is implemented in {\bf Cluster Manager} and its GDR/RDMA semantics are implemented in {\bf Model Manager}'s transfer module. 
All the key RDMA P2P transfer APIs (e.g., RDMA queue pair establishment and RDMA read operation) are exposed to Python via {\small\texttt{Pybind11}}~\cite{pybind11}. 

\section{Evaluation}
\label{sec:evaluation}
We evaluate \SysName by addressing three key questions:  
{\bf (1)} How fast can \SysName distribute model blocks across a GPU cluster (\S\ref{subsec:model_transfer_performance})?
{\bf (2)} How does \SysName scale LLM inference performance compared to state-of-the-art baselines under concurrent, stress-test workloads ~(\S\ref{subsec:throughput}, \S\ref{subsec:latency})? 
{\bf (3)} How elastic and cost-effective is \SysName under a bursty real-world LLM workload (\S\ref{subsec:e2e})?
{\bf (4)} How do different design choices and configurations impact \SysName's performance?~(\S\ref{subsec:ablation_study}) 

\subsection{Experiment Setup}
\label{sec:evaluation:setup}

\PHM{Testbeds.} 
All experiments are conducted on a shared HPC cluster, with an exclusive allocation of up to 24 NVIDIA H800 GPUs and 12 nodes per user. 
%
We configure two testbeds to evaluate \SysName under different scalability scenarios (Table~\ref{tab:testbed_configurations}). Testbed1 is used when a single GPU is sufficient to host the entire model (e.g., Llama-2 7B), 
allowing to stress the scalability test via inter-node communication. 
Testbed2 is used when a single GPU is not large enough to host a model (e.g., Llama-2 70B), 
requiring multiple GPUs per node to validate scalability while involving inter-node communication (e.g. model parallel inference).



\begin{table}[t]
\centering
\caption{Testbed Configurations.
~\textit{\textmd{Each node is equipped with 1TB RAM and 4TB of NVMe SSD local storage.
}}
}
\label{tab:testbed_configurations}
\vspace{-10pt}
\resizebox{\columnwidth}{!}{%
\begin{tabular}{lcccccc}
\hline
\textbf{Testbed} & \textbf{GPU} & \textbf{NIC} & \textbf{Memory} & \textbf{SSD} & \textbf{\#Nodes} \\ 
\hline
Testbed1        & 1$\times$H800       & 1$\times$400Gb/s IB & 64GB/s & 5GB/s & 12             \\ 
Testbed2        & 4$\times$H800       & 1$\times$400Gb/s IB  & 64GB/s & 5GB/s &  6              \\  
\hline
\end{tabular}%
}
\end{table} 


\noindent{\textbf{Models and configurations.}}
We consider two primary experimental parameters: model size and \emph{k} (refer to \emph{k}-way transmission in~\S\ref{subsec:model_multicast} for details).
We test \SysName with the Llama-2~\cite{llama2} series LLMs with 7B, 13B, and 70B parameters and a \emph{k} value ranging from \{1, 2, 4\}. 
Experiments for Llama-2 7B and 13B are conducted on Testbed1, while Llama-2 70B tests run on Testbed2. Unless explicitly stated, all configurations follow these defaults. 

\PHM{Measurement metrics.} 
Our evaluation focuses on key metrics including throughput, latency, and cost-effectiveness.
(1) Throughput (tokens per second) measures \SysName's ability to sustain high-load inference requests.
(2) Latency (time-to-first-token) reflects \SysName's efficiency in generating the first token quickly,
a critical metric for low-latency LLM serving.
(3) Cost-effectiveness (GPU time) evaluates how elastically \SysName provisions and releases GPU resources.

\noindent{\textbf{Baselines.}}
We compare \SysName against three baselines, covering both industry-standard and state-of-the-art research systems as discussed in \S\ref{sec:background-limitations}. \\
\textit{ServerlessLLM}~\cite{fu_serverlessllm_2024}: The state-of-the-art serverless LLM inference system designed for dynamic scaling. We implement ServerlessLLM and remove Ray Serve’s~\cite{rayserve} cluster management overhead to isolate its inference performance. \\
\textit{FaaSNet}~\cite{wang_faasnet_nodate}: An industry-adopted serverless function container provisioning system that optimizes P2P transfer topology for auto-scaling. We use its default binary tree topology and extend it to support GDR-based model loading. 
\textit{NCCL}~\cite{nccl}: An industry-standard communication library for multi-GPU training and inference developed by NVIDIA, optimizing collective communication primitives such as all-reduce and broadcast using GDR. 
Since \textit{NCCL} lacks native multicast support, we adapt its broadcast primitive by dynamically forming  process groups and transmitting model blocks to designated groups, effectively enabling multicast.

\begin{figure}[t]
\centering
\includegraphics[scale=0.1]{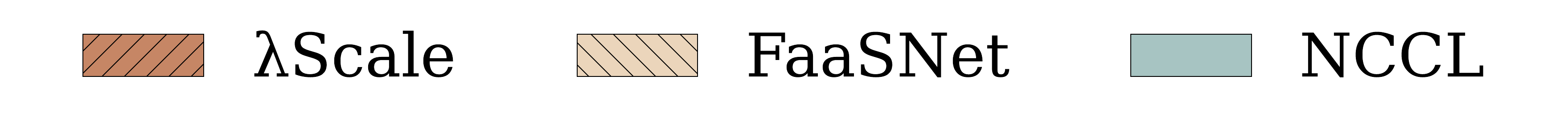} \\
\includegraphics[scale=0.1]
{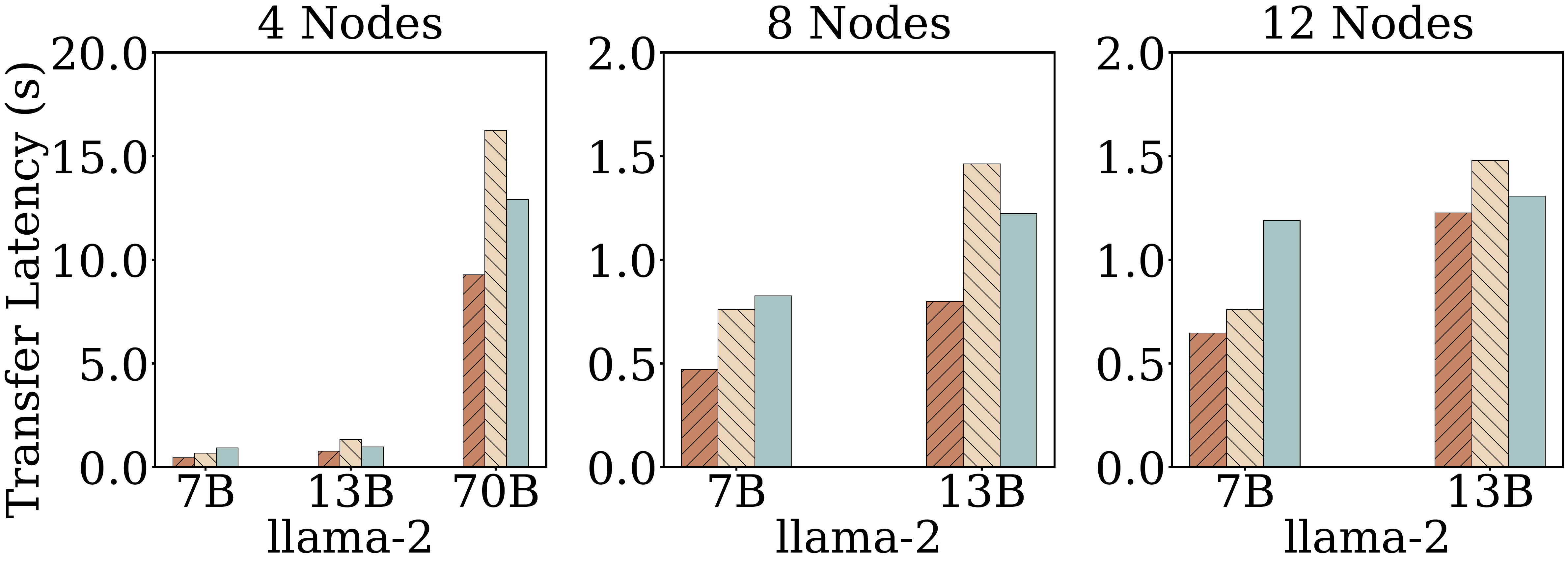}\label{fig:transfer_bar}
\vspace{-10pt}
\caption{End-to-end model multicast latency.
\textit{\textmd{The 4-node test involves 16 GPUs, with each node containing 4 GPUs. The 8-node and 12-node tests use 8 GPUs and 12 GPUs, respectively. All tests have $k = 1$ (a single source).}} 
}
\label{fig:transfer_latency}
\vspace{-5pt}
\end{figure}

\subsection{Multicast Performance}
\label{subsec:model_transfer_performance}

We first evaluate \SysName's raw block transmission latency achieved under binomial pipeline multicast. 

\PHM{Model transfer latency.}
Fig.~\ref{fig:transfer_latency} shows the end-to-end multicast latency with different GPU cluster settings. 
Overall, \SysName achieves up to a $1.82\times$ and $1.53\times$ speedups over \textit{FaaSNet} and \textit{NCCL}, respectively. 
We observe that \SysName's multicast performance advantage increases with both model size and cluster scale. For smaller models on fewer nodes (e.g., 7B on 4 nodes), \SysName is only modestly faster than the other systems; however, with larger models and more nodes (e.g., 70B on 12 nodes), \SysName’s performance benefit expands considerably. 
This improvement comes from \SysName's binomial pipeline, %
which splits a model into blocks and utilizes the entire cluster bandwidth resource to transmit blocks, maximizing link utilization and GPU-level parallelism. We explain the performance differences next when investigating fine-grained block arrival latencies. 

\begin{figure}[t]
\centering
\includegraphics[scale=0.1]{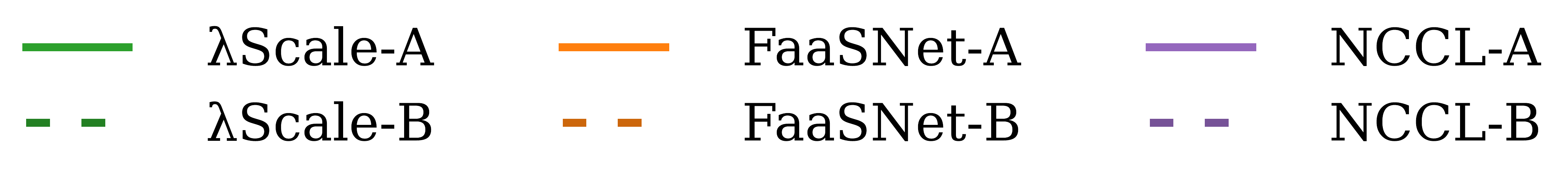} \\
\vspace{-0.5em}
\subfigure[{\small\texttt{4 Nodes}}] {
    \includegraphics[scale=0.1]
    {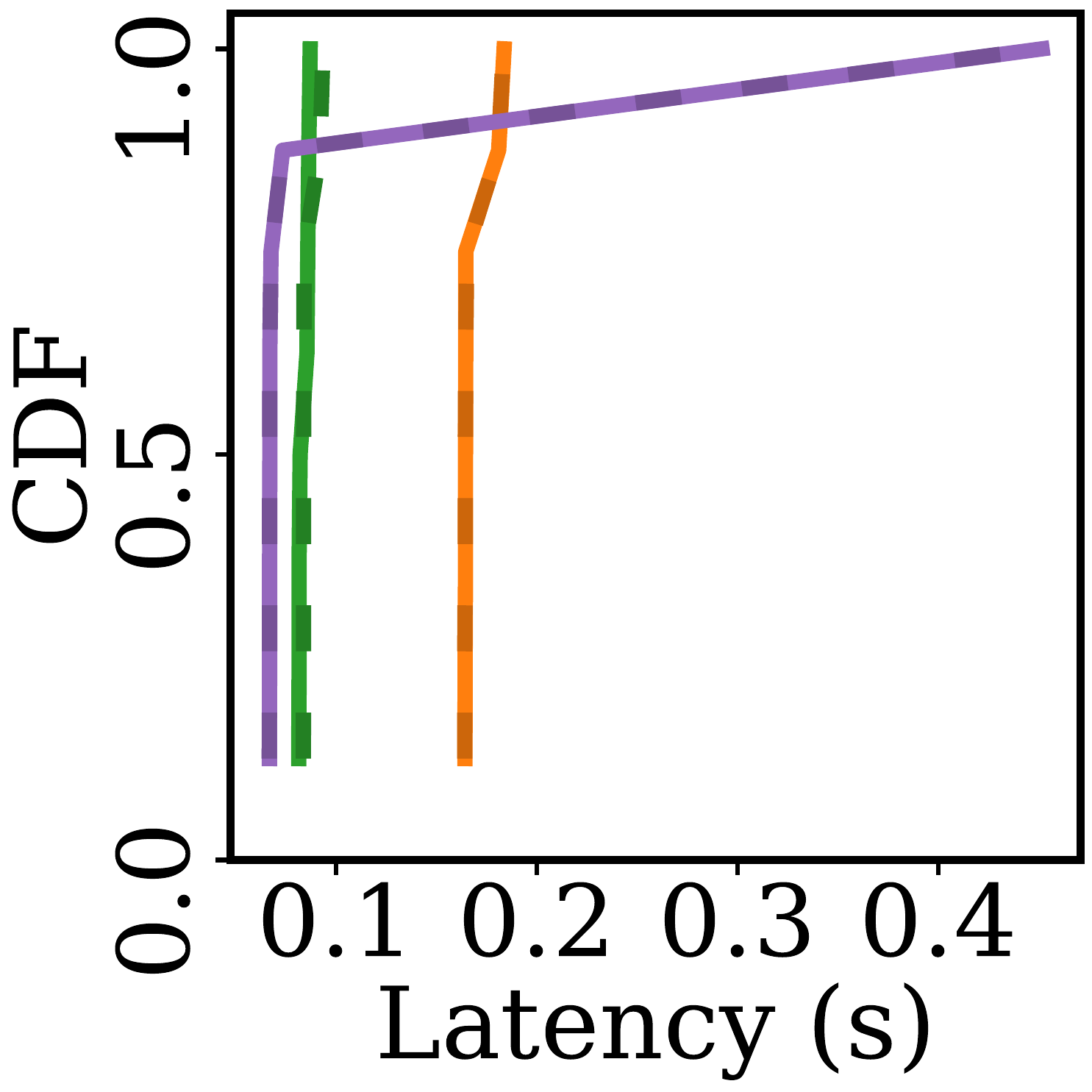}
}
\subfigure[{\small\texttt{8 Nodes}}] {
    \includegraphics[scale=0.1]
    {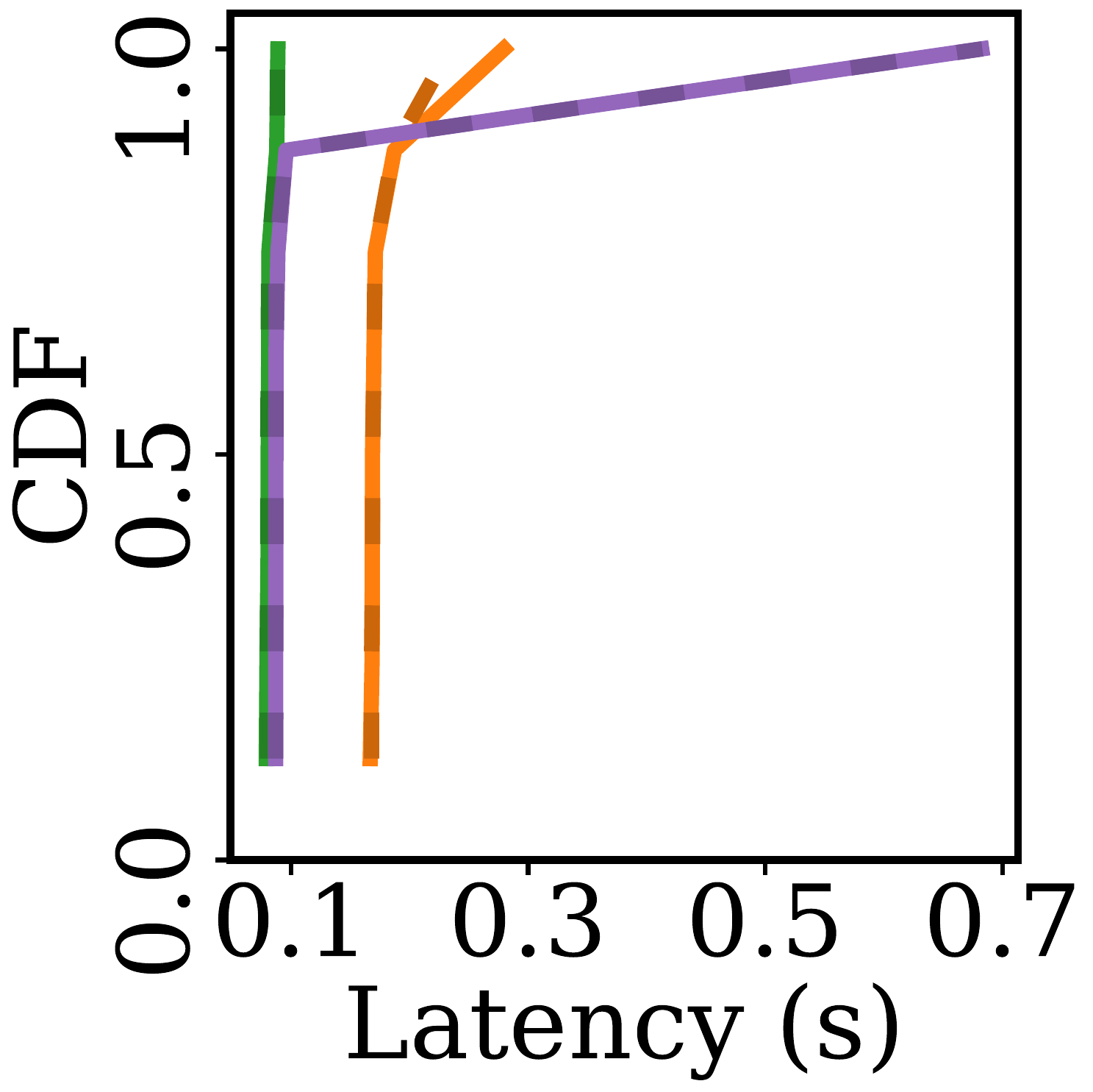}
}
\subfigure[{\small\texttt{12 Nodes}}] {
    \includegraphics[scale=0.1]
    {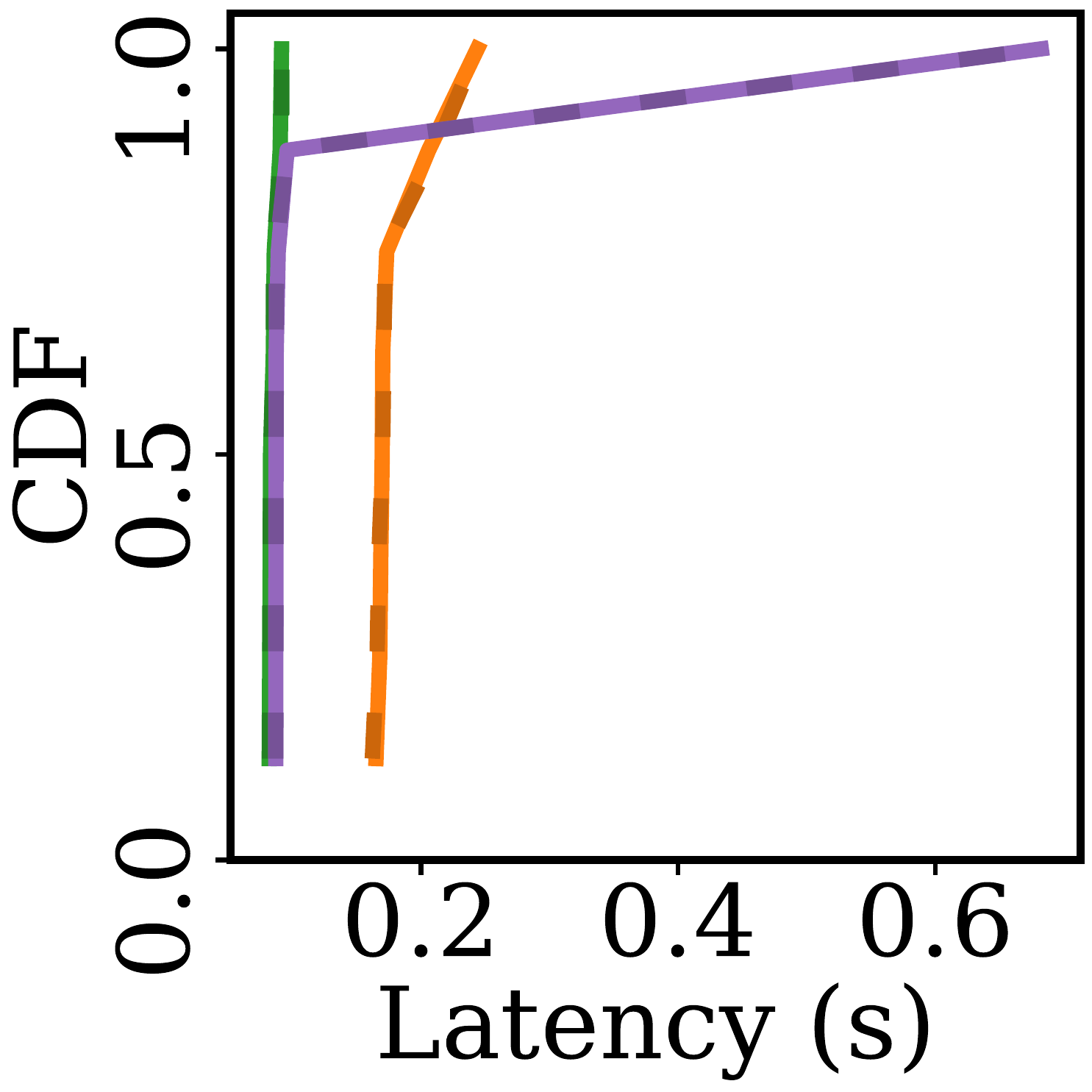}
}
\vspace{-14pt}
\caption{Model block transfer latency CDF. \textit{\textmd{We randomly select two nodes ({\small\texttt{A}} and {\small\texttt{B}}) from the cluster and report the arrival latency for each block. Other nodes show similar pattern. 
}}}
\label{fig:blk_latency_cdf}
\vspace{-5pt}
\end{figure}

\begin{figure*}[ht]
\centering
\includegraphics[scale=0.1]{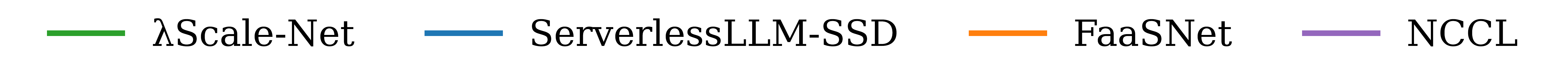} \\
\vspace{-0.5em}
\subfigure[{\small\texttt{Llama-2 7B}}] {
    \includegraphics[scale=0.12]
    {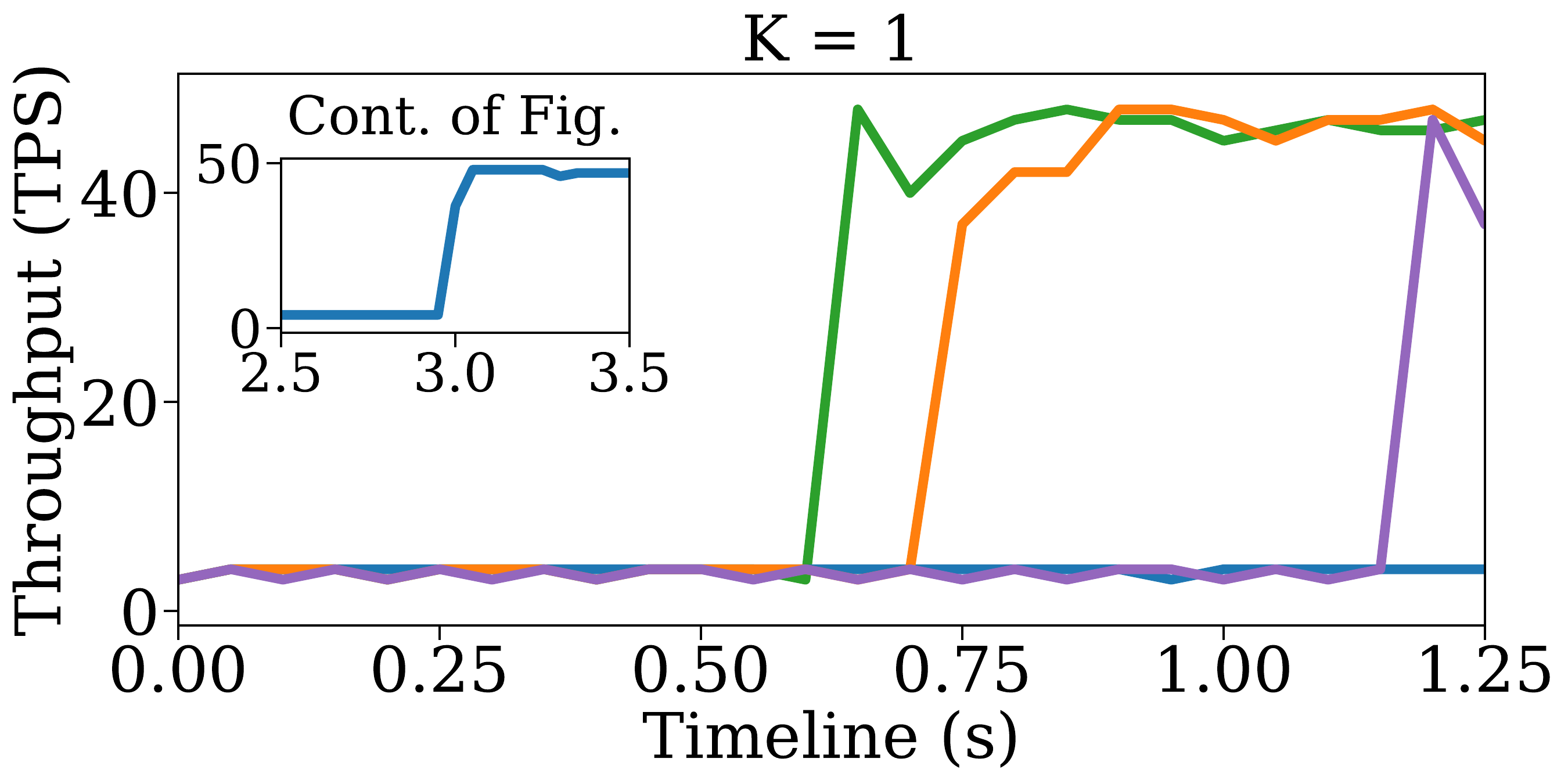}
    \includegraphics[scale=0.12]
    {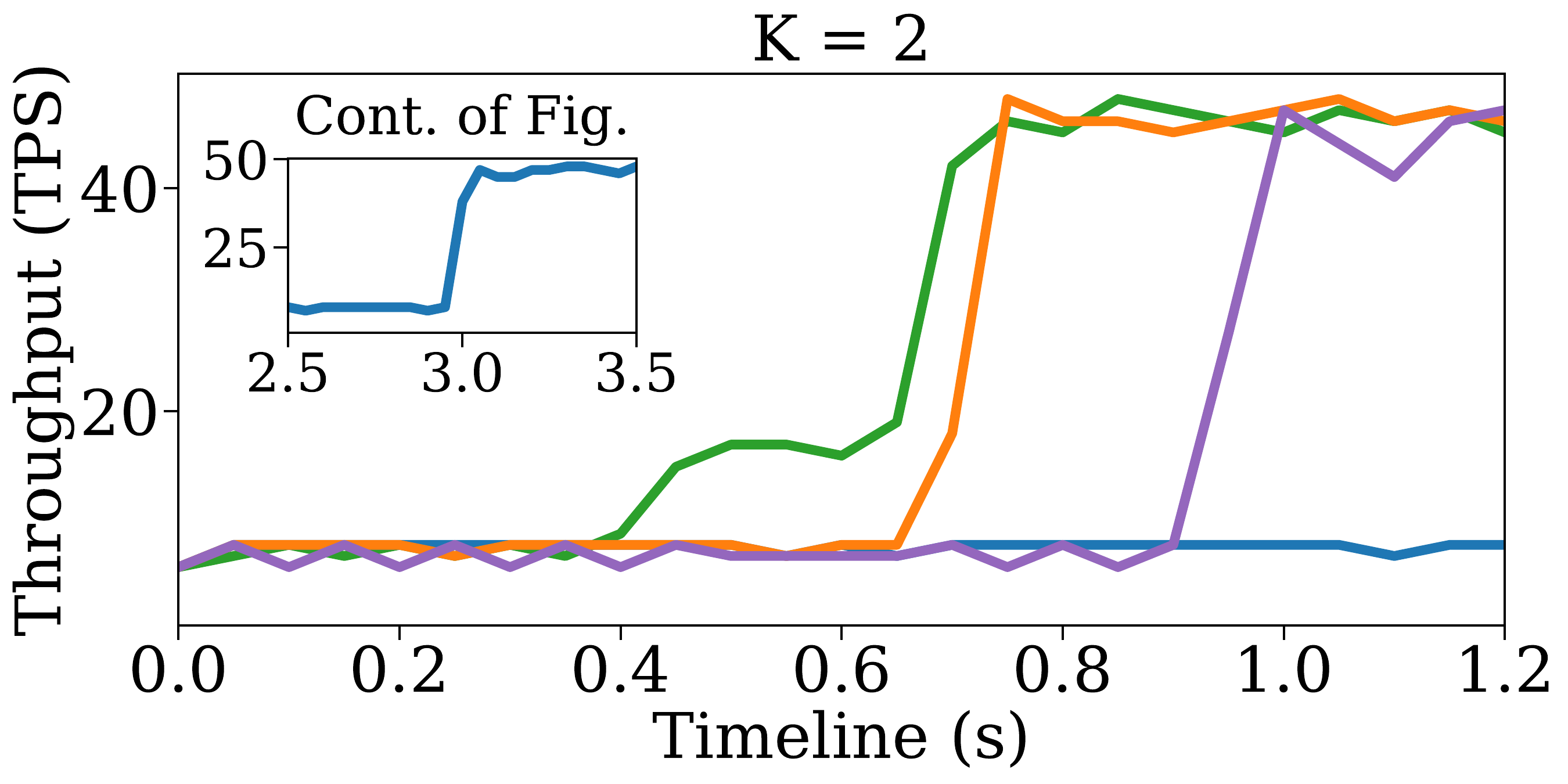}
    \includegraphics[scale=0.12]
    {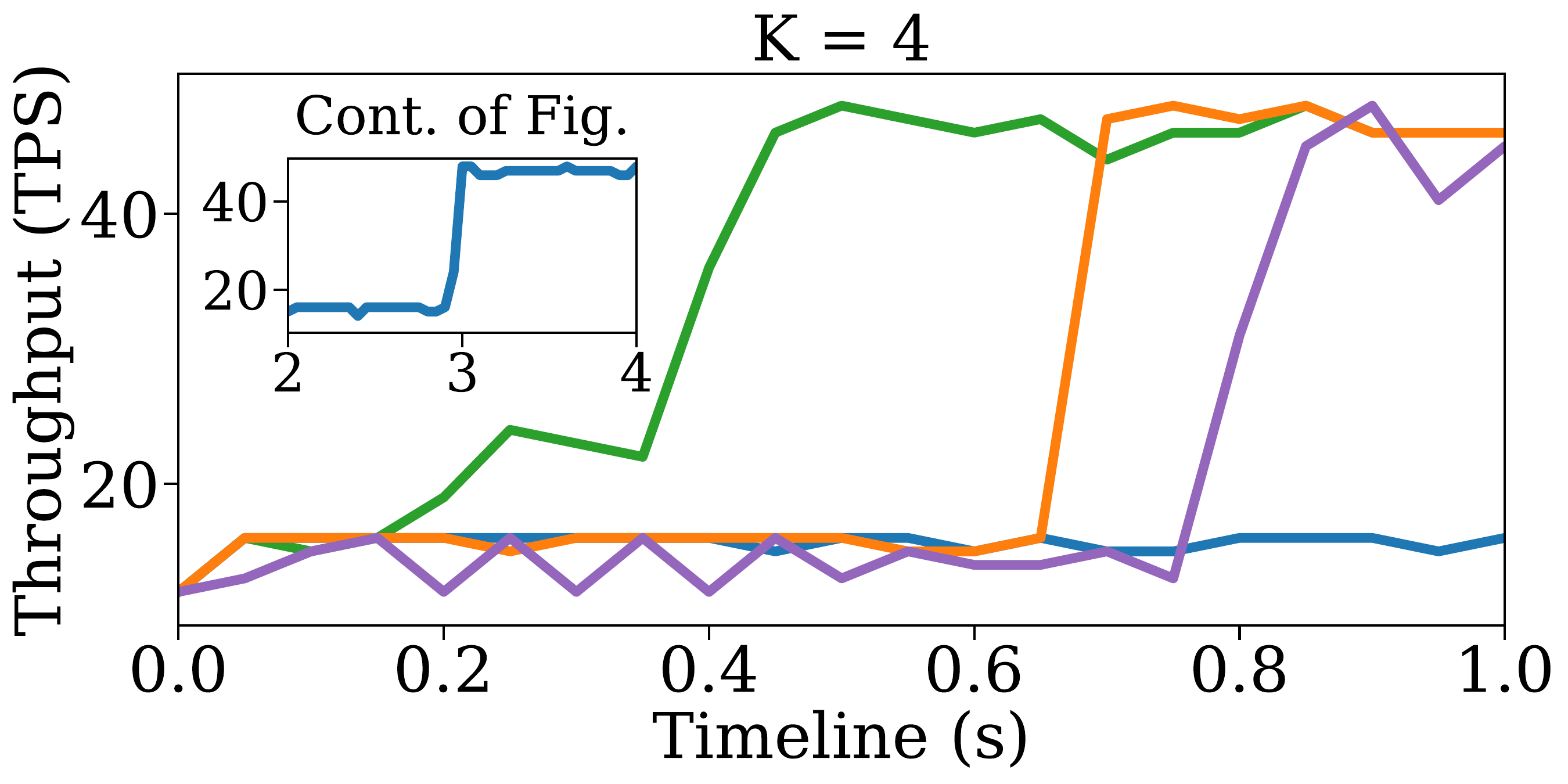}
    \label{fig:throughput_7b}
}
\vspace{-1em}
\subfigure[{\small\texttt{Llama-2 13B}}] {
    \includegraphics[scale=0.12]
    {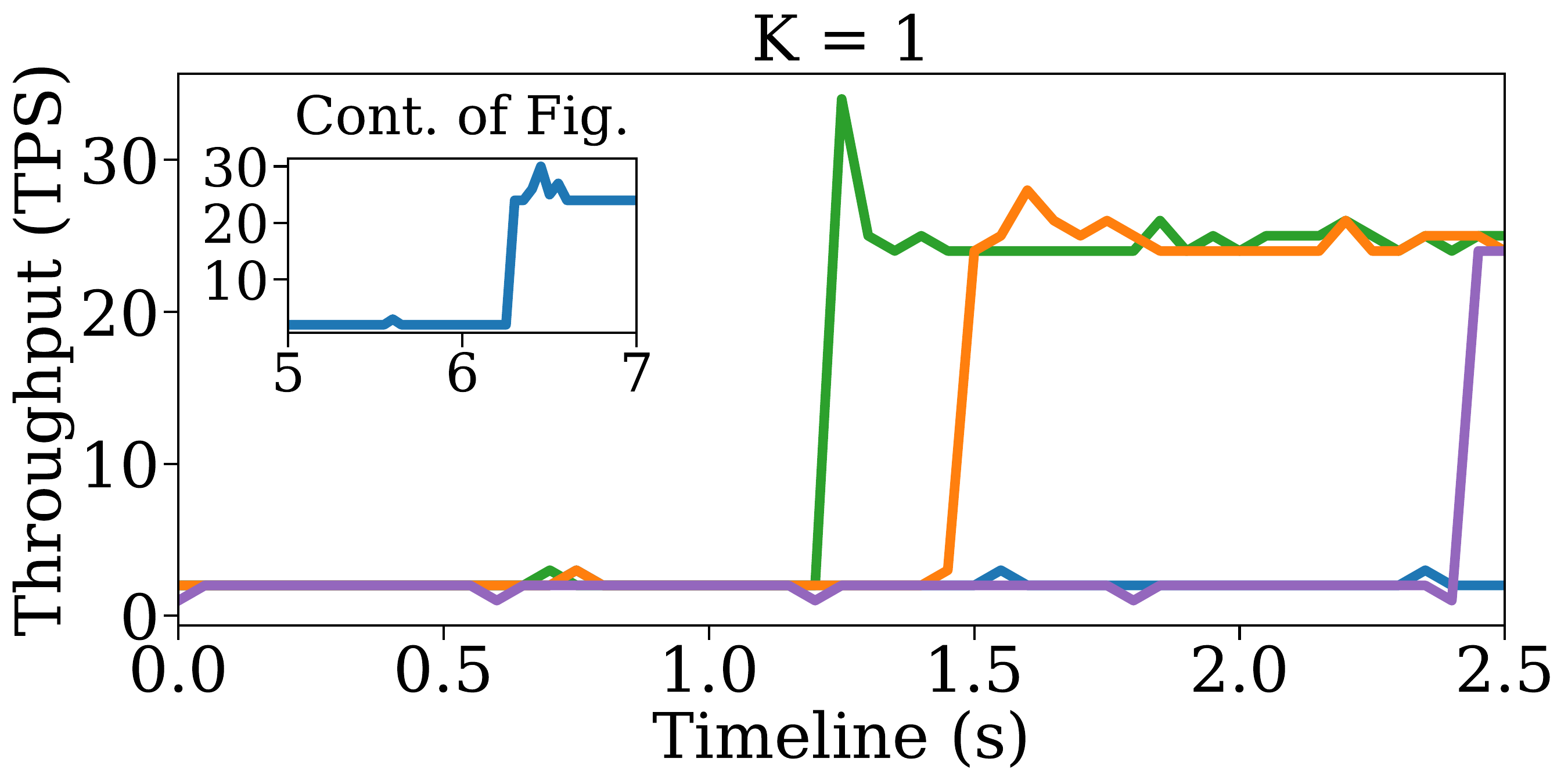}
    \includegraphics[scale=0.12]
    {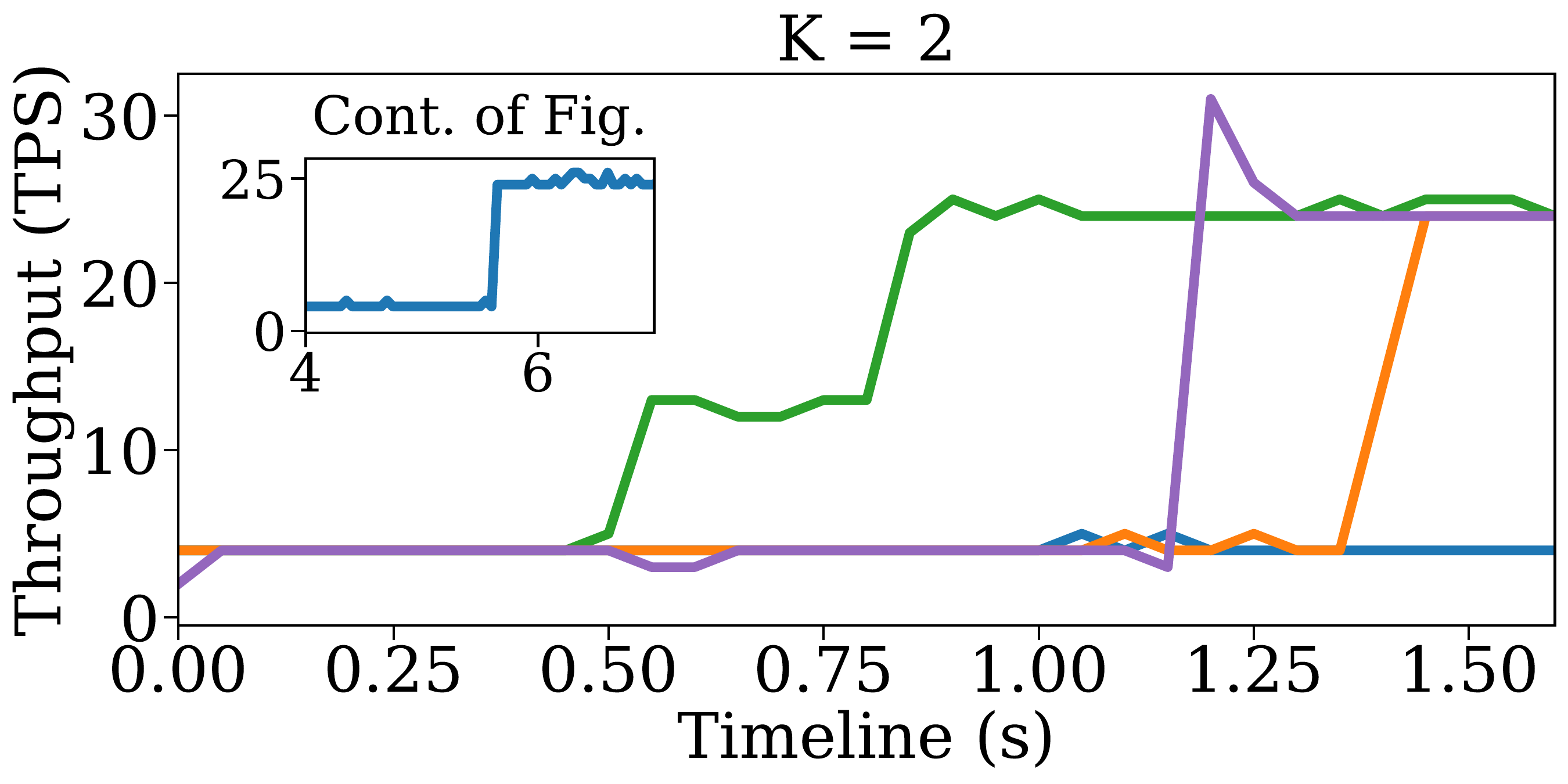}
        \includegraphics[scale=0.12]
    {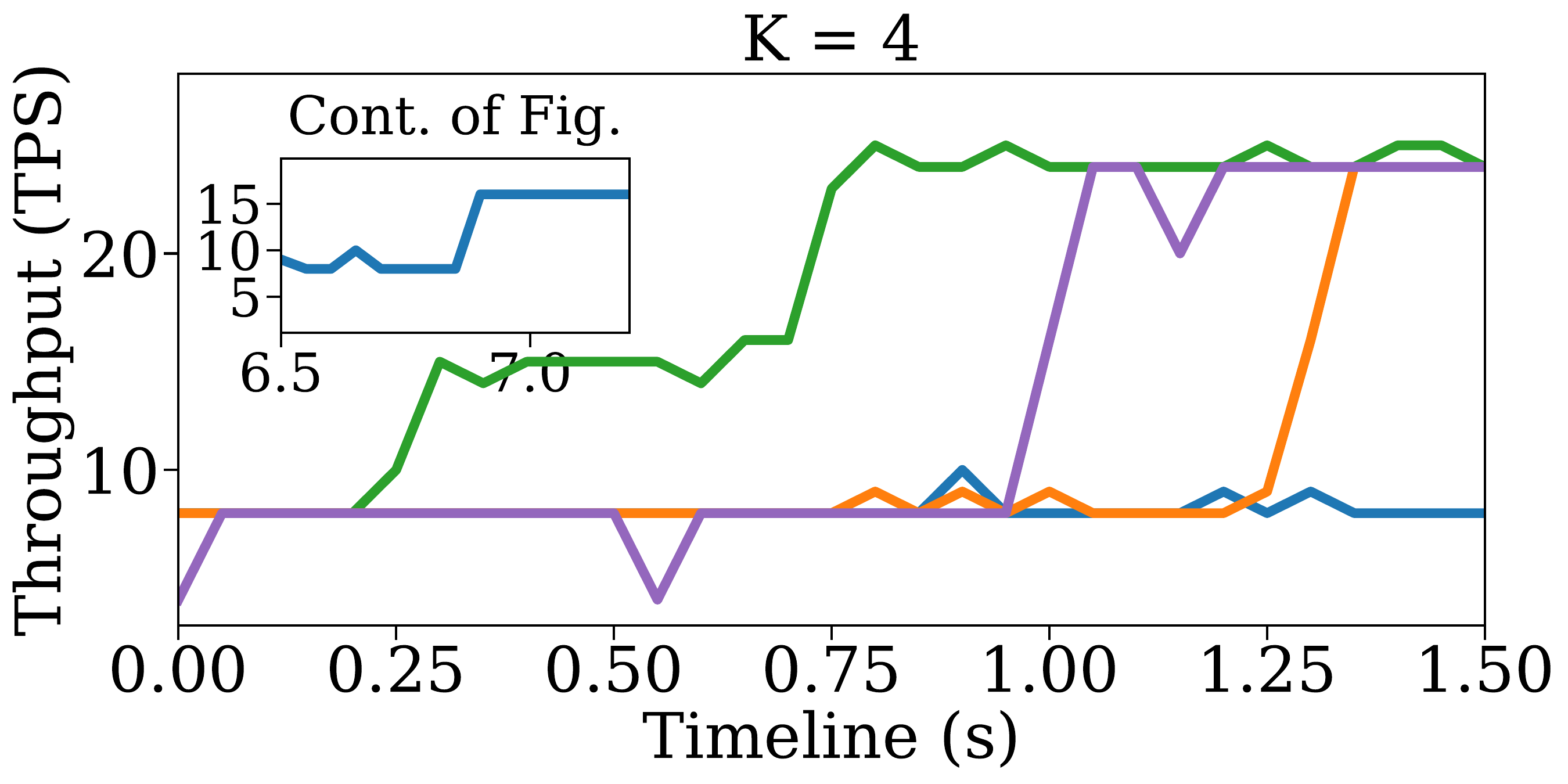}
    \label{fig:throughput_13b}
}
\vspace{-1em}
\subfigure[{\small\texttt{Llama-2 70B}}] {
    \includegraphics[scale=0.12]
    {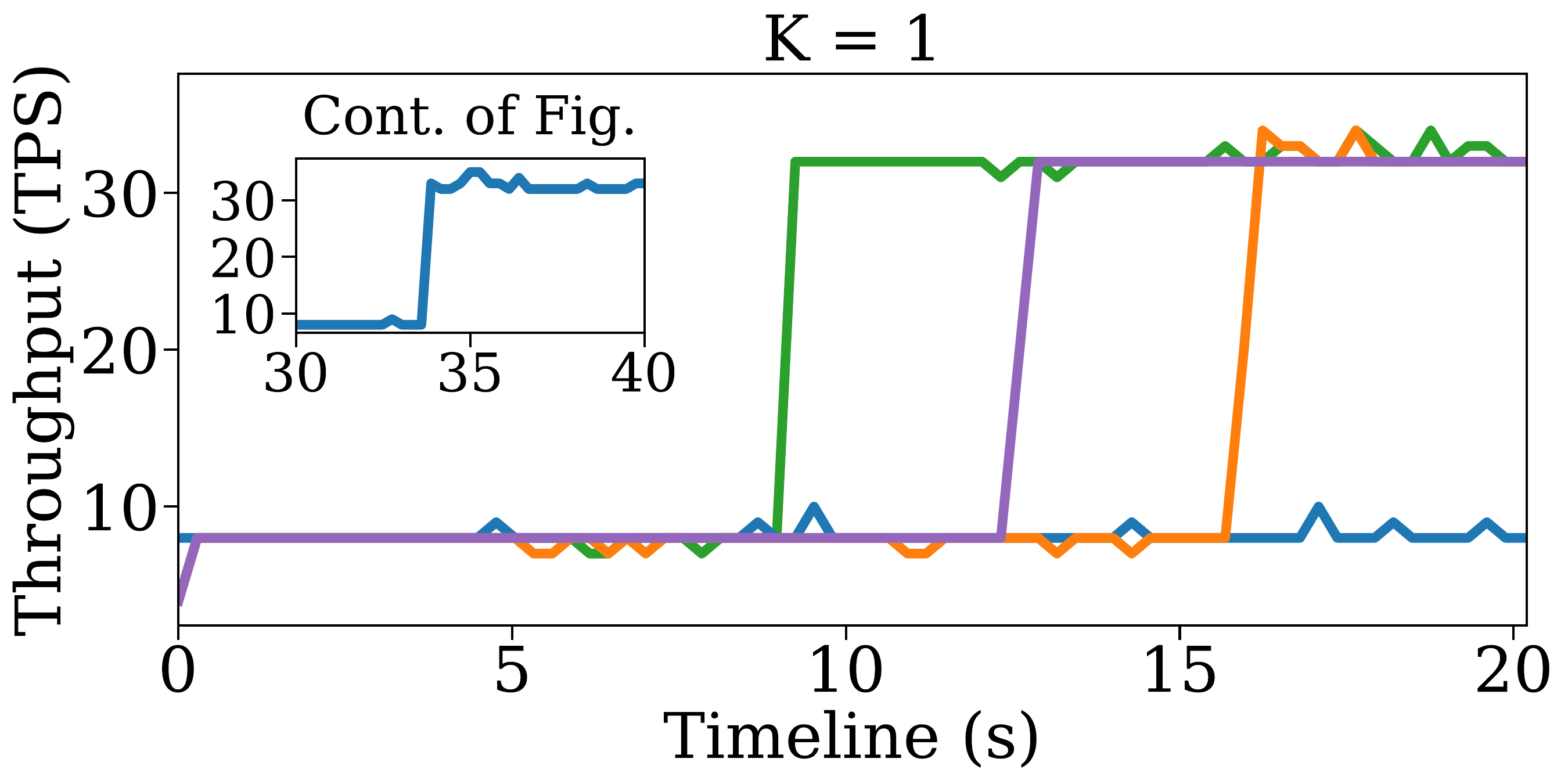}
        \includegraphics[scale=0.12]
    {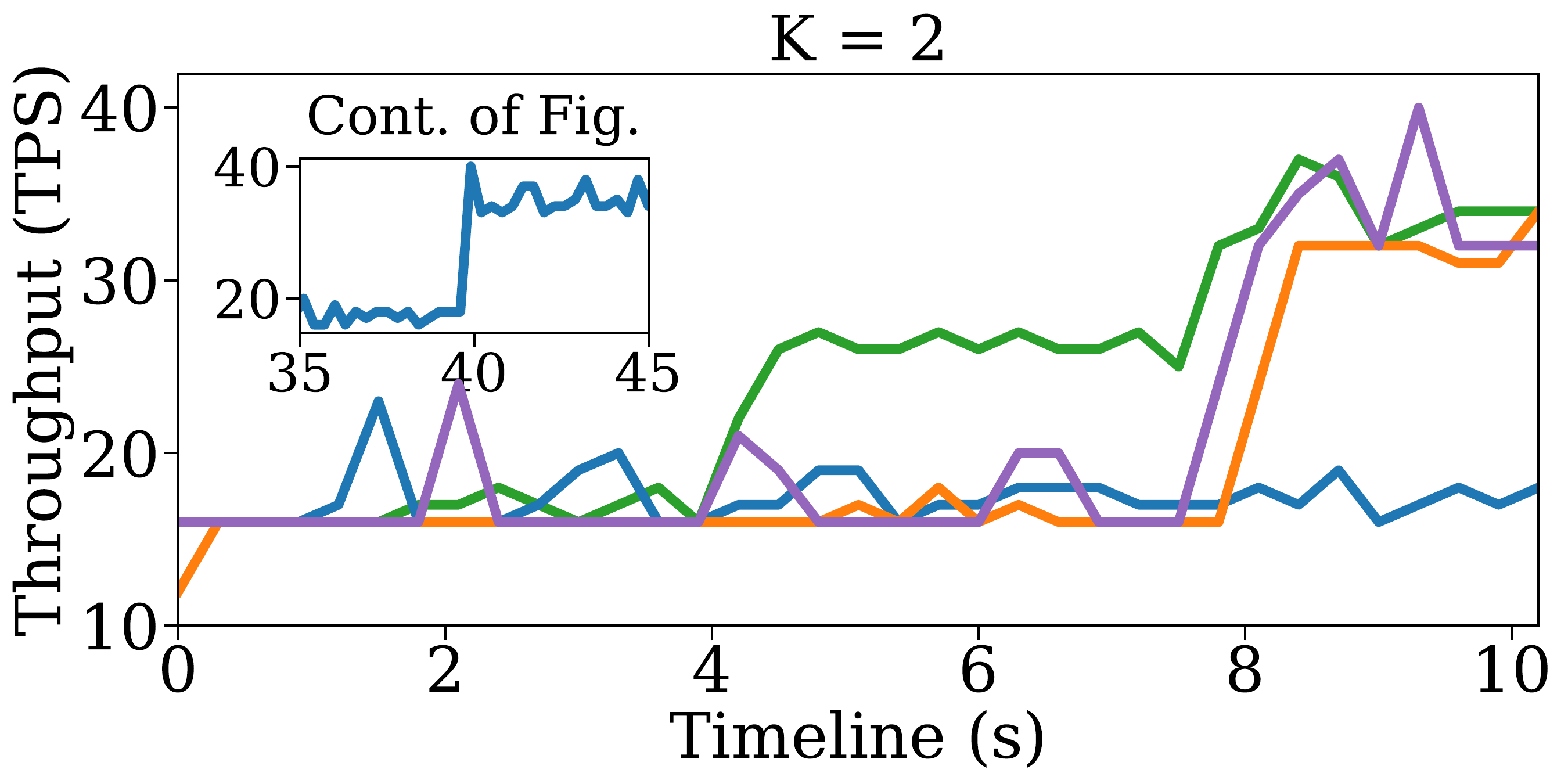}
    \label{fig:throughput_70b} 
}
\vspace{-0.5em}
\caption{Throughput scaling via GDR with varying model sizes.
\textit{\textmd{ServerlessLLM relies on local SSDs during scaling, while all other systems use GDR for inter-node communication. 
The $k$ configurations apply only to \SysName.
The mini-plots show the extended timeline for ServerlessLLM.
}}}
\label{fig:throughput_gdr}
\vspace{-5pt}
\end{figure*}

\noindent{\textbf{Block transfer latency.}
To zoom in, we measure the fine-grained block arrival latency for the Llama-2 13B model (Fig.~\ref{fig:blk_latency_cdf}). 
We observe that \SysName receives the first and last model blocks almost at the same time across all cluster sizes. 
Interestingly, while \SysName and \textit{NCCL} perform comparably for most blocks, \textit{NCCL} experiences a significantly high tail latency for the \emph{first} block, due to high group initialization overhead~\cite{nccl_cold_start_issue}: \textit{NCCL} requires creating a group for target processes before performing the broadcast. However, establishing a new group incurs high cost of up to hundreds of milliseconds as we observe in our tests. 
Another trend is that \textit{FaaSNet}'s tail latency grows as the cluster size increases, while \SysName maintains consistently low latency for all blocks. This is because \textit{FaaSNet}'s tree-based multicast algorithm (binary tree) limits parallelism to a fixed number of children nodes at the bottom of the topology~\cite{wang_faasnet_nodate}, while \SysName's binomial pipeline effectively improves the network I/O parallelism (\# of active senders) at each step.

\subsection{Throughput Performance}
\label{subsec:throughput}

We stress-test \SysName's application-level throughput.

\PHM{Scaling via GDR.} 
We measure \SysName's throughput scaling ability under high-stress loads by varying $k$ and compare it against \textit{ServerlessLLM}, \textit{FaaSNet}, and \textit{NCCL}.  
\SysName (green) consistently outperforms the baselines by achieving peak throughput significantly faster across various \emph{k} levels. 
One key observation is that, while \textit{FaaSNet}, \textit{NCCL}, and \textit{ServerlessLLM} steadily scale their throughput on a similar timeline (e.g., 0.6s for \textit{FaaSNet} for Llama-2 7B), 
\SysName effectively halves its ramp-up time when its \emph{k} increases. For the Llama-2 13B example, \SysName begins scaling at around 1.2s when $k=1$, whereas with $k=4$, scaling starts significantly earlier, at around 0.25s. 
This is because \AlgoName's block transfer and opportunistic execution pipeline allows GPUs to collaboratively load and serve requests as soon as sufficient model blocks are loaded into GPU memory, rather than waiting for the full model to be transferred (see \S\ref{subsec:model_multicast}). 
Meanwhile, \textit{ServerlessLLM-SSD} suffers from a dramatically longer ramp-up period for throughput scaling, primarily because: 
(1) it lacks GDR/RDMA multicast support, relying solely on local SSD, which slows down model loading; 
(2) it waits until the entire model is loaded into a GPU before serving requests; 
and (3) even for local SSD-based loading, it does not incorporate LLM-specific optimizations like \SysName's $k$-way transmission and optimized transfer order. 
\textit{FaaSNet} and \textit{NCCL} leverage GDR in multicast, making them more efficient than \textit{ServerlessLLM}. However, they lack support for collaborative distributed execution and have lower transmission parallelism, which limits their scaling efficiency. 
Another notable observation is that, when $\emph{k}\geq2$, \SysName exhibits a staircase-shaped performance plateau (e.g., between 0.25s and 0.75s for $k=4$ with Llama-2 13B). This behavior is likely due to implementation overhead, which may not be fully optimized. Addressing this performance plateau is part of our future work. 

\begin{figure}[t]
\centering
\includegraphics[scale=0.1]{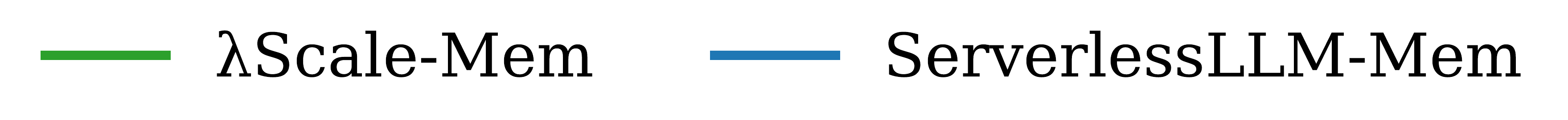} \\

\subfigure[{\small\texttt{Llama-2 7B}}] {
    \includegraphics[scale=0.1]
    {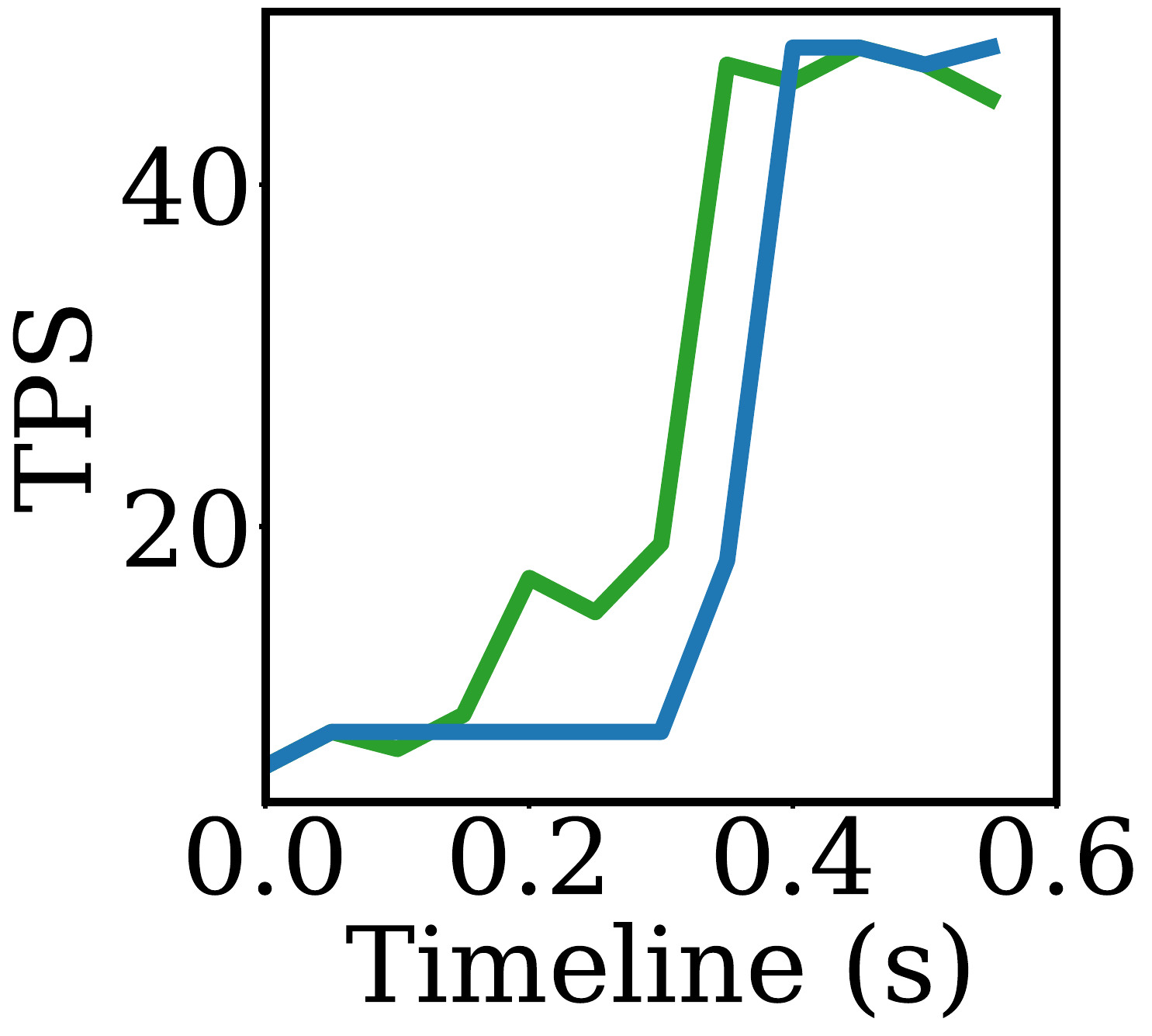}\label{fig:throughput_cache_7b}
}
\subfigure[{\small\texttt{Llama-2 13B}}] {
    \includegraphics[scale=0.1]
    {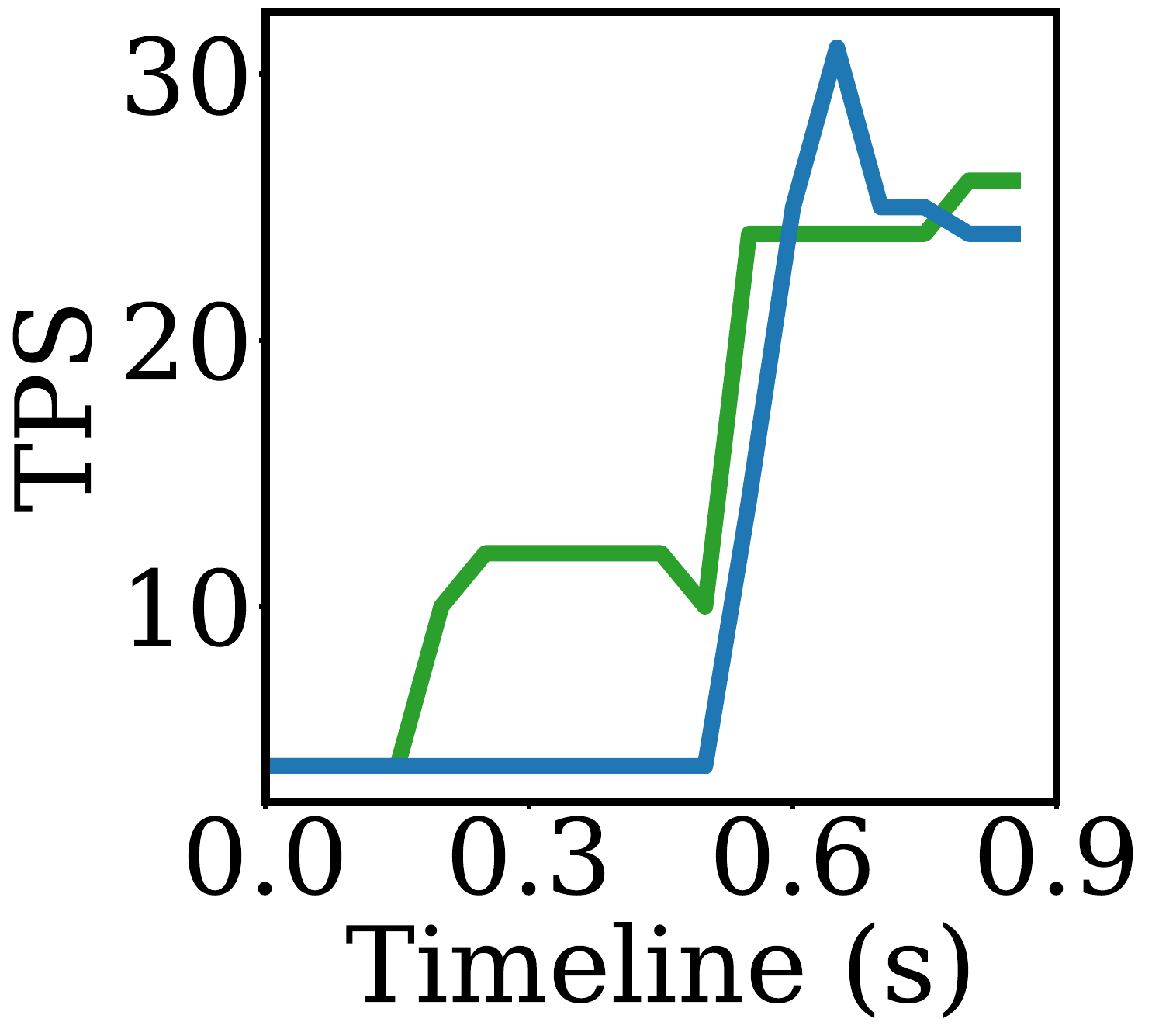}\label{fig:throughput_cache_13b}
}
\subfigure[{\small\texttt{Llama-2 70B}}] {
    \includegraphics[scale=0.1]
    {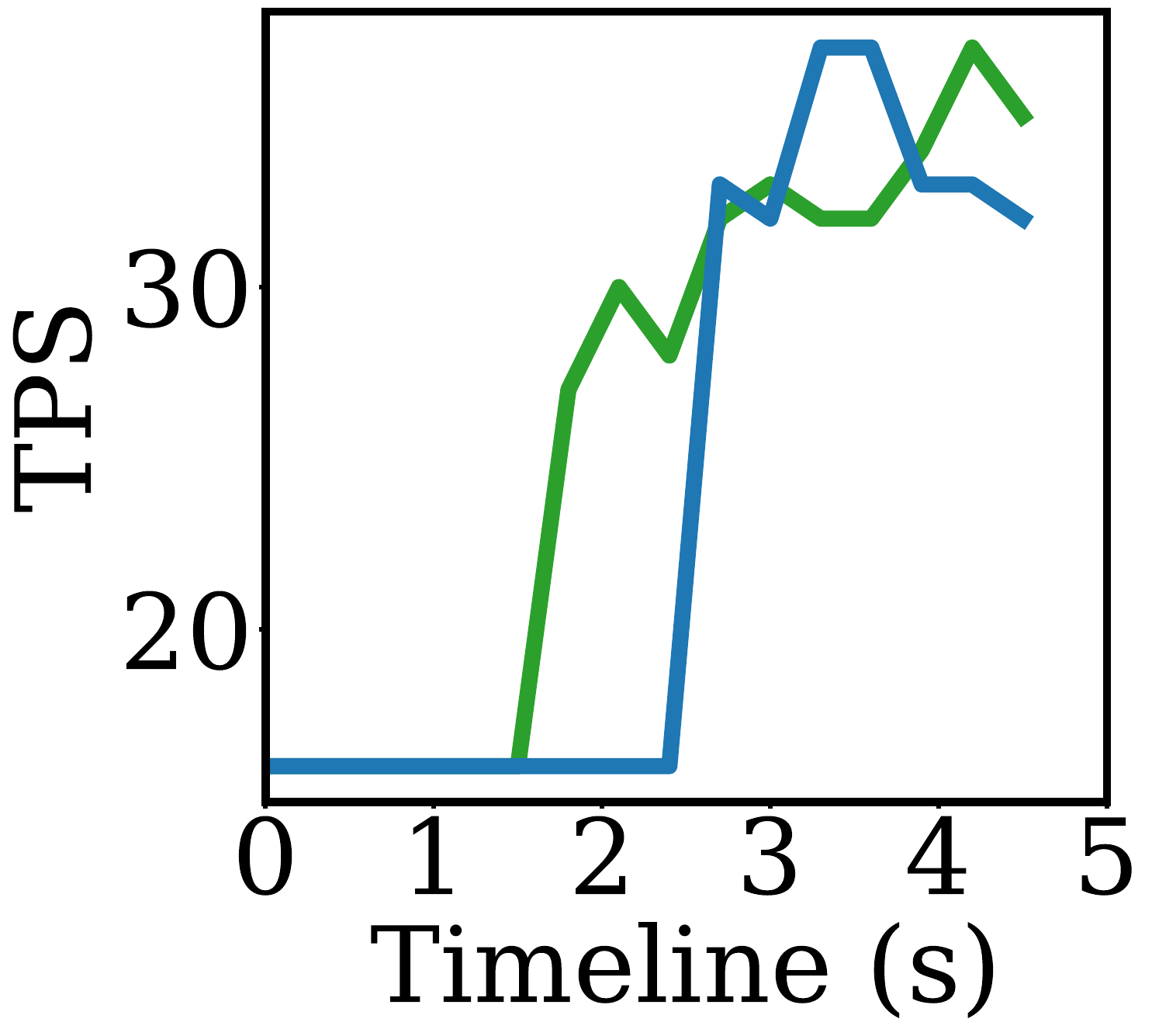}\label{fig:throughput_cache_70b}
}
\vspace{-10pt}
\caption{Throughput scaling via local cache with varying model sizes.} 
\label{fig:throughput_cache}
\vspace{-10pt}
\end{figure}

\PHM{Scaling via local cache. }
A key design of \textit{ServerlessLLM} is its use of host memory caching to improve performance by loading models from local host memory.  
For fair comparison, \SysName is configured to utilize local memory caching similarly to \textit{ServerlessLLM}.
Specifically, both systems are configured such that $R$ out of $N$ nodes have models loaded into their GPUs,
while the remaining $k$ nodes in the cluster load the model from the local host memory cache (\emph{N}= \emph{R} + \emph{k}).  For this evaluation, \emph{k} is fixed at $8$ for Llama-2 7B and 13B, and at $2$ for Llama-2 70B. Fig.~\ref{fig:throughput_cache} shows that \SysName scales $2\times$ to over $4\times$ faster than \textit{ServerlessLLM} in many scenarios. 
A key difference lies in that \SysName's \AlgoName enables collaborative inference during the loading phase, allowing requests to be served across \emph{k} nodes as soon as the first model block becomes available in GPU.   
In contrast, \textit{ServerlessLLM} waits until the model is fully loaded before processing requests.

\begin{figure}[t]
\centering
\includegraphics[scale=0.1]{figures/eval/throughput/shared_legend_k2_cache.pdf} \\
\vspace{-0.5em}
\subfigure[{\small\texttt{Llama-2 7B}}] {
    \includegraphics[scale=0.1]
    {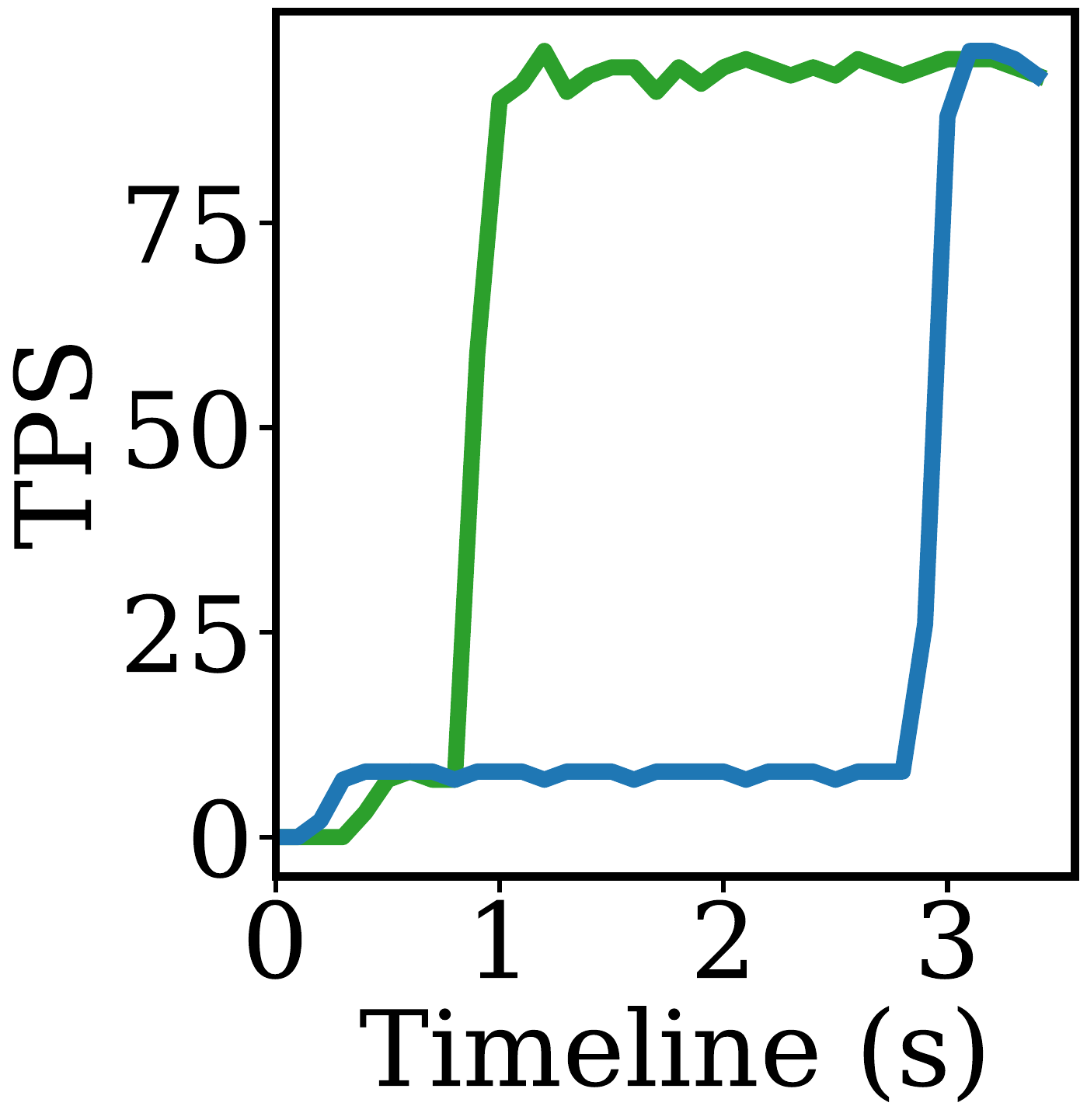}\label{fig:throughput_cold_start_7b}
}
\subfigure[{\small\texttt{Llama-2 13B}}] {
    \includegraphics[scale=0.1]
    {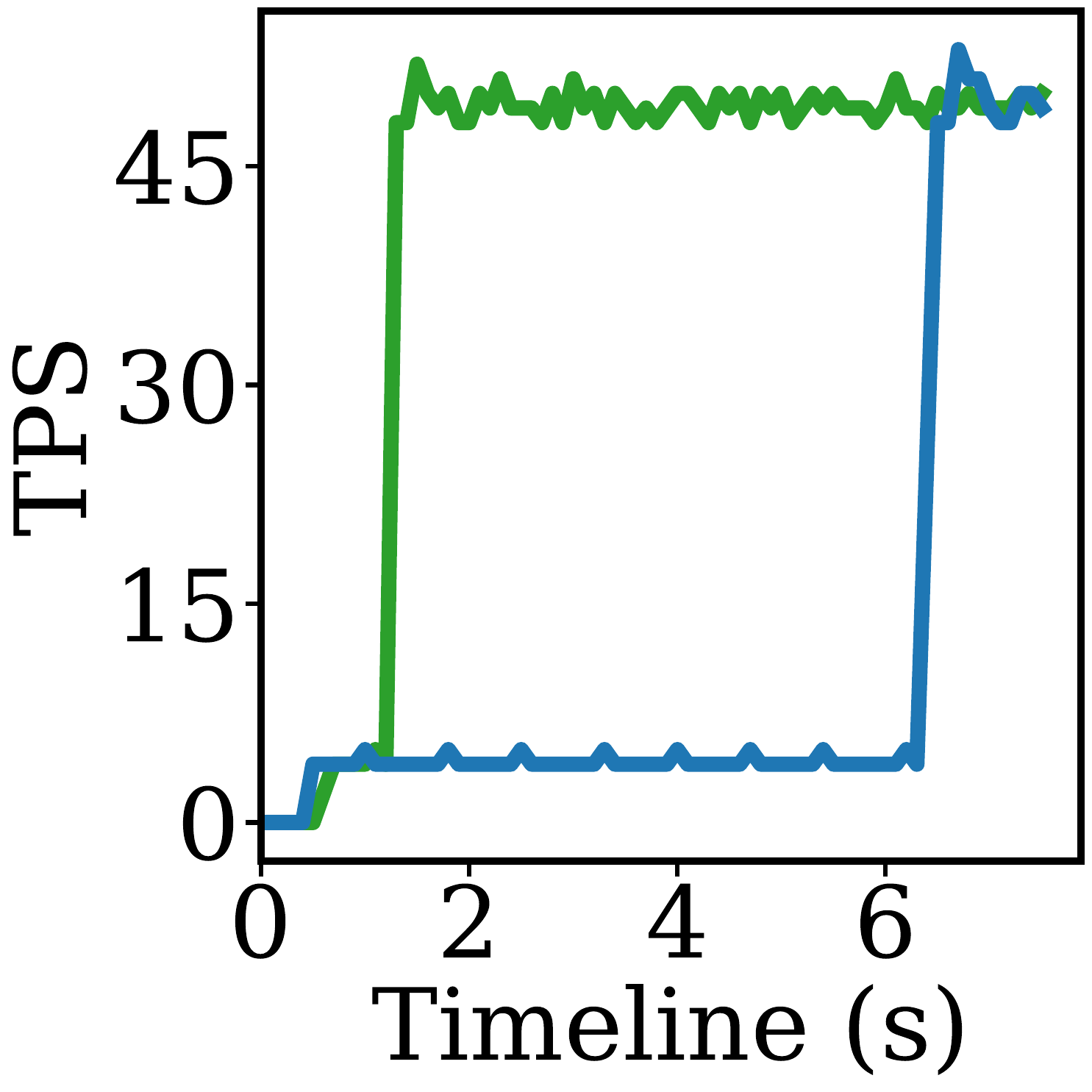}\label{fig:throughput_cold_start_13b}
}
\subfigure[{\small\texttt{Llama-2 70B}}] {
    \includegraphics[scale=0.1]
    {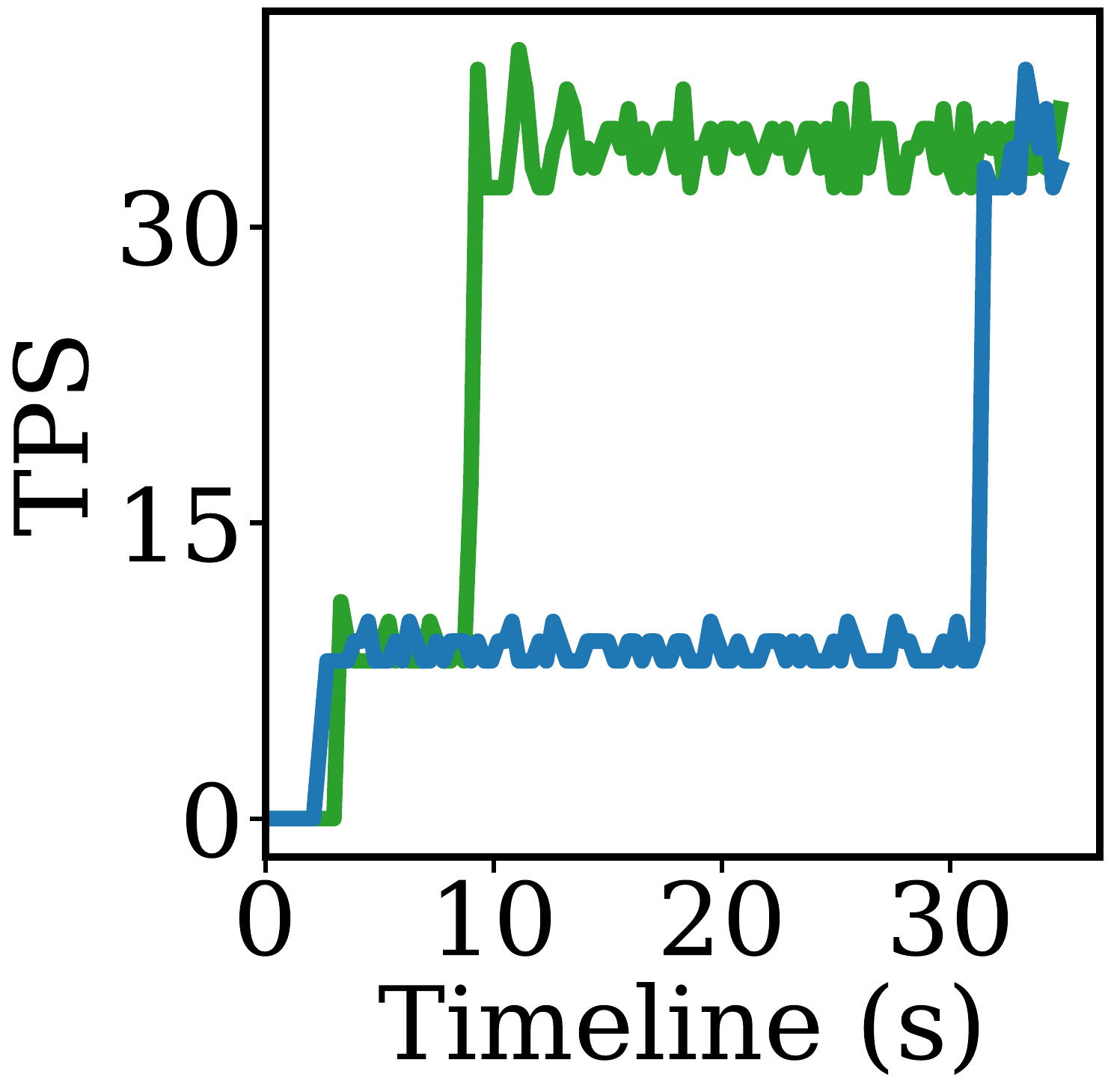}\label{fig:throughput_cold_start_70b}
}
\vspace{-10pt}
\caption{Cold-start throughput (scaling via GDR without models preloaded to GPUs, with $k=1$). 
}
\label{fig:throughput_cold_start}
\vspace{-5pt}
\end{figure}

\PHM{Cold-start comparison. } 
We evaluate throughput performance under a cold-start scenario, where no model instances are preloaded into GPUs across the cluster. Instead, one node stores the model in its host memory cache. 
In \textit{ServerlessLLM}, the remaining nodes load the model from SSD, while in \SysName, the cached model is loaded into local GPU(s) and multicasted across the cluster via GDR.
To ensure fairness, both systems use \emph{k} = 1. 
As shown in Fig.~\ref{fig:throughput_cold_start},  \SysName significantly outperforms \textit{ServerlessLLM} by $3.75\times$ to $11.4\times$ across the three model sizes. This improvement is due to \SysName's \AlgoName, which allows immediate request serving as soon as the first block is loaded into GPU, eliminating unnecessary delays.


\subsection{Latency Performance}
\label{subsec:latency}

Next, we measure the TTFT latency across varying loads.  

\noindent{\textbf{Scaling via GDR. }
Fig.~\ref{fig:latency_gdr} (a)-(c) plots the TTFT latency and Fig.~\ref{fig:latency_gdr} (d)-(f) shows the zoomed-in CDF given a specific RPS (requests per second) level. 
We observe that \SysName starts serving all 50 requests in 1.1s, which is $2\times$, $1.4\times$, and $8\times$ faster than \textit{FaaSNet}, \textit{NCCL}, \textit{ServerlessLLM}, respectively. 
The 7B and 70B models exhibit a similar trend.
\textit{ServerlessLLM}-SSD suffers from a long-tail TTFT latency, caused by: 
(1) slow SSD I/Os during on-demand loading, 
and (2) delayed inference execution due to waiting for the entire model to be fully loaded into GPUs.

\PHM{Scaling via local cache. }
Similarly, Fig.~\ref{fig:latency_cache} (a)-(c) shows the TTFT latency of \SysName and \textit{ServerlessLLM} when scaling with local memory cache and Fig.~\ref{fig:latency_cache} (d)-(f) shows the zoomed-in CDF, under the same setup used in Fig.~\ref{fig:throughput_cache} (\S\ref{subsec:throughput}). 
As shown in Fig.~\ref{fig:latency_cache_13b}, \SysName is $1.63\times$ faster than \textit{ServerlessLLM} at the $90^{th}$ percentile of requests. 
Even in the best-case scenario for \textit{ServerlessLLM}, where local memory caching is used, \SysName still outperforms \textit{ServerlessLLM} in TTFT. 
This is, once again, due to \textit{ServerlessLLM}'s lack of transmission and execution pipeline.

\begin{figure}[t]
\centering
\includegraphics[scale=0.1]{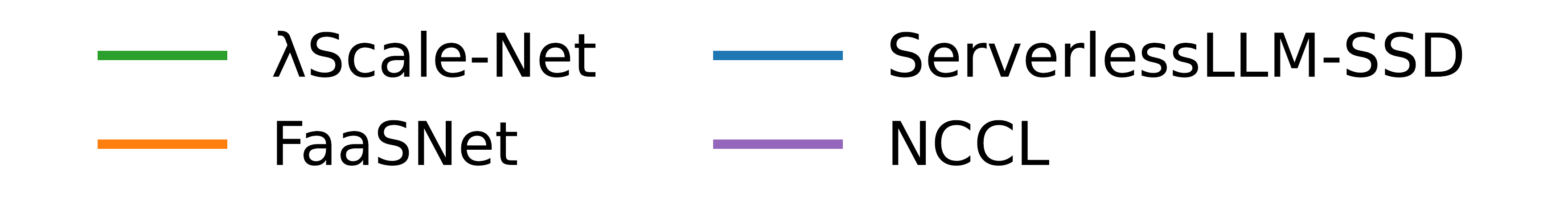} \\
\vspace{-5pt}
\subfigure[{\small\texttt{Llama-2 7B }}] {
    \includegraphics[scale=0.1]
    {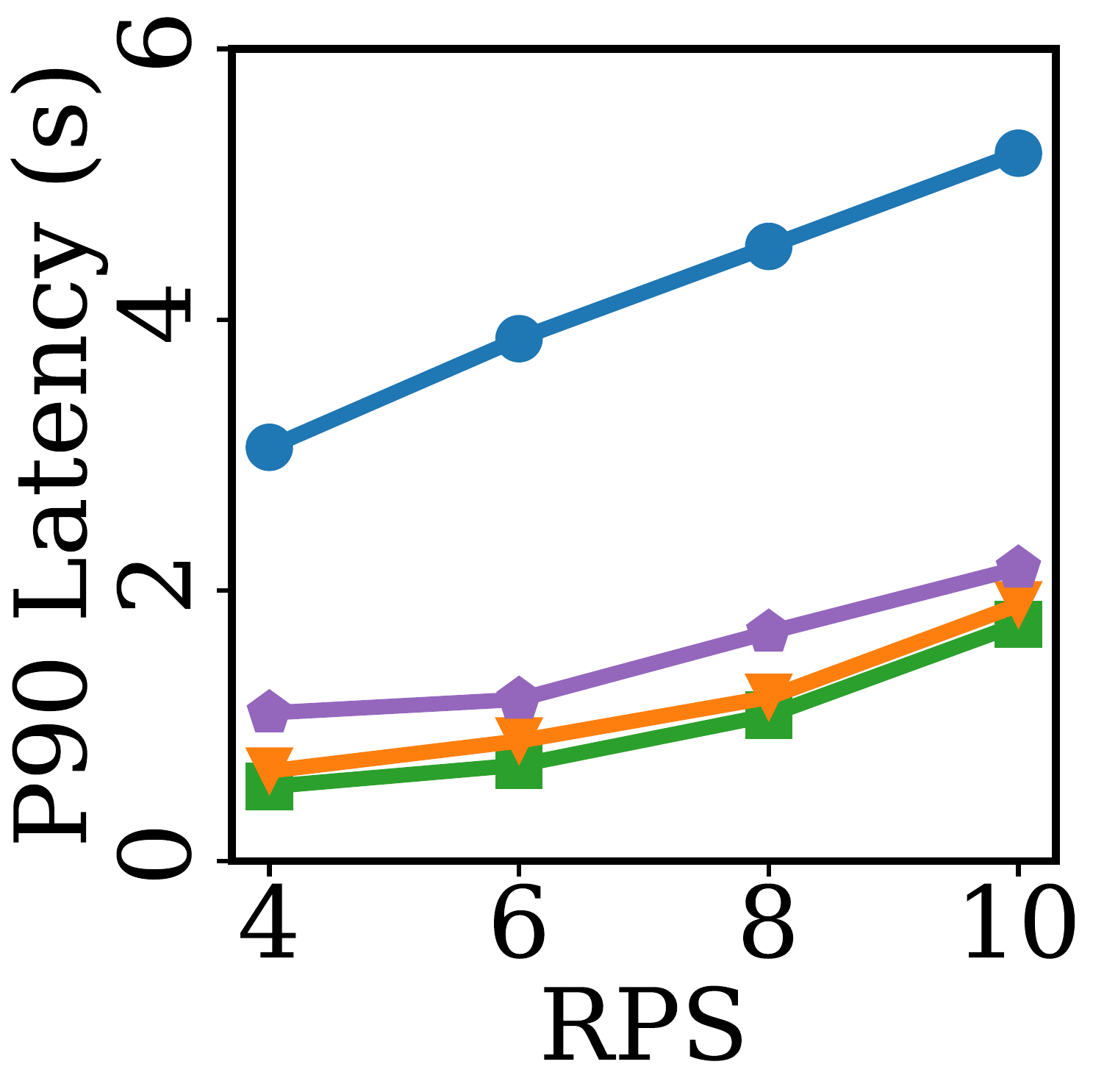}
    \label{fig:latency_gdr_7b}
}
\subfigure[{\small\texttt{Llama-2 13B}}] {
    \includegraphics[scale=0.1]{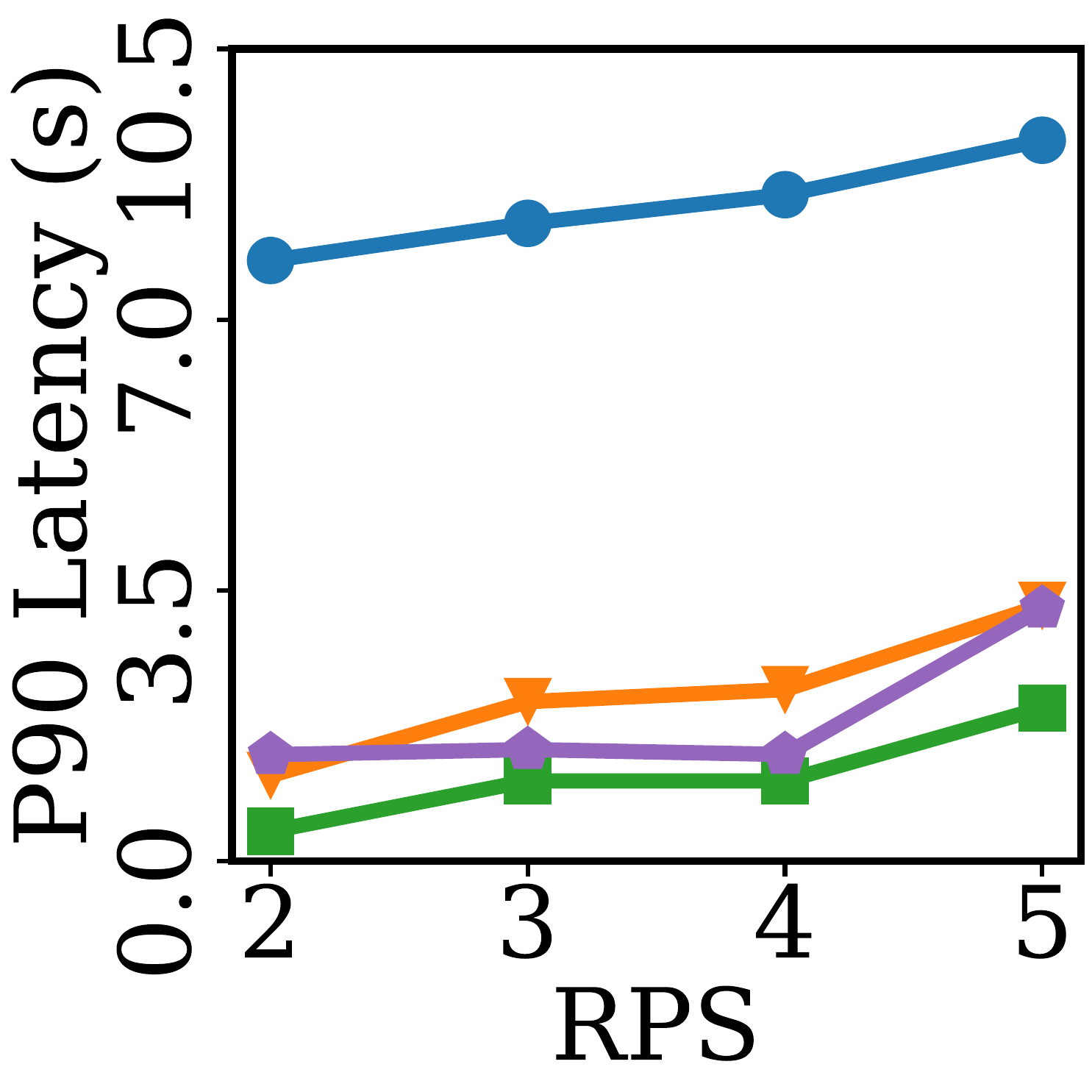}\label{fig:latency_gdr_13b}
}
\subfigure[{\small\texttt{Llama-2 70B}}] {
    \includegraphics[scale=0.1]
    {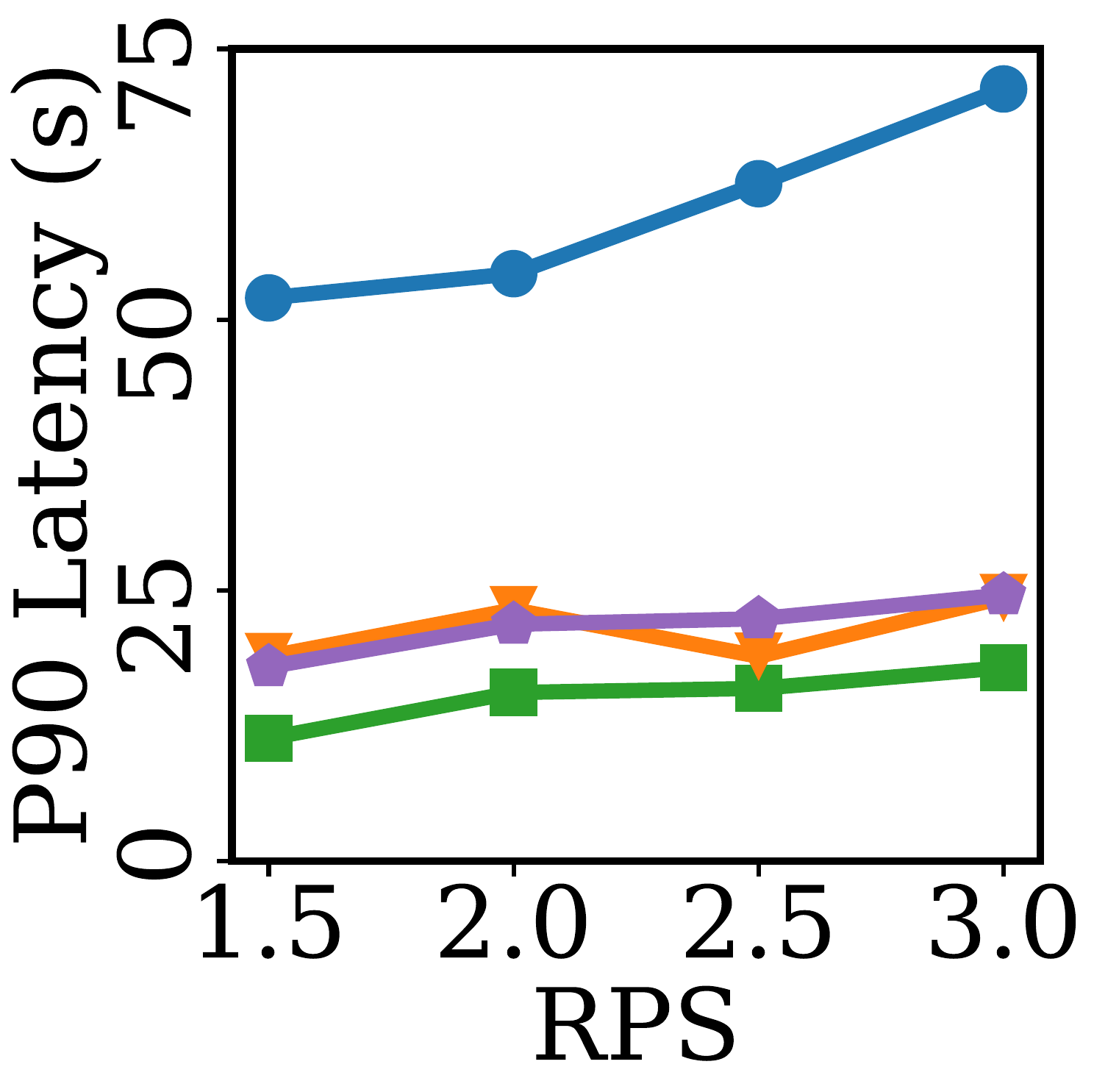}\label{fig:latency_gdr_70b}
}
\subfigure[\small
\begin{tabular}{c}
\texttt{Llama-2 7B}\\
\texttt{RPS=6}
\end{tabular}] {
    \includegraphics[scale=0.1]
    {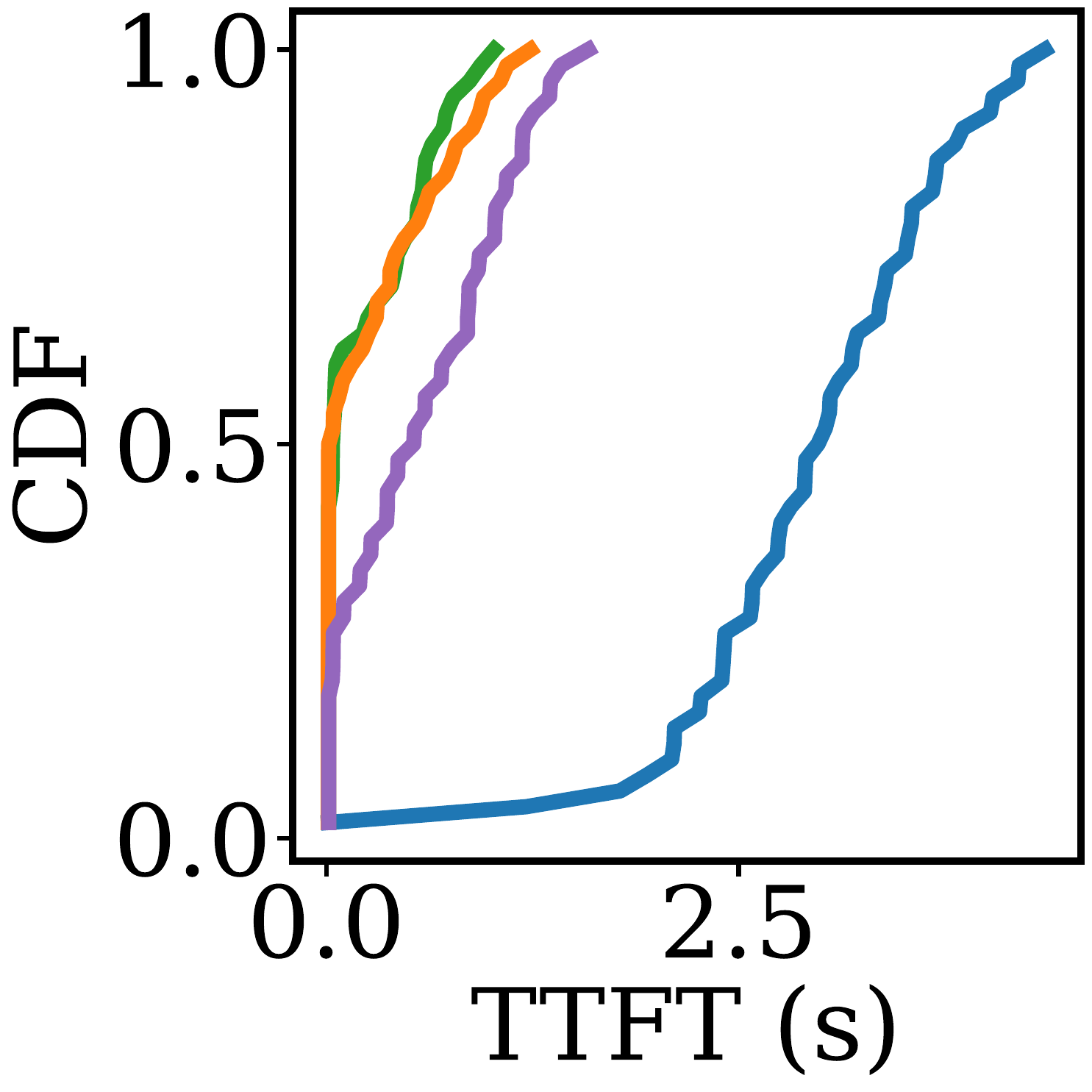}\label{fig:latency_gdr_7b_rps6}
}
\subfigure[\small
\begin{tabular}{c}
\texttt{Llama-2 13B}\\
\texttt{RPS=3}
\end{tabular}] {
    \includegraphics[scale=0.1]
    {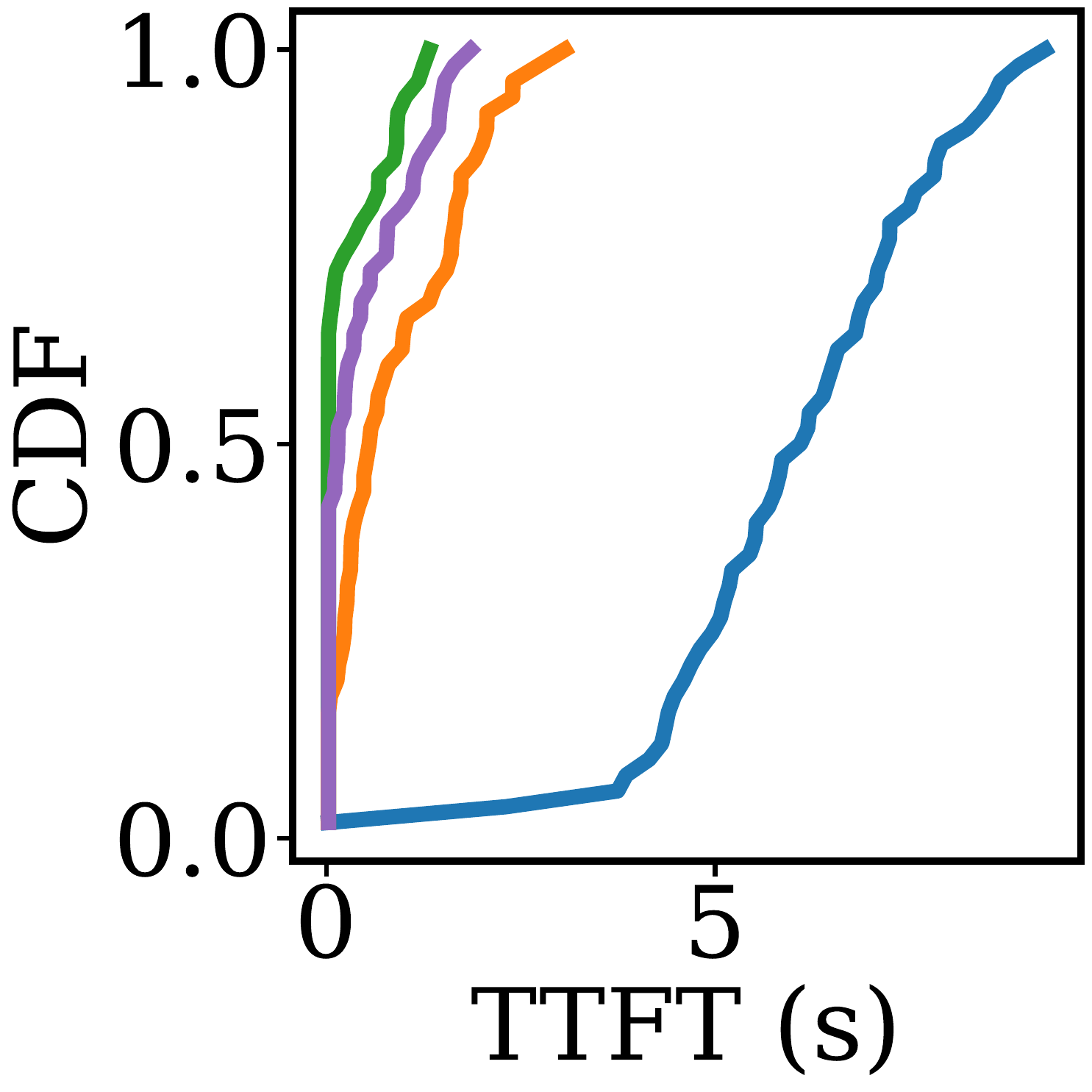}\label{fig:latency_gdr_13b_rps3}
}
\subfigure[\small
\begin{tabular}{c}
\texttt{Llama-2 70B}\\
\texttt{RPS=1.5}
\end{tabular}] 
{
    \includegraphics[scale=0.1]
    {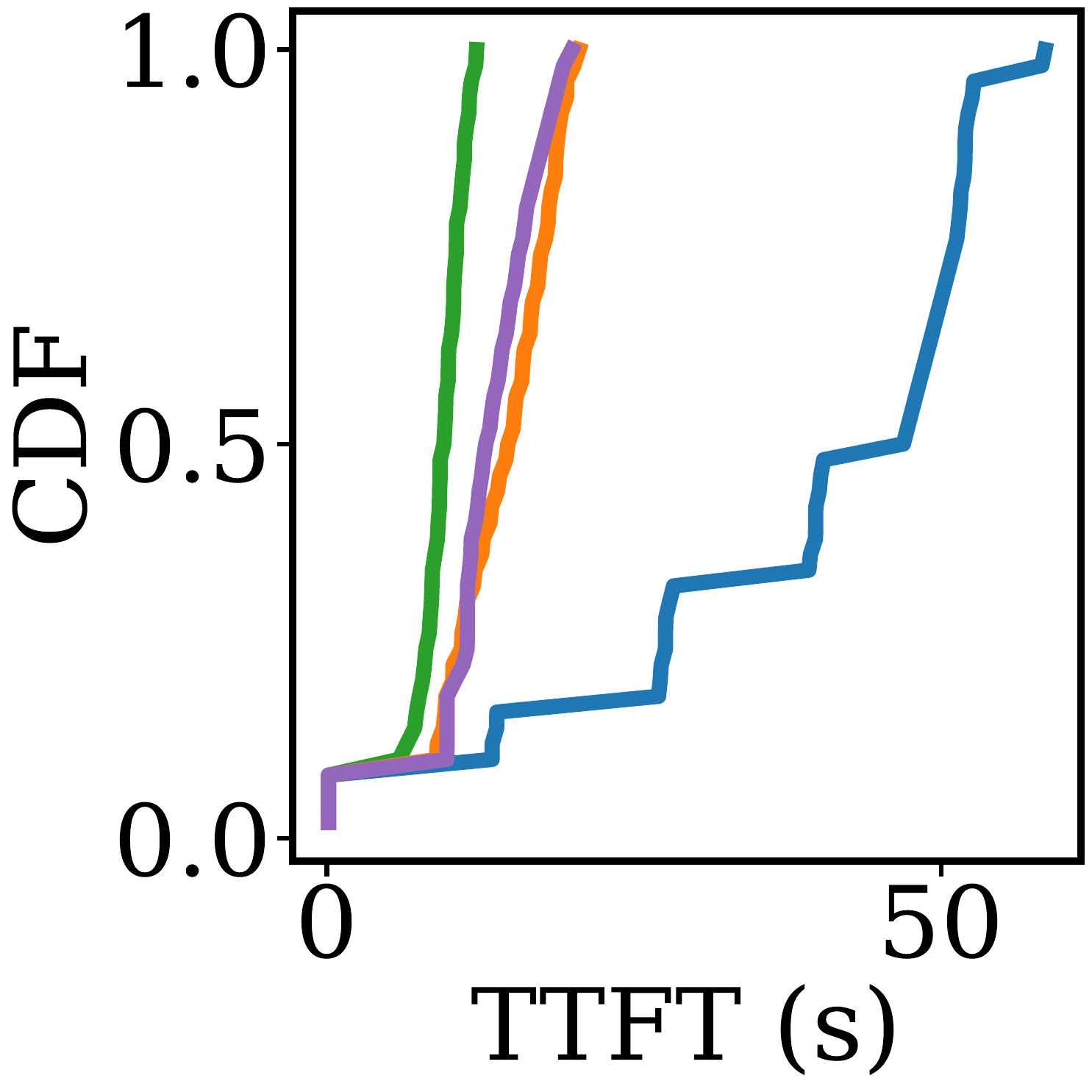}\label{fig:latency_gdr_70b_rps1.5}
}
\vspace{-1.5em}
\caption{Latency scaling via GDR.
}
\label{fig:latency_gdr}
\vspace{-10pt} 
\end{figure}

\begin{figure}[t]
\centering
\includegraphics[scale=0.1]{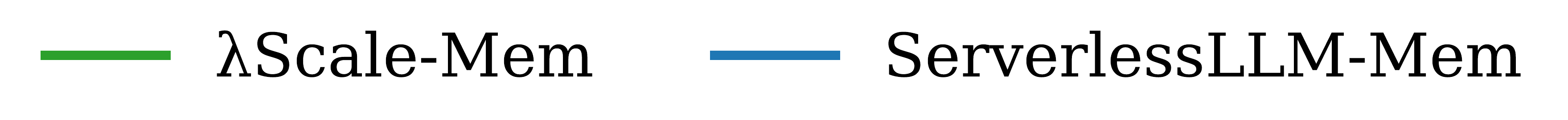} \\
\vspace{-0.5em}
\subfigure[{\small\texttt{Llama-2 7B}}] {
    \includegraphics[scale=0.1]
    {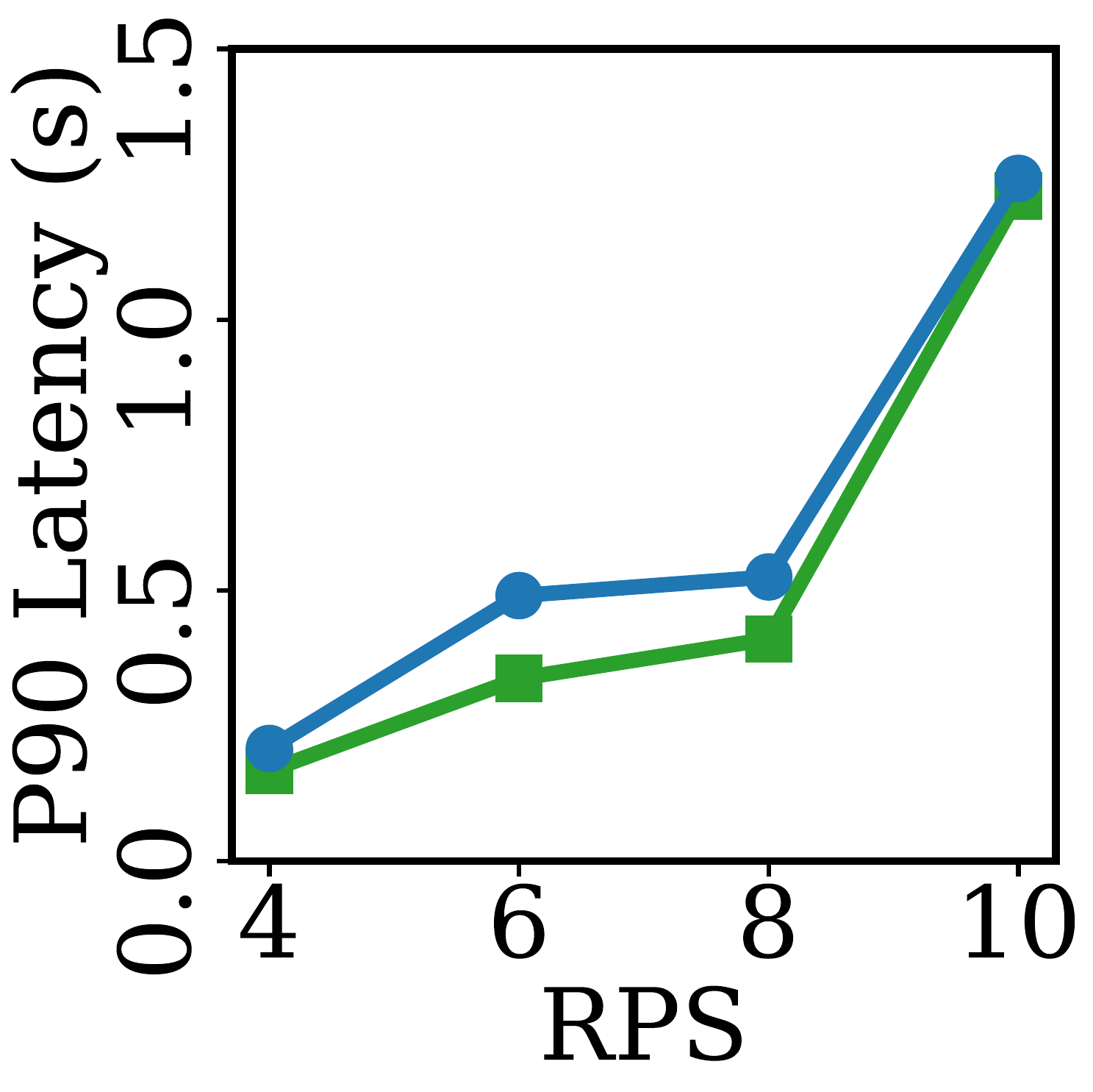}\label{fig:latency_cache_7b}
}
\subfigure[{\small\texttt{Llama-2 13B}}] {
    \includegraphics[scale=0.1]
    {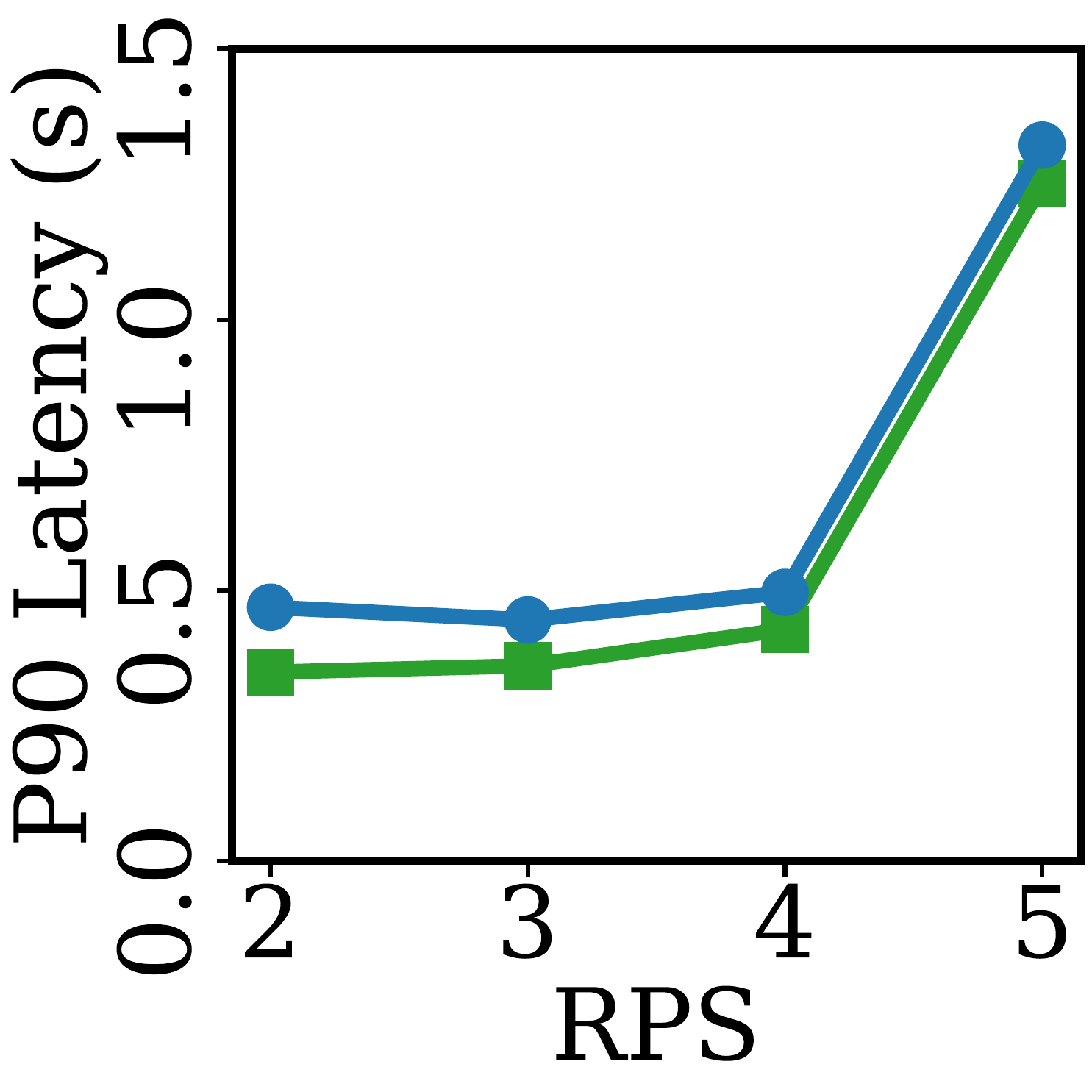}\label{fig:latency_cache_13b}
}
\subfigure[{\small\texttt{Llama-2 70B}}] {
    \includegraphics[scale=0.1]
    {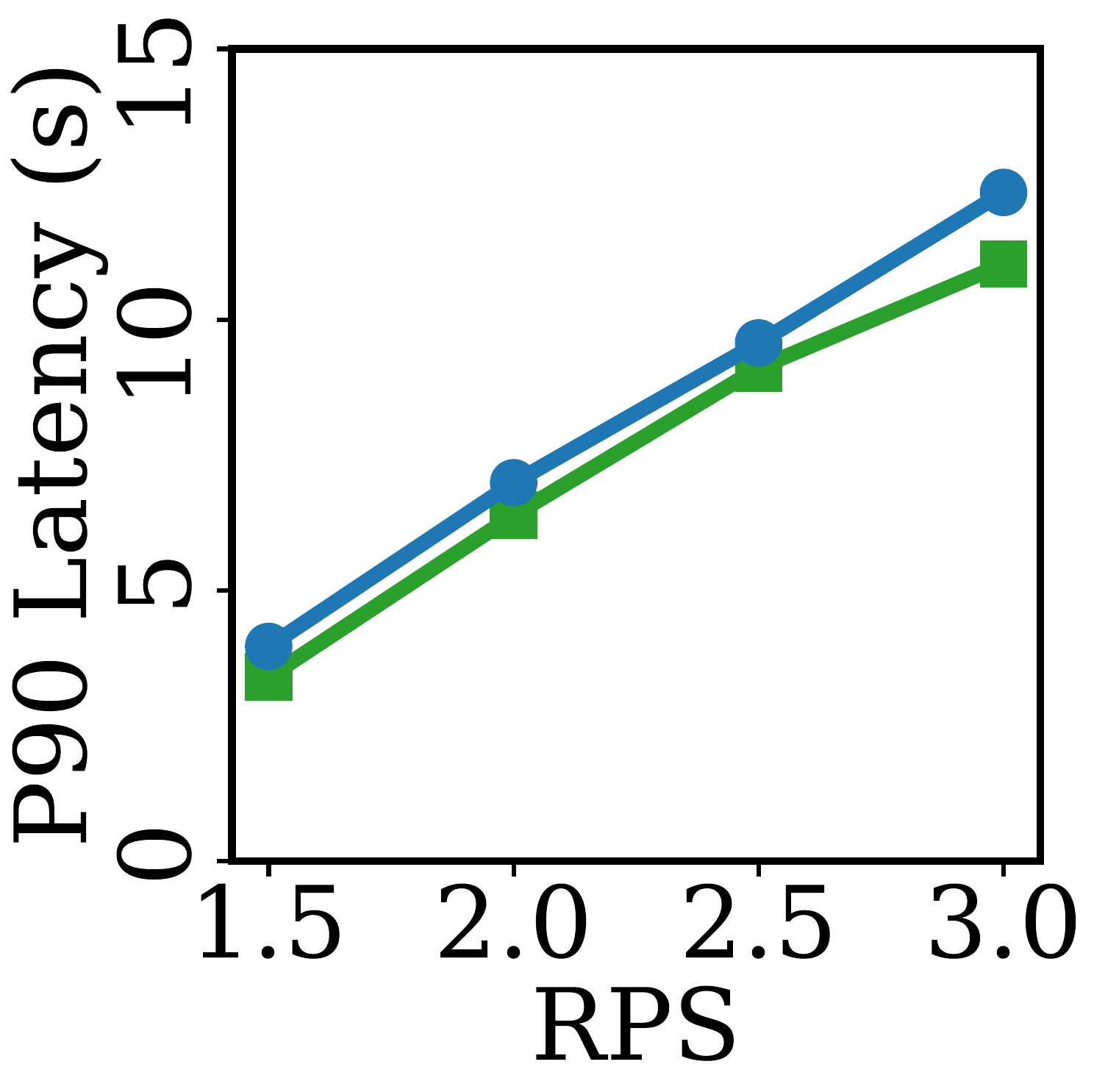}\label{fig:latency_cache_70b}
}
\subfigure[\small
\begin{tabular}{c}
\texttt{Llama-2 7B}\\
\texttt{RPS=6}
\end{tabular}] {
    \includegraphics[scale=0.1]
    {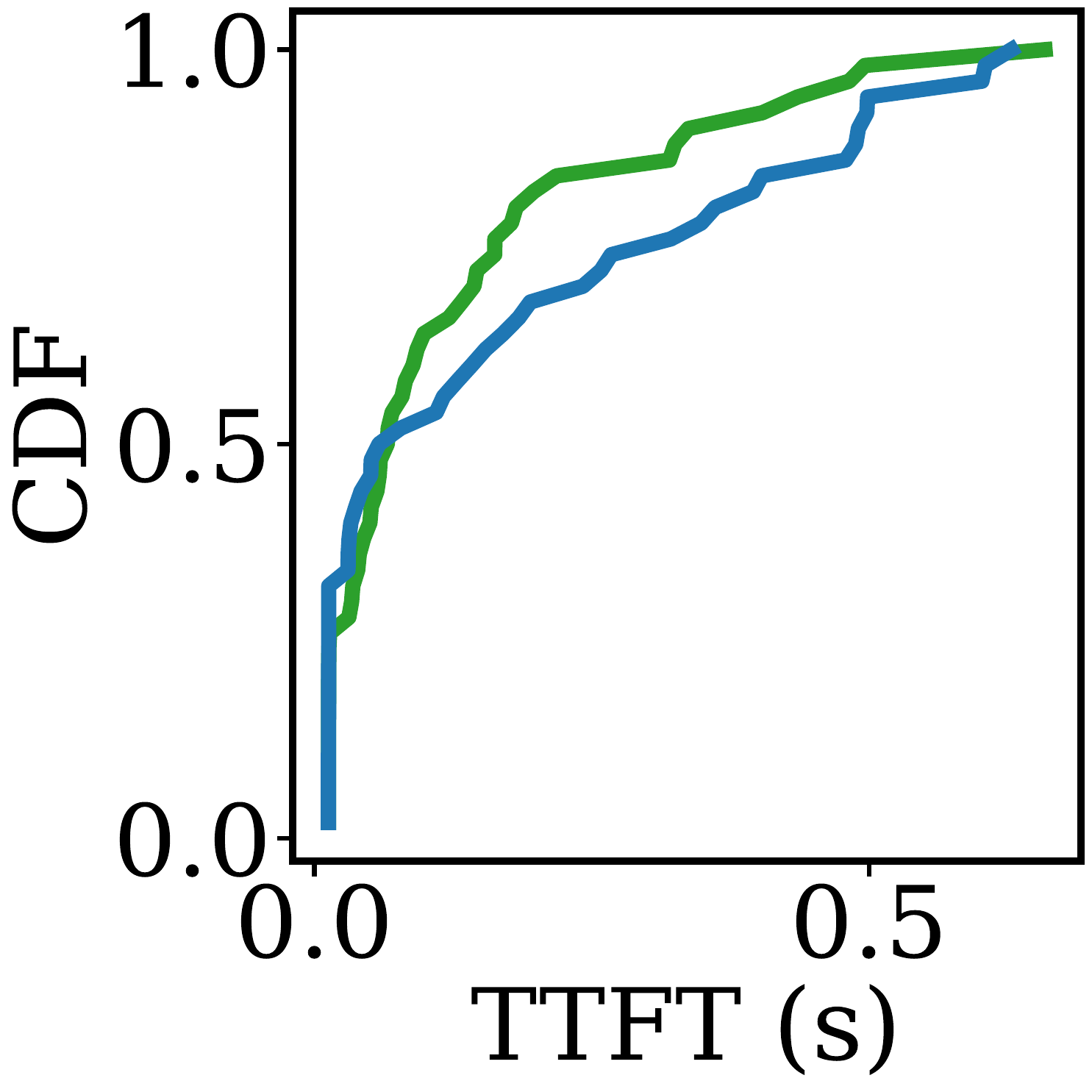}\label{fig:latency_cache_7b_rps6}
}
\subfigure[\small
\begin{tabular}{c}
\texttt{Llama-2 13B}\\
\texttt{RPS=3}
\end{tabular}] {
    \includegraphics[scale=0.1]
    {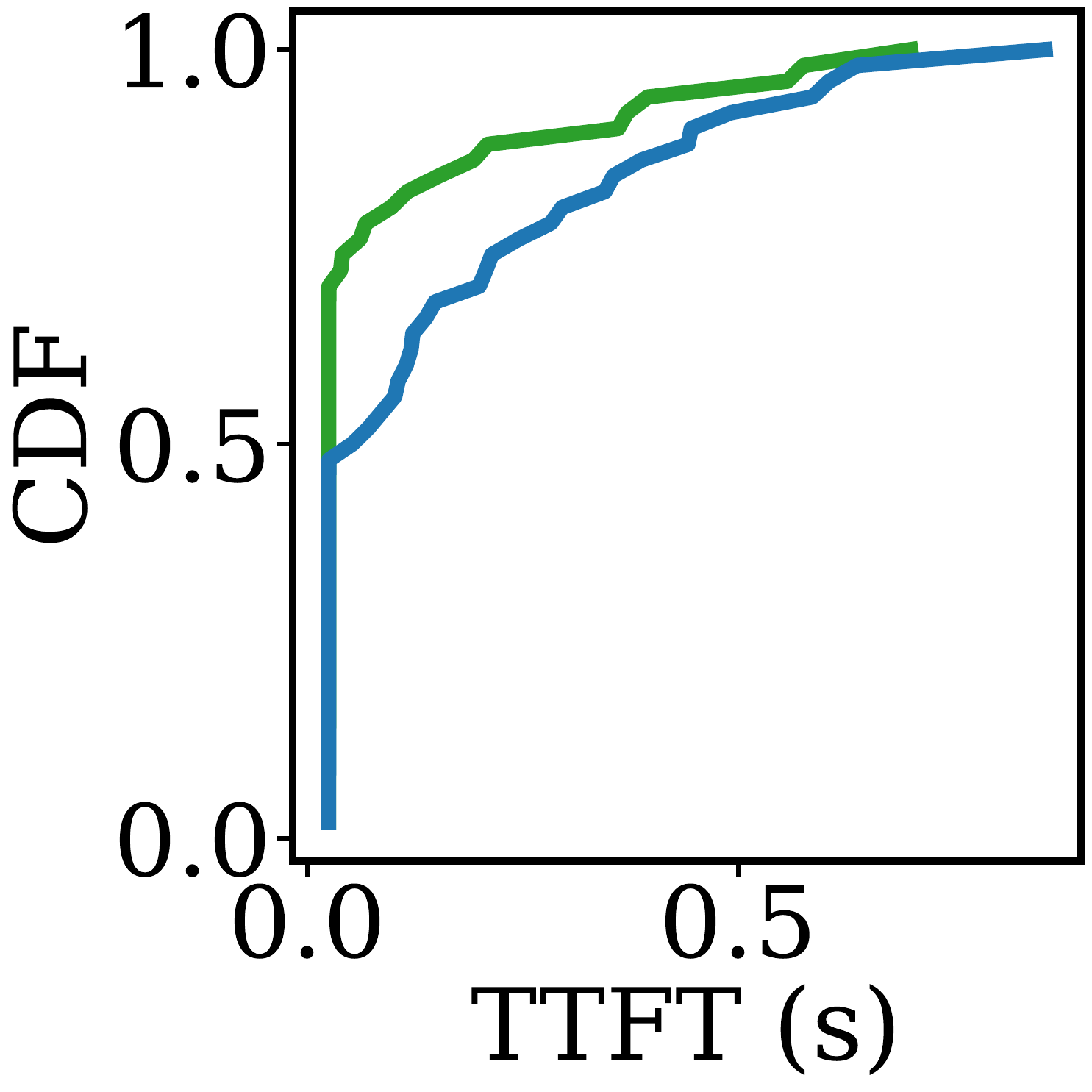}\label{fig:latency_cache_13b_rps3}
}
\subfigure[\small
\begin{tabular}{c}
\texttt{Llama-2 70B}\\
\texttt{RPS=1.5}
\end{tabular}] 
{
    \includegraphics[scale=0.1]
    {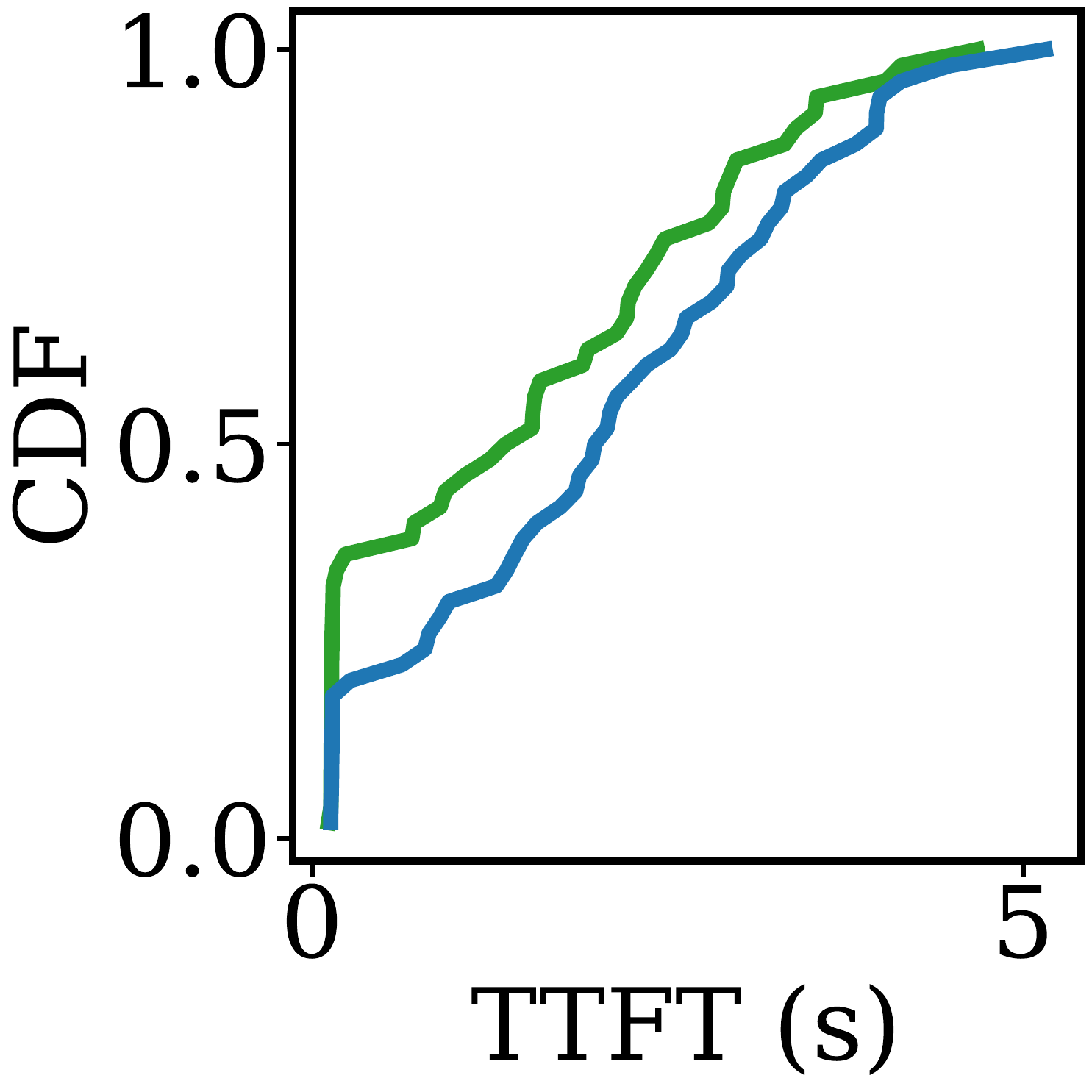}\label{fig:latency_cache_70b_rps1.5}
}
\vspace{-1.5em}
\caption{Latency scaling via local cache. 
}
\label{fig:latency_cache}
\vspace{-1.5em}
\end{figure}

\begin{figure*}[t]
\centering
\includegraphics[scale=0.1]{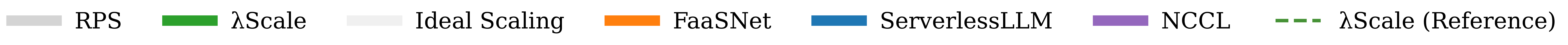} \\
\vspace{-0.5em}
\subfigure[{\small\texttt{Llama-2 7B}}] {
    \includegraphics[scale=0.123]
    {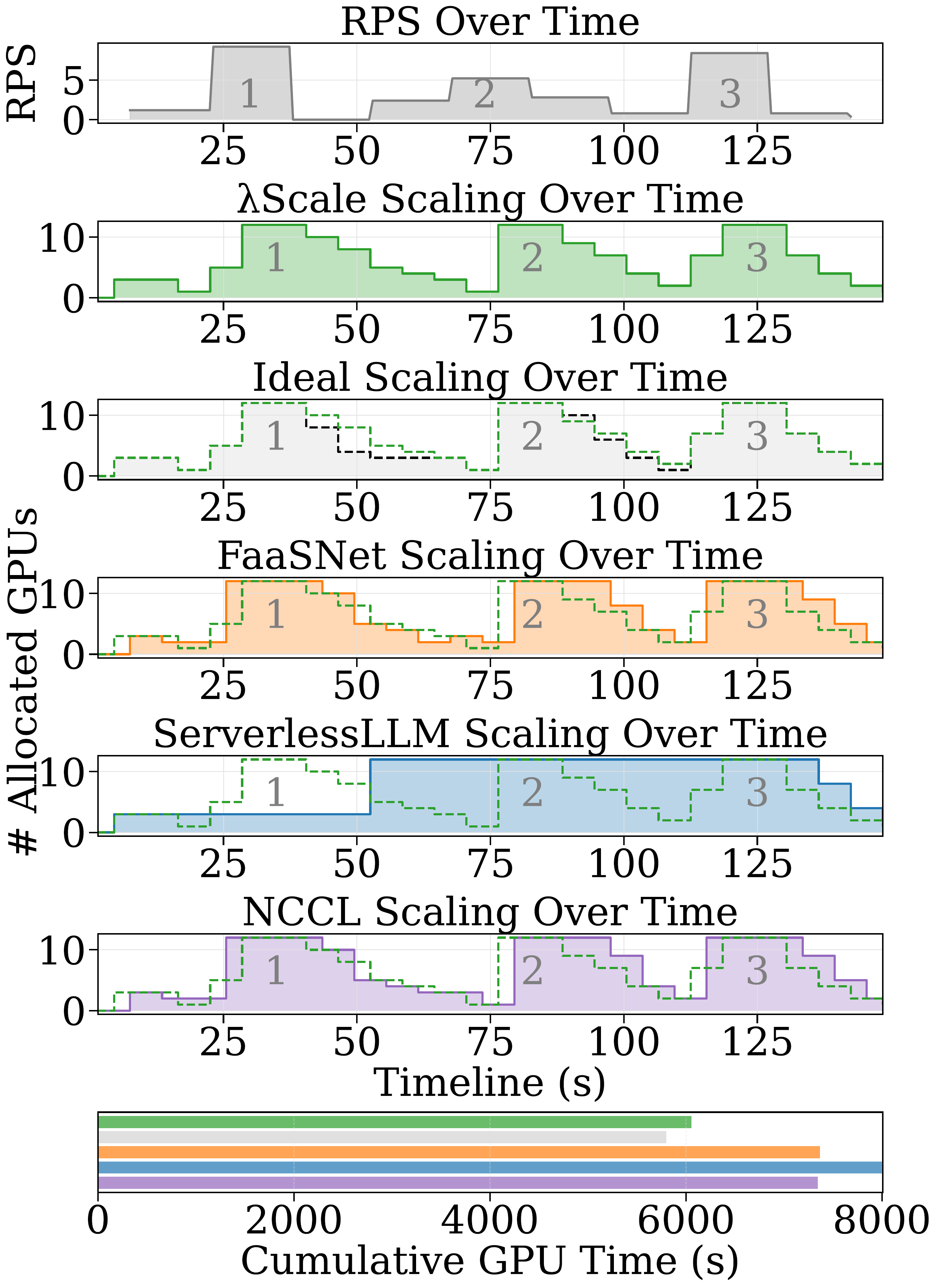}\label{fig:e2e_sys_behavior_7b}
}
\subfigure[{\small\texttt{Llama-2 13B}}] {
    \includegraphics[scale=0.123]
    {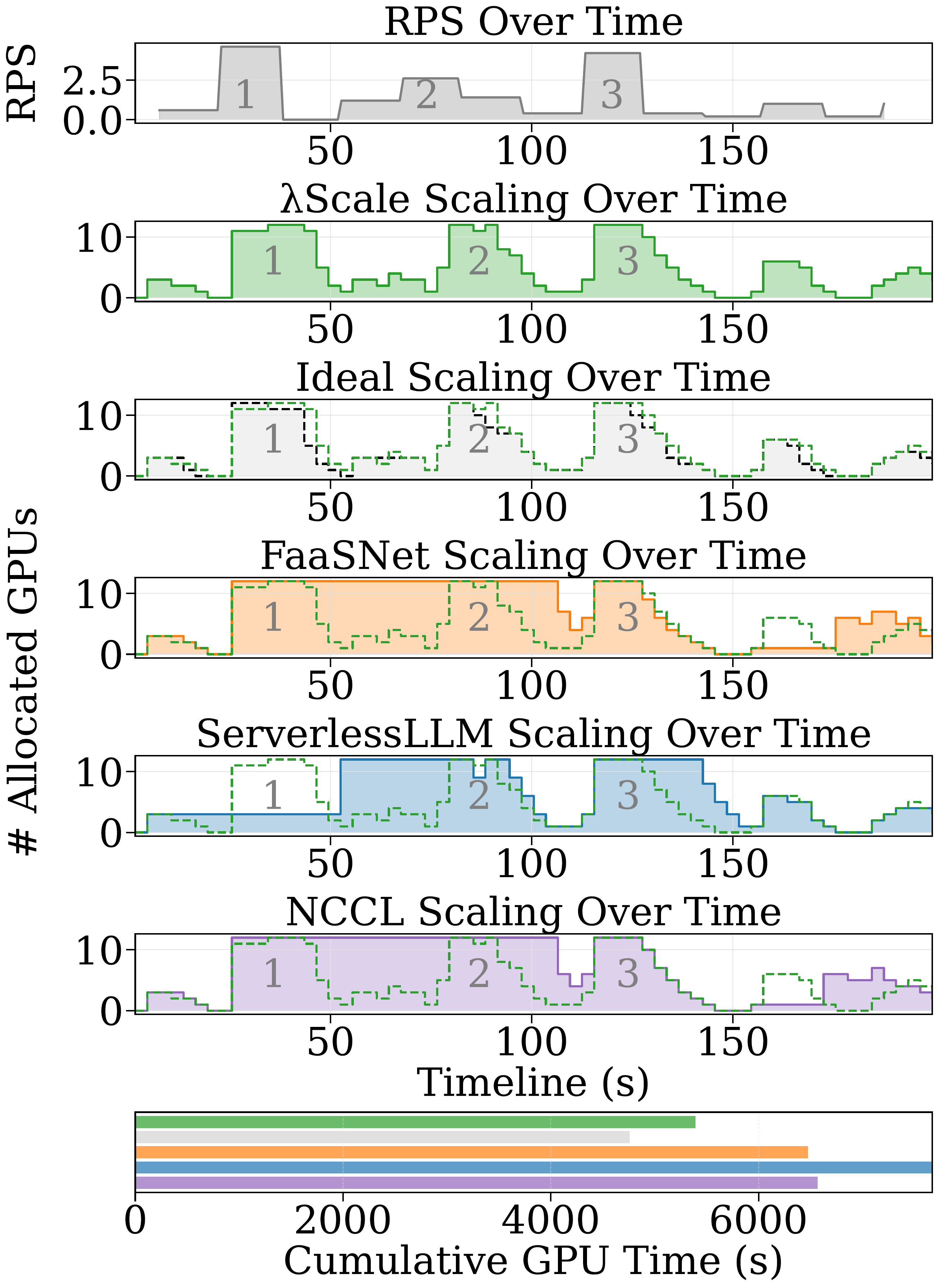}\label{fig:e2e_sys_behavior_13b}
}
\subfigure[{\small\texttt{Llama-2 70B}}] {
    \includegraphics[scale=0.123]
    {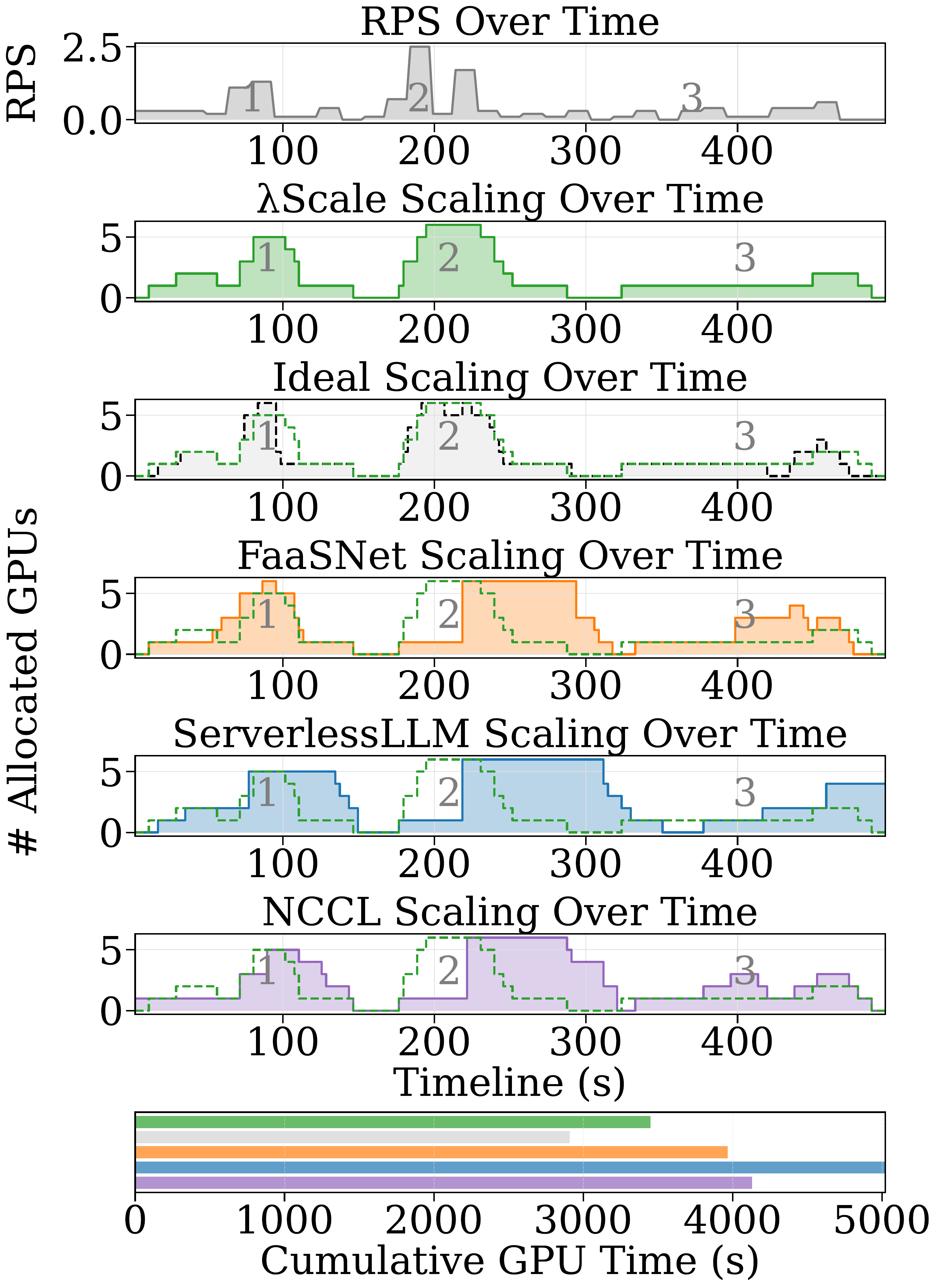}\label{fig:e2e_sys_behavior_70b}
}
\vspace{-10pt}
\caption{GPU allocation over time under a 30-minute BurstGPT workload.  
~\textit{\textmd{Top: RPS over time. 
Middle (Row 2-5): System-specific results sharing the same timeline. 
Bottom: Cumulative GPU consumption of all systems.  
The RPS timeline include labeled spikes (Top). 
The green dotted line shadows \SysName's scaling behavior, allowing direct comparison against other systems.   
}}}
\label{fig:e2e_sys_behavior}
\vspace{-5pt} 
\end{figure*}

\subsection{Real-World LLM Workload}
\label{subsec:e2e}

In this section, we evaluate \SysName's performance using BurstGPT~\cite{burstGPT_arxiv24}, a real-world LLM workload trace collected from regional Azure OpenAI GPT services. The original workload is highly bursty and we select a 30-minute trace snippet from the workload for evaluation.
We make the following assumptions:
\textit{NCCL} and \textit{FaaSNet} prioritize loading models from remote GPUs using GDR and only fall back to local SSD load if none of the GPUs in the cluster have an available model instance. 
\textit{ServerlessLLM} relies solely on local-cache-based loading---it loads models from host memory on a cache hit and from SSD on a cache miss.
As \textit{ServerlessLLM}, \SysName supports best-effort local host memory caching but falls back to \AlgoName multicast if the requested model is not in host memory. 

\PHM{Scaling behaviors.} 
Fig.~\ref{fig:e2e_sys_behavior} shows the dynamic GPU allocation timeline in response to fluctuating RPS. 
\textit{Ideal Scaling} assumes zero model-loading overhead, where a model can be instantly loaded into and swapped out of GPU(s) without delay. This is not practically achievable due to real-world constraints such as resource limitation and data transfer cost. 
\SysName scales out and in significantly faster than other systems across three model sizes. 
All three baselines experience delayed scaling out and delayed scaling in when responding to workload spikes. 
To quantify the cost effectiveness of GPU resources, we measure the cumulative GPU time for each system. \SysName consumes up to $17.8\%$, $18.1\%$, and $31.3\%$ less GPU resource than \textit{FaaSNet}, \textit{NCCL}, and \textit{ServerlessLLM}, respectively. 
While \textit{Ideal Scaling} achieves the lowest cumulative GPU time, \SysName maintains the closest GPU consumption to this ideal case, with a small gap ranging from $4.3\%$ and $18.6\%$, demonstrating \SysName's superior auto-scaling capability. 

\begin{figure}[h]
\centering
\includegraphics[scale=0.1]{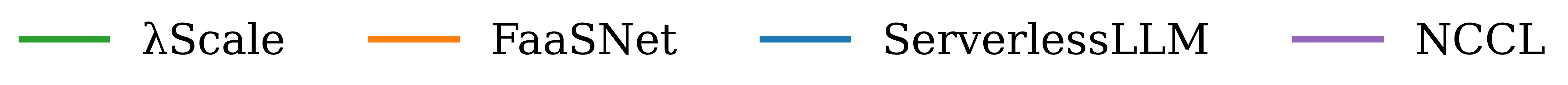} \\
\subfigure[{\small\texttt{Llama-2 7B}}] {
    \includegraphics[scale=0.1]
    {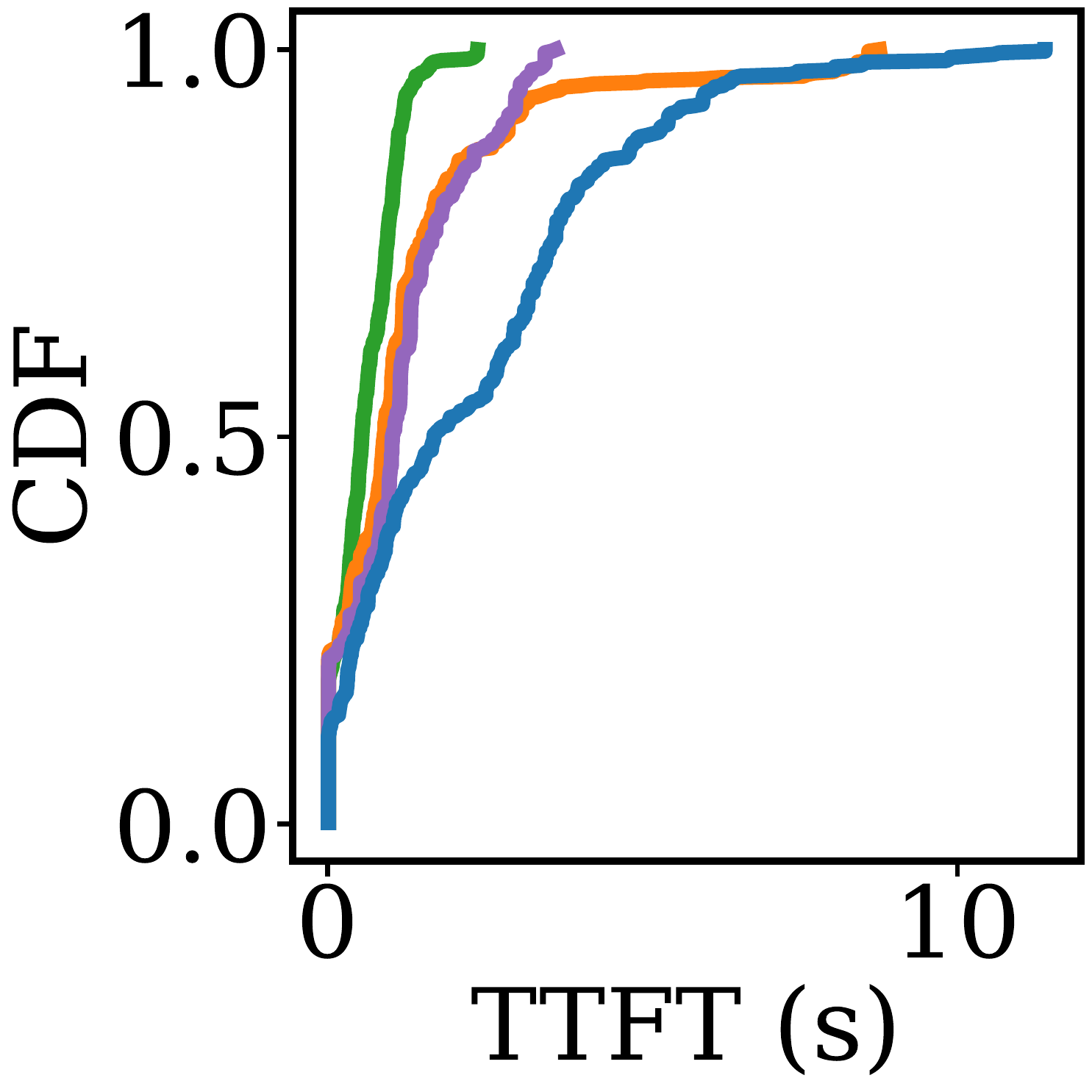}\label{fig:e2e_ttft_7b}
}
\subfigure[{\small\texttt{Llama-2 13B}}] {
    \includegraphics[scale=0.1]
    {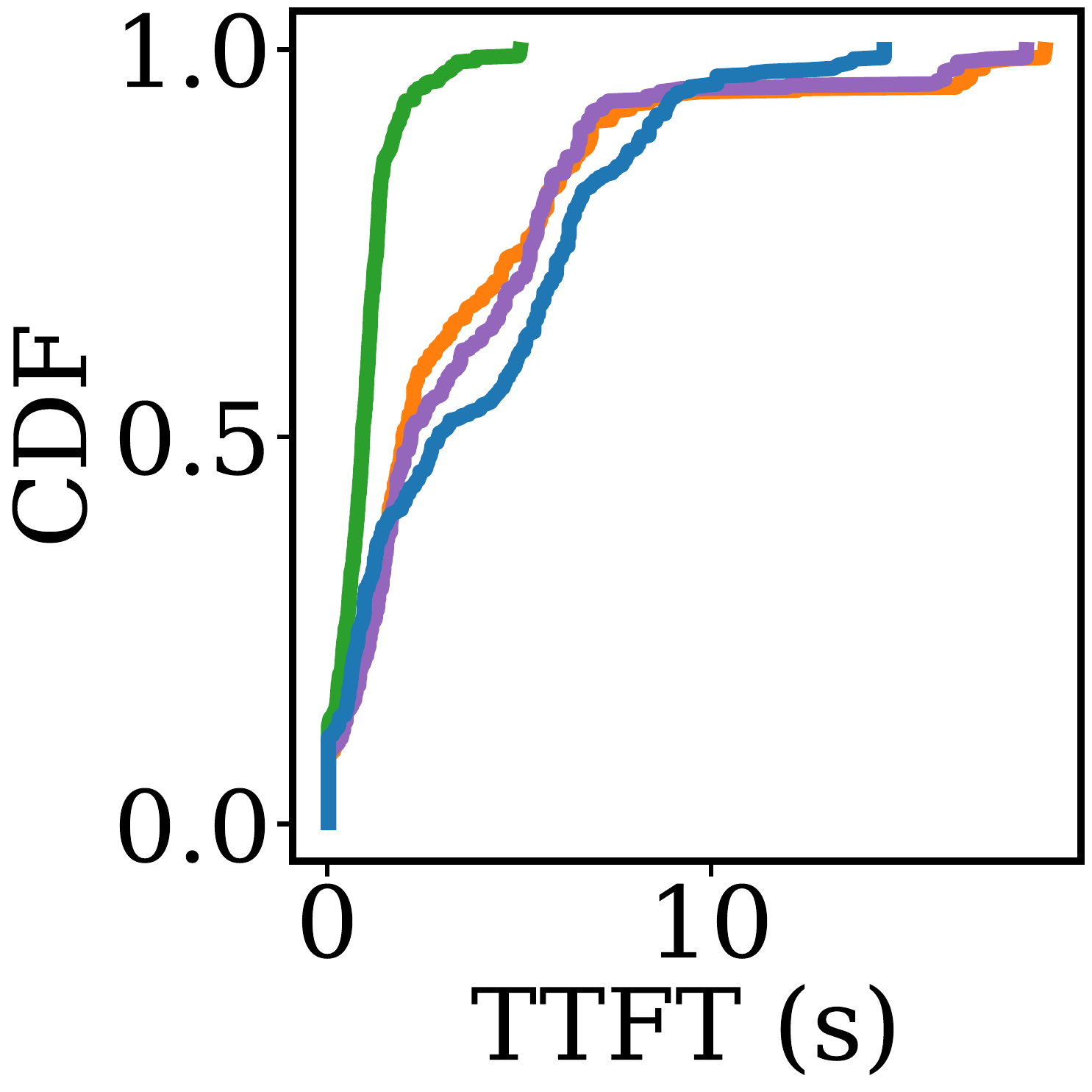}\label{fig:e2e_ttft_13b}
}
\subfigure[{\small\texttt{Llama-2 70B}}] {
    \includegraphics[scale=0.1]
    {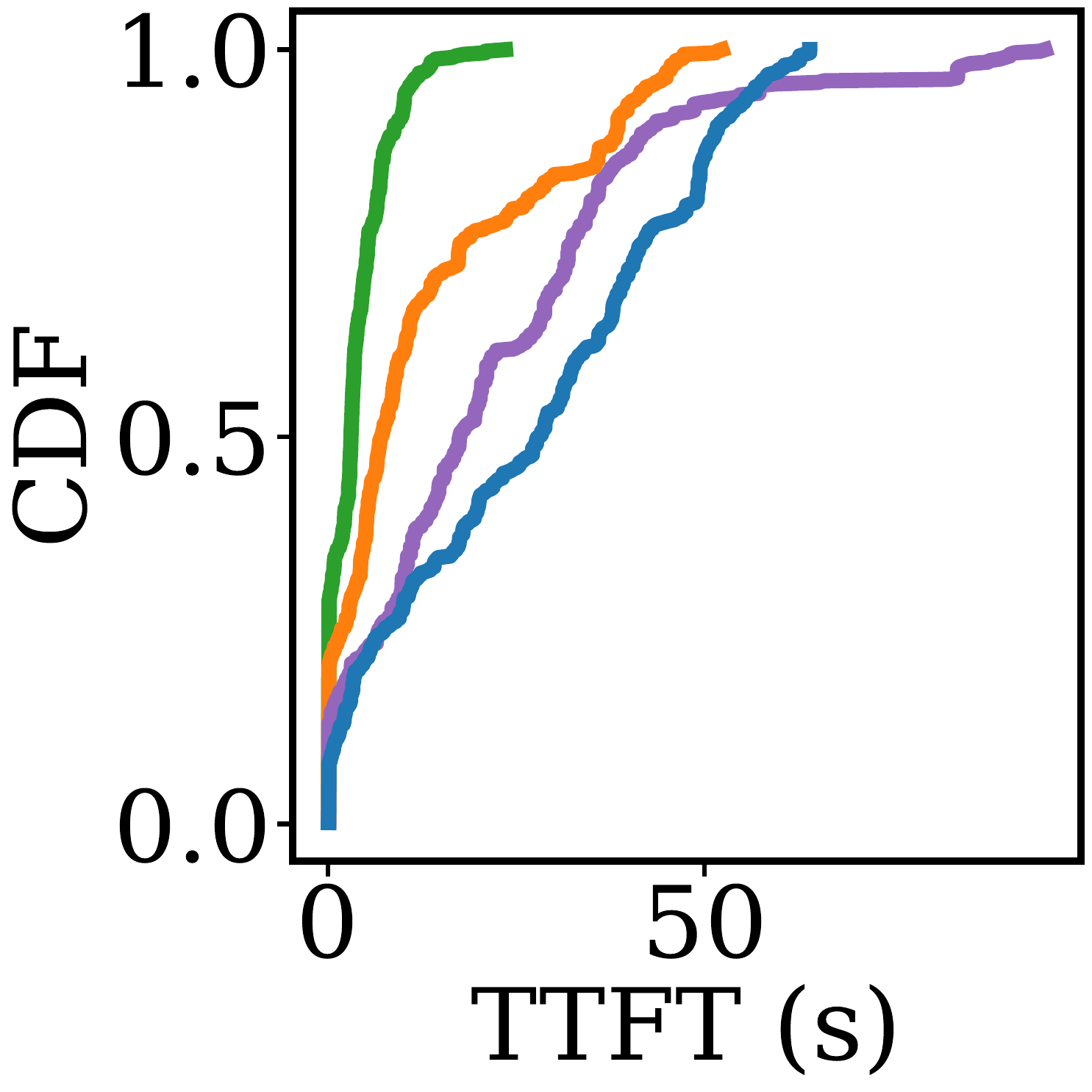}\label{fig:e2e_ttft_70b}
}
\vspace{-1em}
\caption{CDF of TTFT under the BurstGPT workload.
}
\label{fig:e2e_ttft}
\vspace{-10pt}
\end{figure}

\PHM{TTFT comparison.} 
Fig.~\ref{fig:e2e_ttft} shows the TTFT distribution, \SysName outperforms all baselines across the three models
Compared to previous TTFT results reported under static workloads (\S\ref{subsec:latency}), two key observations emerge. 
First, the CDF curves of \textit{NCCL} and \textit{FaaSNet} shift rightward, due to frequent model loading from SSD, leading to longer tail latency. 
Second, the CDF curve of \textit{ServerlessLLM} shifts leftward, likely due to a high cache hit rate when loading models into GPUs.

\subsection{Sensitivity Analysis and Ablation Study}
\label{subsec:ablation_study}

We conduct further experiments to evaluate the impact
of the number of transfer blocks, k-way transmission, and
system optimizations on \SysName’s performance. We fix
$k=1$ to isolate the effects of the number of blocks and system optimizations on transfer latency. For $k$-way transmission analysis, we vary $k$ to quantify its impact on LLM request serving.

\PHM{Impact of \emph{k}-way transmission on throughput. }
Fig.~\ref{fig:ablation_blk_ordering} compares 
\SysName's throughput performance under different \emph{k}-way transmission levels. \textit{\SysName-Net} (\emph{k}=4) achieves the fastest scaling, demonstrating that a higher \emph{K}-way transmission levels significantly enhance performance. In constract,  \textit{\SysName-Half-Reorder} (\emph{k}=2), scales more slowly, showing a moderate performance drop. Finally, \textit{\SysName-Non-Reorder} (\emph{k}=1) performs the worst, characterized by the slowest scaling and the lowest overall throughput.

\begin{figure}[ht]
\centering
\includegraphics[scale=0.125]
{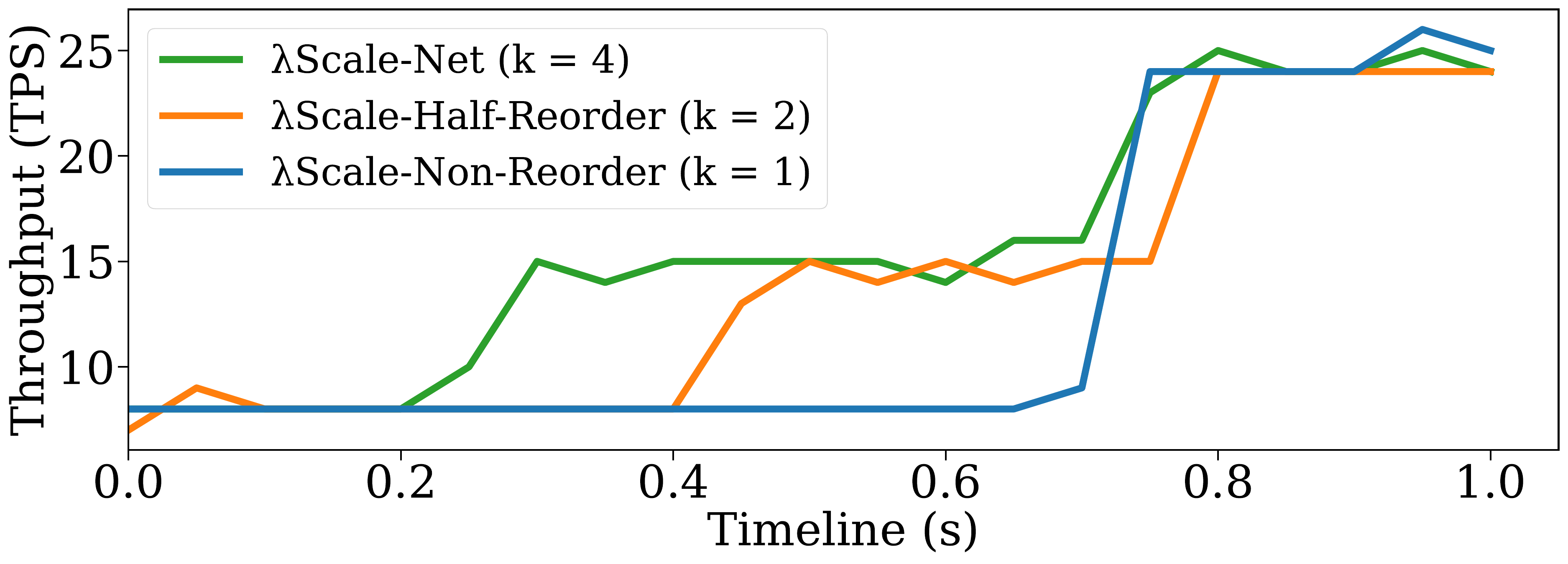}\label{fig:throughput_cache_13b}
\vspace{-10pt}\caption{Impacts of \emph{k}-way transmission on throughput performance.
}
\label{fig:ablation_blk_ordering}
\vspace{-10pt}
\end{figure}
\PHM{Effect of system optimizations on transfer latency. }
Fig.~\ref{fig:ablation_breakdown} shows the impact of various optimizations on transfer latency in \SysName.
The``None'' configuration, without any optimizations, shows the highest transfer latency, exceeding $20$ ms. 
Adding pre-allocation (``+Pre-alloc'') significantly reduces the latency, demonstrating its importance in minimizing allocation overhead.
Further improvement is observed with tensor packing (``+Tensor-pack''), which reduces latency by optimizing the data layout for efficient transfers. The lowest latency is achieved with the addition of  (``+Host-mem RDMA''). 
This progression underscores the cumulative benefits of these optimizations, with each step contributing to a substantial reduction in transfer latency.

\PHM{Impact of the number of model blocks on transfer latency. }
Fig.~\ref{fig:ablation_num_blocks} shows that 16 blocks achieve the lowest transfer latency, likely due to a balance between RDMA request processing overhead and efficient data transfer. Fewer blocks (e.g., 8) result in higher latency because larger block sizes increase RDMA request processing overhead. More blocks (e.g., 24–48) also increase latency, likely due to the additional overhead of managing and transferring a larger number of smaller blocks. Latency decreases up to 16 blocks but rises beyond this point, highlighting diminishing returns and added complexity.

\begin{figure}
    \centering
    \begin{minipage}{0.225\textwidth}
        \centering
        \includegraphics[scale=0.125]{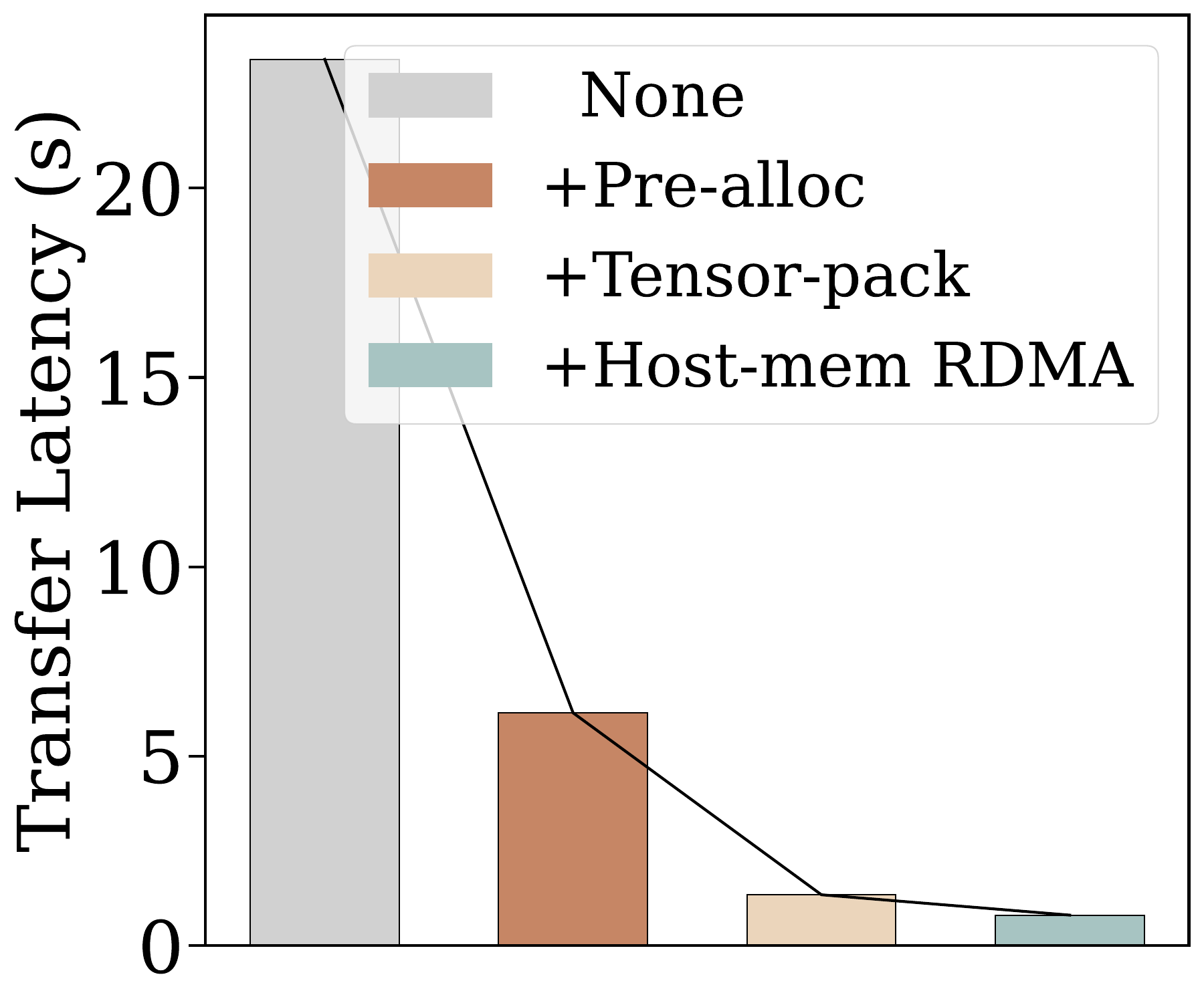}
        \vspace{-10pt}\caption{Performance breakdown of transfer latency.}
        \label{fig:ablation_breakdown}
    \end{minipage}\hfill
    \begin{minipage}{0.225\textwidth}
        \centering
        \includegraphics[scale=0.125]{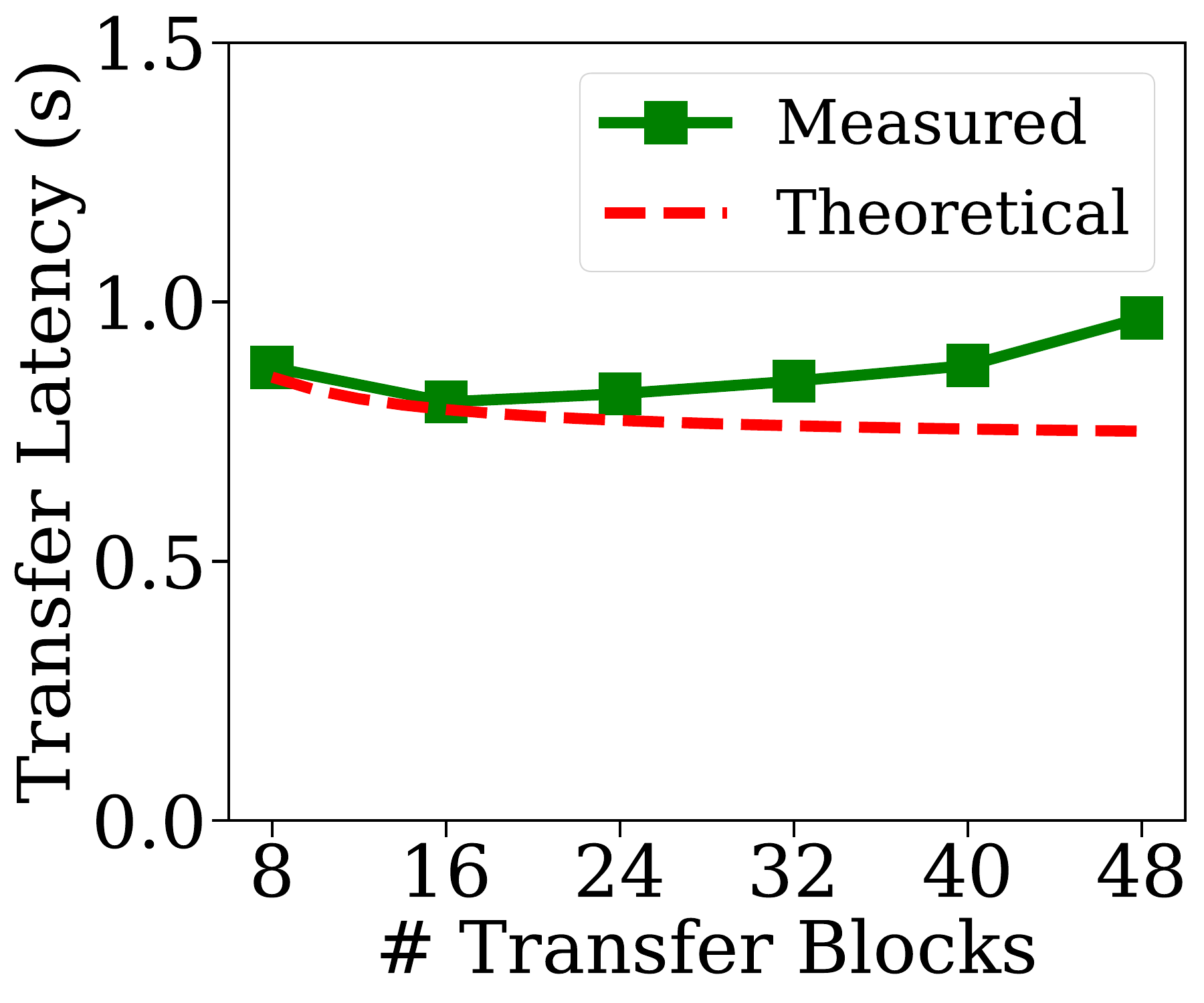}
        \vspace{-10pt}\caption{Latency with varying number of transfer blocks}
        \label{fig:ablation_num_blocks}
    \end{minipage}
    \vspace{-5pt}
\end{figure}

\section{Related Work and Discussion}
\label{sec:discussion}

\PHM{Pipelined inference execution.}
Prior works on pipelined inference execution typically rely on static resource configurations~\cite{crankshaw_clipper,shen_nexus_2019,dhakal_gslice_2020,bai_pipeswitch_nodate,li_alpaserve_2023}, whereas \SysName focuses on dynamic pipeline construction and execution using the ``execute-while-load'' approach.
BlitzScale~\cite{BlitzScale} is a close work to \SysName, which utilizes chain-based model scaling and primarily targets Prefill/Decoding (P/D) disaggregated LLM inference settings~\cite{zhong_distserve_2024}. 
\revise{\SysName targets a more general serverless environment and extends binomial pipeline~\cite{rdmc, binomial-pipe} for rapid LLM scaling.}

\PHM{Support for tensor parallelism.} 
In addition to Pipeline Parallelism (PP), \SysName can inherently support Tensor Parallelism (TP) or Hybrid Parallelism~\cite{SpotServe,hybrid_parallel_survey} by extending its model partitioning strategies.
By tracking block-level execution dependencies as a DAG, \SysName can effectively coordinate inference tasks under TP-based partitioning strategies, which we leave for future work.

\PHM{Support for other models.} 
Current \SysName supports a variety of mainstream models, including LLMs, image classification models, and vision transformers~\cite{dpt-beit-large-512,clip-vit-large-patch14,metainf}.
Enabling fast scaling for smaller models is much easier; therefore, they are excluded from our evaluation. 
Additionally, \SysName can be easily extended to support new models, provided that their structures are available for model partitioning.

\section{Conclusion}
\label{sec:conclusion}

This paper presents \SysName, a serverless inference system that achieves fast model scaling.
\SysName leverages high-speed RDMA networks for fast model multicast across GPU nodes and enables cross-node, collaborative inference execution during model transfer. 
\SysName realizes this ``execute-while-load'' approach through \AlgoName, a novel model scaling scheme that supports low-latency model loading and dynamic, pipelined inference execution. 
Combined with efficient model management, \SysName delivers superior scaling performance, effectively sustaining load spikes and achieving up to 5$\times$ tail-latency speedup over state-of-the-art systems.
%

\section*{Acknowledgement}


\revise{We thank the Derecho/RDMC team for their binomial pipeline multicast implementation and the guidance for our
work.}

\newpage


\bibliographystyle{plain}
\bibliography{references}
\clearpage

\end{document}